\documentclass[11pt,a4paper]{article}
\pdfoutput=1
\usepackage{jheppub}

\usepackage{xspace}
\usepackage{bbm} 

\newcommand{\eVdist}{\kern-0.06em}

\newcommand{\gev}{\:\text{Ge\eVdist V}}

\newcommand{\Z}[1]{\ensuremath{\mathbbm{Z}_{#1}}} 

\newcommand{\be}{\begin{equation}}
\newcommand{\ee}{\end{equation}}
\newcommand{\bea}{\begin{eqnarray}}
\newcommand{\eea}{\end{eqnarray}}

\newcommand{\SARAH}{{\tt SARAH}\xspace}
\newcommand{\SPheno}{{\tt SPheno}\xspace}
\newcommand{\MO}{{\tt MicrOmegas}\xspace}

\newcommand{\SSP}{{\tt SSP}\xspace}
\newcommand{\CalcHep}{{\tt CalcHep}\xspace}

\allowdisplaybreaks[1]

\title{A precision study of the fine tuning in the DiracNMSSM}

\preprint{\parbox{5cm}{\flushright DESY-13-256\\ OUTP-13-25p \\ CERN-PH-TH-2014-002 \\ BONN-TH-2014-01}}

\author[a]{Anna Kaminska,}
\author[b]{Graham G.~Ross,}
\author[c]{Kai Schmidt-Hoberg,}
\author[d]{and Florian Staub}
\affiliation[a]{Deutsches Elektronen-Synchrotron DESY, \\
  Notkestra\ss e 85, D-22607 Hamburg, Germany}
\affiliation[b]{Rudolf Peierls Centre for Theoretical Physics, University of Oxford,\\
 1 Keble Road, Oxford OX1 3NP, UK}
\affiliation[c]{Theory Division, CERN, 1211 Geneva 23, Switzerland}
\affiliation[d]{Bethe Center for Theoretical Physics \& Physikalisches Institut der 
Universit\"at Bonn, \\
Nu{\ss}allee 12, 53115 Bonn, Germany}

\emailAdd{anna.kaminska@fuw.edu.pl} 
\emailAdd{g.ross1@physics.ox.ac.uk} 
\emailAdd{kai.schmidt-hoberg@cern.ch} 
\emailAdd{fnstaub@th.physik.uni-bonn.de} 

\abstract{
Recently the DiracNMSSM has been proposed as a possible solution to reduce the fine tuning in supersymmetry. 
We determine the degree of fine tuning needed in the DiracNMSSM with and without non-universal gaugino masses 
and compare it with the fine tuning in the GNMSSM. To apply reasonable cuts on the allowed parameter regions 
we perform a precise calculation of the Higgs mass. In addition, we include the limits from 
direct SUSY searches and dark matter abundance. We find that both models are comparable in terms of fine tuning,
with the minimal fine tuning in the GNMSSM slightly smaller.
}

\begin{document}

\maketitle

\section{Introduction}
The discovery of the Higgs boson with a mass of about 125~GeV \cite{Atlas:2012gk,CMS:2012gu} has a strong impact on the parameter range of supersymmetric models. In particular in the Constrained Minimal Supersymmetric Standard Model (CMSSM) large regions of the parameter space are not consistent with this mass range and a large mass splitting in the stop sector is needed to push the tree level mass $m_h \leq M_Z$ to that level. The fine tuning needed to achieve this mass is large, requiring a cancellation between uncorrelated parameters of order 1 part in 300. In the more general context of the MSSM one still requires 1\% fine tuning even for an extremely low messenger scale of 10 TeV~\cite{Hall:2011aa}\footnote{Note that this definition of fine tuning differs from ours in the choice of the measure and the fact that the parameters are taken to be low-scale parameters.}. In addition, it has recently been pointed out that those regions in parameter space which could explain the Higgs mass by a rather light 
SUSY spectrum together with maximal mixing in the stop sector have only a metastable electroweak vacuum while in the global vacuum charge and color are broken \cite{Camargo-Molina:2013sta,Blinov:2013fta,Chowdhury:2013dka}. 

To accommodate a heavier Higgs while avoiding very large fine tuning and the need of radiative corrections of about 35~GeV requires new structure.  The most widely studied solution is an enhancement of the Higgs mass already at tree level by new F- or D-term contributions \cite{Ellwanger:2009dp,Ellwanger:2006rm,Ma:2011ea,Zhang:2008jm,Hirsch:2011hg}. In singlet extensions the mass of the SM-like Higgs is at tree-level roughly given by $m^2_h \simeq M^2_Z \left(\cos^22\beta + \frac{\lambda^2}{g^2}\sin^22\beta\right)$ and becomes maximal for $\tan\beta \sim 2$ and for a large coupling $\lambda$ between the singlet and Higgs fields. The most common singlet extension is the  Next-to-Minimal-Supersymmetric-Standard-Model (NMSSM, see e.g.~\cite{Ellwanger:2006rm} for a review) which assumes an underlying $\mathbbm{Z}_3$ symmetry. As expected the fine tuning in the NMSSM gets significantly reduced in comparison to the MSSM  \cite{BasteroGil:2000bw,Dermisek:2005gg,Dermisek:2006py,Dermisek:2007yt,Ellwanger:2011mu}. 

However, other singlet extensions fare even better in terms of fine tuning. Based on an operator analysis it could be expected that singlet extensions leading to certain operators are favoured \cite{Dine:2007xi, Cassel:2009ps}. One option to generate the necessary operators
is an underlying $R$ symmetry, $\Z{4}^R$ or $\Z{8}^R$.  After supersymmetry breaking, both the singlet mass and the $\mu$ term are generated but both are constrained to be of order the supersymmetry breaking mass \cite{Lee:2010gv, Lee:2011dya}. The resulting model is therefore a generalised version of the NMSSM (GNMSSM)~\cite{Ross:2011xv}. Indeed it was found that the fine tuning in the GNMSSM becomes even better than in 
the NMSSM \cite{Ross:2011xv,Ross:2012nr,Kaminska:2013mya}, see also \cite{Delgado:2010uj,Delgado:2012yd}. In addition, the symmetry underlying the GNMSSM has the appealing feature that it forbids dangerous dimension 5 proton decay operators and does not lead to the domain wall problem of the NMSSM \cite{Abel:1995wk}.
Other phenomenologically interesting aspects of the GNMSSM include a possible enhancement of the diphoton decay rate of the Higgs boson \cite{SchmidtHoberg:2012yy} as well as a potential simultaneous explanation of the Fermi line at 130~GeV \cite{SchmidtHoberg:2012ip}. Such signals however would require $\lambda$ to become non-perturbative well below the scale of a grand unified theory (GUT), making an interpretation in terms of an underlying GUT model difficult, see however \cite{Hardy:2012ef}. 
In this article we will assume an underlying GUT structure and a fully perturbative extrapolation to the GUT scale.

An idea to reduce the fine tuning even further has recently been proposed in Ref.~\cite{Lu:2013cta}, 
where it has been argued that a very natural extension of the MSSM is the DiracNMSSM with two additional singlets.
The main motivation for the second singlet ${\bar{S}}$ was a possible mixed ('Dirac') mass term, $M_s S {\bar{S}}$ which allows for very heavy singlets without a suppression
of the tree-level F-term contribution to the Higgs mass while keeping the soft SUSY breaking terms small. As the soft mass squared of the NMSSM singlet
feeds into the soft Higgs masses, this was argued to eliminate this source of fine tuning.
The first study of the fine tuning in the DiracNMSSM was based on a rough fine tuning measure including only parameters at the electroweak scale which takes the impact of the RGEs only crudely into account. In addition, the estimate of the Higgs mass was subject to large theoretical uncertainties and the constraints from SUSY searches as well as dark matter abundance were not included. In this work we perform a full numerical study of the fine tuning in the DiracNMSSM using state of the art computer tools.
To this end we implemented the DiracNMSSM in \SARAH to produce a corresponding version
of \SPheno -- a state of the art spectrum calculator. 
Our estimate of the fine tuning is based on a full two-loop running of the renormalisation group equations and we perform a precise mass calculation in the Higgs sector. The dark matter abundance is calculated with \MO. 

We proceed as follows: in sec.~\ref{sec:model} we introduce the DiracNMSSM and discuss the Higgs sector in some detail. In sec.~\ref{sec:FT} we give details about the fine tuning calculation and present our numerical results in sec.~\ref{sec:results}. We conclude in sec.~\ref{sec:conclusion}. In the appendix we present all renormalisation group equations, mass matrices and vertices which are changed in comparison to the MSSM and explain in great detail the renormalisation of the CP even Higgs sector in the DiracNMSSM.

\section{The DiracNMSSM } 
\label{sec:model}
\subsection{The superpotential and soft-breaking terms}
In the DiracNMSSM one adds two chiral singlet superfields $S$ and $\bar{S}$ to the MSSM with superpotential
\begin{equation}
 \mathcal{W} =  \mathcal{W}_\text{MSSM} + \lambda S H_u H_d + M_s S \bar{S}  + \xi_s S + \xi_{\bar{s}}\bar{S} \;.
\end{equation}
The general soft SUSY breaking  terms associated with the Higgs and singlet sectors are
\begin{align}
 V_\text{soft} 
   &=  m_s^2 |s|^2 + m_{\bar{s}}^2 |{\bar{s}}|^2 + m_{h_u}^2 |h_u|^2+ m_{h_d}^2 |h_d|^2 \nonumber \\
   &+ \left(b\mu \, h_u h_d + \lambda A_\lambda s h_u h_d +  b_s s {\bar{s}}  + t_s s + t_{\bar{s}} {\bar{s}} + h.c.\right) \;.
\label{soft}
\end{align}
Since the renormalisation group equations (RGEs) for the DiracNMSSM have not been
given in the literature before we list the $\beta$-functions for all superpotential and soft-breaking parameters 
as well as all gauge couplings and vacuum expectation values (VEVs) up to two loop in Appendix~\ref{app:RGEs}.

\subsection{Particle content after EWSB}
After electroweak symmetry breaking (EWSB) the complex scalars in the Higgs sector acquire VEVs and they 
are decomposed in their neutral components as
\begin{align} 
\label{eq:EWSB1}
h_d^0 = & \, \frac{1}{\sqrt{2}} \left(v_d + \phi_{d}   + i  \sigma_{d}\right) \, , \hspace{1cm}
h_u^0 =  \, \frac{1}{\sqrt{2}} \left(v_u + \phi_{u}  + i \sigma_{u}\right) \, , \\ 
\label{eq:EWSB2}
s = & \, \frac{1}{\sqrt{2}} \left(v_s + \phi_s   + i  \sigma_s\right) \, , \hspace{1cm}
\bar{s} =  \, \frac{1}{\sqrt{2}} \left( v_{\bar{s}}  + \phi_{\bar{s}}  +  i \sigma_{\bar{s}}\right) \;. 
\end{align}
As usual, the ratio of the two Higgs VEVs is given by $\frac{v_u}{v_d} = \tan\beta$ and $v=\sqrt{v_d^2+v_u^2} \simeq 246$~GeV. 
The charged Higgs sector is very similar to the MSSM and contains one physical charged Higgs with mass 
\begin{equation}
M_{H^+} =  \frac{1}{4} v^2 (g_2^2 - 2 \lambda^2) + \frac{1+\tan^2\beta}{\tan\beta} B_\text{eff} \;.
\end{equation}
Here, we defined 
\begin{equation}
B_\text{eff} = b\mu + \frac{\lambda}{\sqrt{2}}(M_s v_{\bar{s}} + v_s A_\lambda) \, .
\end{equation}

In the neutral Higgs sector there are four CP even states and three CP odd ones. These fields come together with in total six neutralinos. 
The neutralino mass matrix is given in the basis $\left(\lambda_{\tilde{B}}, \tilde{W}^0, \tilde{H}_d^0, \tilde{H}_u^0, \tilde{S},\tilde{\bar{S}}\right)$ by 
\begin{equation} 
m_{\tilde{\chi}^0} = \left( 
\begin{array}{cccccc}
M_1 &0 &-\frac{1}{2} g_1 v_d  &\frac{1}{2} g_1 v_u  &0 &0\\ 
0 &M_2 &\frac{1}{2} g_2 v_d  &-\frac{1}{2} g_2 v_u  &0 &0\\ 
-\frac{1}{2} g_1 v_d  &\frac{1}{2} g_2 v_d  &0 &-\mu_\text{eff}  &- \frac{1}{\sqrt{2}} v_u \lambda  &0\\ 
\frac{1}{2} g_1 v_u  &-\frac{1}{2} g_2 v_u  &- \mu_\text{eff}  &0 &- \frac{1}{\sqrt{2}} v_d \lambda  &0\\ 
0 &0 &- \frac{1}{\sqrt{2}} v_u \lambda  &- \frac{1}{\sqrt{2}} v_d \lambda  &0 &M_s\\ 
0 &0 &0 &0 &M_s &0\end{array} 
\right) 
 \end{equation} 
Here, we introduced 
\begin{equation}
\mu_\text{eff} =  \frac{1}{\sqrt{2}} v_{s} \lambda  + \mu \;.
\end{equation}
The matrix which diagonalizes the neutralinos is called $N$ in the following
\begin{equation} 
N^* m_{\tilde{\chi}^0} N^{\dagger} = m^{dia}_{\tilde{\chi}^0}  \;.
\end{equation} 
Generically the lightest neutralino which is a good dark matter candidate can be an admixture of the bino, the Higgsinos (in the case of non-universal gaugino masses also the Wino) 
and the Singlinos. However, we will see that usually large $M_s$ is preferred so that the Singlino component is typically very suppressed.

In the following we will mostly concentrate on the CP even Higgs states. For completeness, all matrices and vertices which are different to the MSSM are listed in 
Appendices~\ref{app:matrices}--\ref{app:vertices}.  

\subsection{The Higgs sector}
\subsubsection{The Higgs mass at tree level}
\label{sec:HiggsTree}
 The four minimum conditions with respect to the CP even scalars introduced in eqs.~(\ref{eq:EWSB1})--(\ref{eq:EWSB2}) read
\begin{align} 
\Theta_d = \frac{\partial V}{\partial \phi_{d}} =& +\frac{1}{8} \Big(g_{1}^{2} + g_{2}^{2}\Big)v_d \Big(v^2_d- v^2_u\Big)+ v_d \Big(m_{h_d}^2  + |\mu|^2  + \sqrt{2} v_{s} \Re(\lambda \mu^*)  + \Big(v_{s}^{2} + v_{u}^{2}\Big)\frac{|\lambda|^2}{2}  \Big)\nonumber \\ 
 &- v_u \Big(\Re(\lambda \xi_s^*) + {\Re\Big(b\mu\Big)}  + \frac{\sqrt{2}}{2} \Big(v_{\bar{s}} \Re(\lambda M_s^* ) +  v_{s} \Re(\lambda A_{\lambda})\Big)\Big)=0\\ 
\Theta_u = \frac{\partial V}{\partial \phi_{u}} =& +\frac{1}{8} \Big(g_{1}^{2} + g_{2}^{2}\Big)v_u \Big(v_{u}^{2}- v_{d}^{2}\Big)+v_u \Big(m_{h_u}^2  + |\mu|^2  + \sqrt{2} v_{s} \Re(\lambda \mu^*)  + \Big(v_{d}^{2} + v_{s}^{2}\Big)\frac{|\lambda|^2}{2} \Big)\nonumber \\ 
 &- v_d \Big(\Re(\lambda \xi_s^*) + {\Re\Big(b\mu\Big)}  + \frac{\sqrt{2}}{2} \Big(v_{\bar{s}} \Re(\lambda M_s^*) + v_{s} \Re(\lambda A_{\lambda})\Big)\Big)=0\\ 
\Theta_s = \frac{\partial V}{\partial \phi_s} =& \Big(v_{d}^{2} + v_{u}^{2}\Big)\Big(v_{s} \frac{|\lambda|^2}{2}  + \frac{\sqrt{2}}{2} \Re(\mu \lambda^*) \Big) + m_{s}^2 v_{s}  + v_{\bar{s}} {\Re(b_s)} + |M_s|^2 v_{s}  +  \sqrt{2} \Re(M_s^* {\bar{\xi}}_s)  \nonumber \\ 
 & +  \sqrt{2}\Re(t_s) - \frac{\sqrt{2}}{2}  v_d v_u \Re(\lambda A_{\lambda})=0\\ 
\Theta_{\bar{s}} = \frac{\partial V}{\partial \phi_{\bar{s}}} =&  |M_s|^2 v_{\bar{s}}  + \sqrt{2} \Re(M_s^*  \xi_s)  -\frac{\sqrt{2}}{2} v_d v_u \Re(M_s^* \lambda) + \frac{2}{\sqrt{2}} \Re(t_{\bar{s}}) + m_{\bar{s}}^2 v_{\bar{s}}  + v_{s} {\Re\Big(b_s\Big)} =0
\end{align} 
where $\Re(a)$ refers to the real part of $a$.
For given input parameters these equations can now in principle be solved for the four vevs $v_d,v_u,v_s,v_{\bar{s}}$.
However, given that the electroweak vev is known, $v=\sqrt{v_u^2+v_d^2}\simeq 246 \gev$, it makes sense to use this information and
solve for some of the input parameters instead. 
There are now many reasonable combinations of parameters which can be fixed by these equations. The easiest choice might be to choose the Higgs and singlet soft masses $m_{h_d}^2$, $m_{h_u}^2$, $m_s^2$, and $m_{\bar{s}}^2$. However, this assumes automatically that the Higgs soft-terms don't unify with the other scalar soft SUSY masses. A set of parameters which can be consistent with the unification of scalars is $\mu,b\mu,t_s, t_{\bar{s}}$. In the limit of real $\mu$ two distinct solutions are found
\begin{align}
\nonumber 
t_s =& \frac{\mp 1}{8 \Big(v^2_d- v^2_u\Big)} \Big(\sqrt{2} \Big(4 v_{s} \Big(v^2_d- v^2_u\Big)(M_{s}^{2} + m_{s}^2)+\Big(v_{d}^{2} + v_{u}^{2}\Big)\sqrt{D} \lambda \Big)\Big)\\ 
& \hspace{1cm} +\frac{1}{2} \Big(v_d v_u \lambda A_{\lambda} - \sqrt{2} v_{\bar{s}} b_s \Big)  \\
t_{\bar{s}} =& \frac{1}{2} \Big(M_s v_d v_u \lambda  - \sqrt{2} \Big(m_{\bar{s}}^2 v_{\bar{s}}  + v_{s} b_s \Big) - \sqrt{2} M_{s}^{2} v_{\bar{s}} \Big)   \\
b\mu =& \frac{1}{4} \Big(2 \Big(m_{h_d}^2- m_{h_u}^2\Big) \frac{2 v_d v_u}{v^2_u- v^2_d } - v_d v_u \Big(g_{1}^{2} + g_{2}^{2} -2 \lambda^{2}  \Big)-2 \sqrt{2} \Big(M_s v_{\bar{s}} \lambda  + v_{s} \lambda A_{\lambda} \Big)\Big) \\
\label{eq:tadMu}
\mu =& -\frac{1}{\sqrt{2}}  \lambda v_s  \pm \frac{\sqrt{D}}{2\sqrt{2}\Big(v^2_d- v^2_u  \Big)}  
\end{align}
with
\begin{equation}
\sqrt{D} = \sqrt{\Big(v^2_u-v^2_d\Big) \Big(\Big(g_1^2+g_2^2\Big) \Big(v_d^4-v_u^4\Big)+8 m_{h_d}^2 v_d^2-8 m_{h_u}^2 v_u^2\Big)}
\end{equation}

The mass matrix of the CP even Higgs fields is given in the basis \( \left(\phi_{d}, \phi_{u}, \phi_s, \phi_{\bar{s}}\right)\) by
\begin{equation}
\label{eq:HiggsTreeMass}
m^2_{h} = \left(\begin{array}{cc} M_D & M_M \\ M^T_M & M_S \end{array}\right) 
\end{equation}
with $2\times 2$ matrices containing the elements involving only doublets ($M_D$) or singlets ($M_S$) as well as the entries linking both sectors:
\begin{align}
M_D = & \left(\begin{array}{cc} 
 m_{\phi_{d}\phi_{d}} &m_{\phi_{u}\phi_{d}} \\
 m_{\phi_{u}\phi_{d}} & m_{\phi_{u}\phi_{u}}
            \end{array} \right) \\
M_S =  & \left(\begin{array}{cc} 
\frac{1}{2} \Big(v_{d}^{2} + v_{u}^{2}\Big)|\lambda|^2  + m_{s}^2 + |M_s|^2 &{\Re\Big(b_s\Big)}\\ 
 {\Re\Big(b_s\Big)} &m_{\bar{s}}^2 + |M_s|^2                
               \end{array} \right) \\ 
M_M = & \left(\begin{array}{cc} 
\frac{1}{\sqrt{2}} \Big(2 \Re(v_d \lambda \mu^*) - v_u {\Re(\lambda A_{\lambda})} \Big) + v_d v_{s} |\lambda|^2  & \quad - \frac{1}{\sqrt{2}} v_u {\Re\Big(\lambda M_s^* \Big)} \\ 
 \frac{1}{\sqrt{2}} \Big(2 \Re(v_u \lambda \mu^*) - v_d {\Re(\lambda A_{\lambda})} \Big) + v_u v_{s} |\lambda|^2 &\quad - \frac{1}{\sqrt{2}} v_d {\Re\Big(\lambda M_s^* \Big)}          \end{array} \right)
\end{align}
with 
\begin{align} 
m_{\phi_{d}\phi_{d}} &=  |\mu|^2  +  \Re(\sqrt{2} v_{s} \lambda \mu^*) + \frac{1}{2}\Big(v_{s}^{2} + v_{u}^{2}\Big)|\lambda|^2 + \frac{1}{8} \Big(g_{1}^{2} + g_{2}^{2}\Big)\Big(3 v_{d}^{2}  - v_{u}^{2} \Big) + m_{h_d}^2\\ 
m_{\phi_{d}\phi_{u}} &= -\Re(\lambda \xi_s^* + b\mu)+ v_d v_u \Big(|\lambda|^2 -\frac{1}{4} (g_{1}^{2} + g_{2}^{2})\Big)  - \frac{\sqrt{2}}{4}\Big(2 v_{\bar{s}} {\Re(\lambda M_s^*)}+2 v_{s} {\Re(\lambda A_{\lambda})} \Big)  \\ 
m_{\phi_{u}\phi_{u}} &= \frac{1}{2} \Big(2 |\mu|^2  + 2 \Re(\sqrt{2} v_{s} \lambda\mu^*) + \Big(v_{d}^{2} + v_{s}^{2}\Big)|\lambda|^2 \Big) -\frac{1}{8} \Big(g_{1}^{2} + g_{2}^{2}\Big)\Big(v_{d}^{2}-3 v_{u}^{2}\Big) + m_{h_u}^2
\end{align}
This matrix is diagonalized by \(Z^H\): 
\begin{equation} 
Z^H m^2_{h} Z^{H,\dagger} = m^{dia}_{2,h} 
\end{equation} 

To gain some insight into the mass-dependence of the lightest, SM-like Higgs boson on the different parameters, we can perform a rotation of the tree level Higgs matrix to the basis $(h,H,S,\bar{S})$ via the rotation matrix
\begin{equation}
\left(\begin{array}{cc} \cos\beta & \sin\beta \\ -\sin\beta & \cos\beta \end{array} \right) \;.
\end{equation}
In first approximation, the $h$--$H$ mixing as well as the $h$--$\bar{S}$ mixing can be neglected and the remaining matrix in the basis $(h,S)$ reads
\begin{equation}
\label{eq:mhS}
m_{hS} =  \left(\begin{array}{cc} M_{hh} & M_{hS} \\ M_{hS} & M_{SS} \end{array} \right)
\end{equation}
with
\begin{align}
M_{hh} =&  M^2_Z \left(\cos^22\beta + \frac{\lambda^2}{g^2}\sin^22\beta\right)\\
M_{hS} =& -\frac{1}{2v}(\lambda \sqrt{D}\sec2\beta + \sqrt{2} v^2 \lambda A_{\lambda}) \sin2\beta \\
M_{SS} =&  M_s^2 + m_s^2
\end{align}
$M_{hh}$ shows already the famous F-term enhancement of the Higgs mass at tree-level. Hence, if we want to make use of this effect to reduce the fine tuning, we have to concentrate on large $\lambda$ and small $\tan\beta$. However, this effect can easily be spoiled by the mixing with the singlet-sector coming from $M_{hS}$ 
which tends to reduce the smaller eigenvalue. Not relying on special cancellations, the natural size of $M_{hS}$ is $\lambda  v  M_{SUSY}$,
which implies that for small mixing $M_{SS}$ should be rather large. The expectation that for the range of interest $M_s$ is in the few TeV range
for the correct Higgs mass will be confirmed in our numerical analysis.

\subsubsection{Radiative corrections to the Higgs mass}
\label{sec:ModelHiggsLoop}
Of course, a tree-level calculation is not sufficient to have a reliable estimate for the Higgs mass. 
The radiative corrections to the Higgs masses in the NMSSM are discussed in detail in the literature \cite{King:1995vk,Elliott:1993uc,Elliott:1993bs,Ellwanger:1993hn,Franke:1995xn,Degrassi:2009yq,Staub:2010ty,Graf:2012hh,Ender:2011qh}. In contrast, studies in the context of extensions of the NMSSM often include just the dominant stop corrections at the 1-loop level but neglect all other, potentially important, effects. However, we are not relying on these approximations but calculate also the Higgs mass in the DiracNMSSM with a precision comparable to the NMSSM: we include all corrections at the one-loop level and the dominant two-loop corrections stemming from (s)tops.  Details of the calculation are given in Appendix~\ref{app:HiggsLoop}. \\

\begin{figure}[hbt]
\begin{center}
\includegraphics[width=0.40\linewidth]{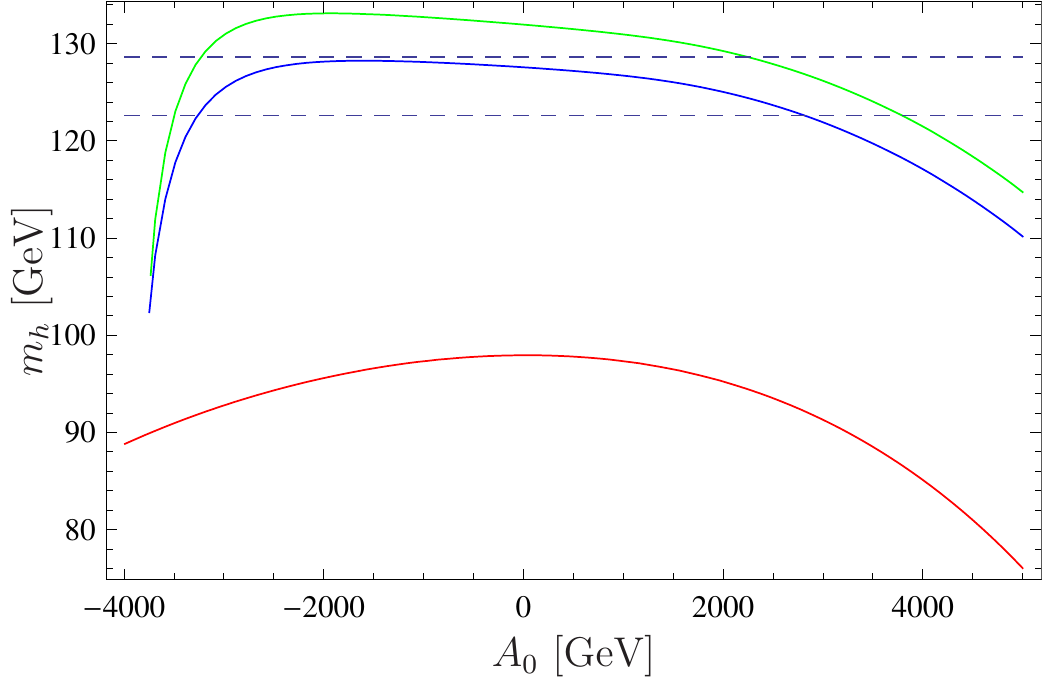} \quad
\includegraphics[width=0.41\linewidth]{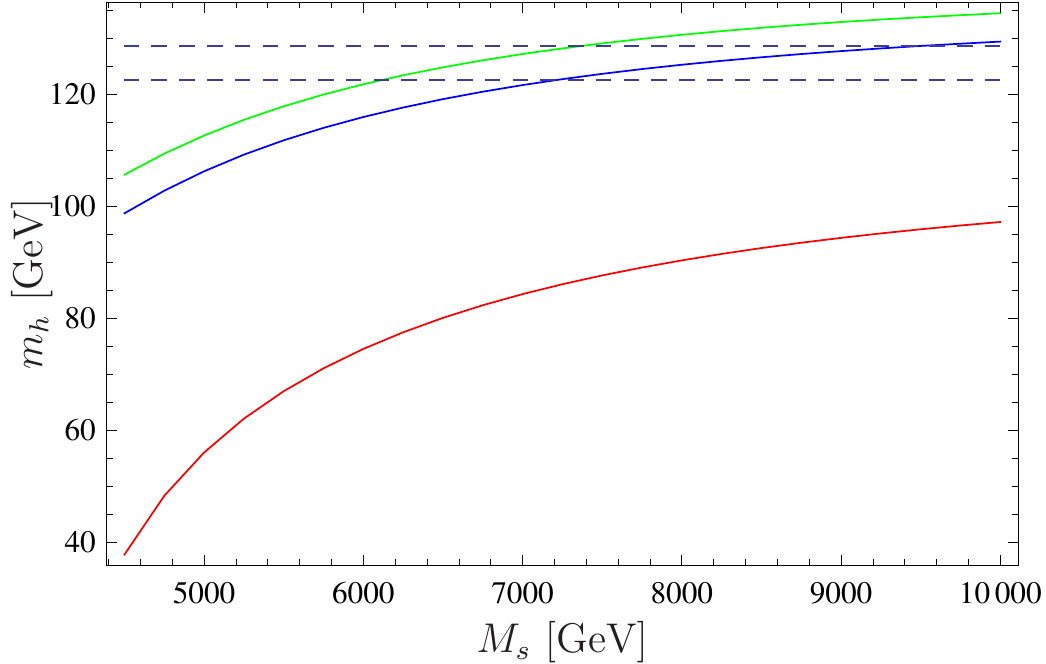} \\
\includegraphics[width=0.40\linewidth]{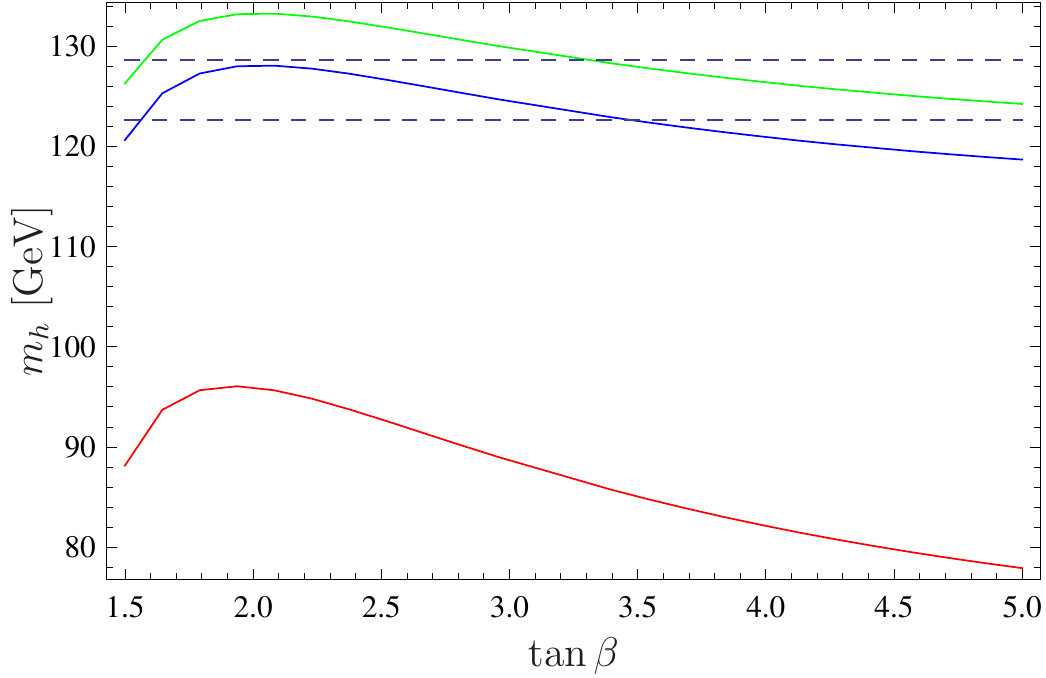} \quad
\includegraphics[width=0.40\linewidth]{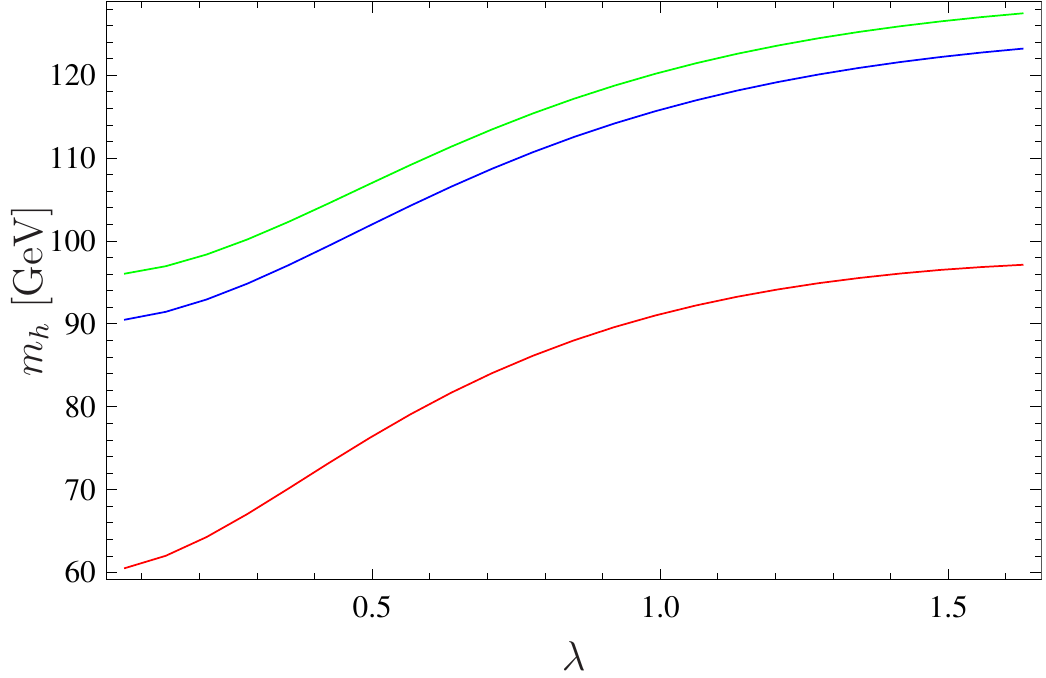} 
\end{center}
\caption{Red line: tree-level mass; blue line: tree-level and full one-loop corrections; green line: tree-level, full one-loop and dominant two loop corrections.}
\label{fig:comparisonLoops}
\end{figure}

To give an impression of the size of the loop corrections and the dependence on the different input parameters we compare in Fig.~\ref{fig:comparisonLoops} the Higgs mass calculated (i) at tree level, (ii) at one-loop, (iii) at two-loop including stop corrections. 
For these plots we have solved the tadpole equations with respect to $\{\mu,b\mu,t_s, t_{\bar{s}}\}$ and have used in addition 
\begin{eqnarray*}
&m_0 = 300~\text{GeV}\,,\hspace{0.3cm} m_{1/2} = 800~\text{GeV}\,,\hspace{0.3cm} \tan\beta=2.3\,,\hspace{0.3cm} A_0 = -2500~\text{GeV}\,,&\\
&\lambda = 1.6\,,\hspace{0.3cm} A_\lambda=-100~\text{GeV}\,,\hspace{0.3cm}v_s=8.3~\text{GeV}\,,\hspace{0.3cm} v_{\bar{s}}=1.5~\text{GeV}\,,&\\
&M_s=9000~\text{GeV}\,,\hspace{0.3cm}b_s=3\cdot 10^6~\text{GeV}^2\,,\hspace{0.3cm} m_{\bar{s}}^2 = 7\cdot 10^{12}\gev^2\,.&
\end{eqnarray*}
There are some features visible in these plots: (i) as expected, the Higgs mass becomes maximal for $\tan\beta \sim 2$ and large values of $\lambda$. (ii) since $m_{s}^2$ was chosen to unify with the other scalars, large $M_s$ is needed to get a sufficiently large $m_h$. The observation that $m_{h}$ grows with increasing $M_{s}$ might seem at first glance inconsistent with the approximate tree-level expression for the Higgs mass derived in Ref.~\cite{Lu:2013cta}. This derivation however did not include the mixing effects in the neutral scalar sector, which have a non-negligible impact on the light eigenvalues and are fully included in our numeric computation. (iii) the absolute shift coming from the 1- and 2-loop corrections are rather insensitive to $\lambda$, i.e. the radiative corrections for the given point are completely dominated by the (s)top loops. 

\begin{figure}[hbt]
\centering
\includegraphics[width=0.45\linewidth]{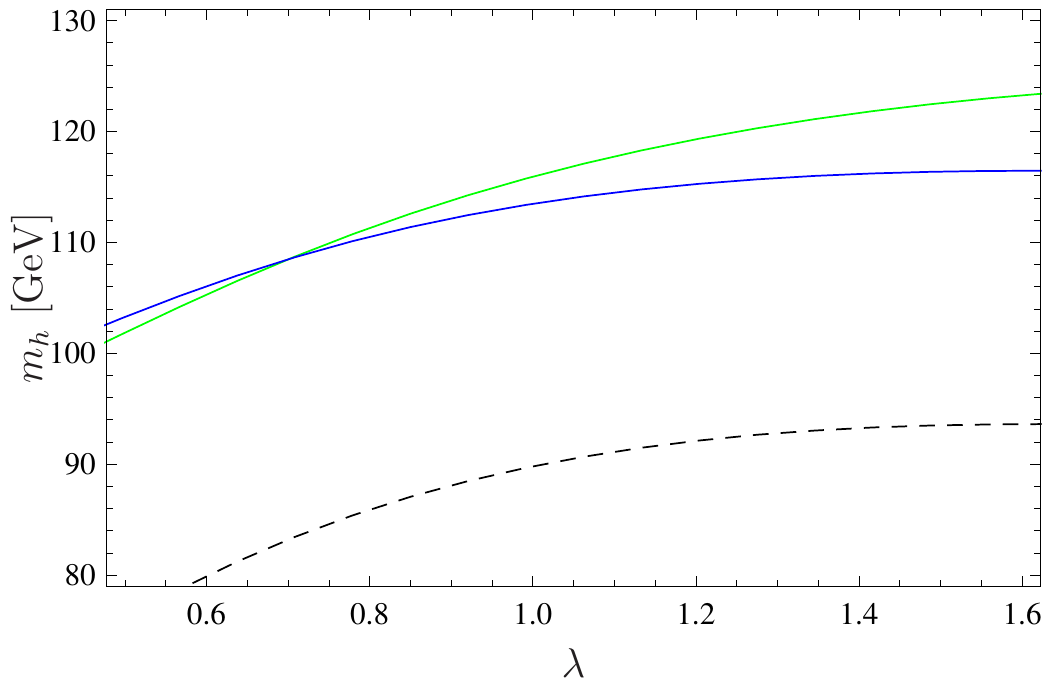} 
\caption{One-loop contributions to the Higgs mass as function of the value of $\lambda$ at the GUT scale. The color code is as follows: black, dashed line: pure tree-level; blue-line: one-loop mass without Higgs/Neutralino/Chargino corrections; 
 green line: full one-loop corrections including Higgs/Neutralino/Chargino contributions.}
\label{fig:loopContributions}
\end{figure}

However, there are also parameter points where other loop corrections beside the (s)top-loops can be very important. In Fig.~\ref{fig:loopContributions} we show the one-loop corrected mass with and without the corrections stemming from Higgs and neutralino/chargino loops. The other parameters have been chosen as
\begin{eqnarray*}
&m_0 = 500~\text{GeV}\,,\hspace{0.3cm} m_{1/2} = 600~\text{GeV}\,,\hspace{0.3cm} \tan\beta=2.3\,,\hspace{0.3cm} A_0 = 2200~\text{GeV}\,,&\\
&\lambda = 1.6\,,\hspace{0.3cm} A_\lambda=1300~\text{GeV}\,,\hspace{0.3cm}v_s=1.0~\text{GeV}\,,\hspace{0.3cm} v_{\bar{s}}=0.1~\text{GeV}\,,&\\
&M_s=6600~\text{GeV}\,,\hspace{0.3cm}b_s=1\cdot 10^4~\text{GeV}^2\,,\hspace{0.3cm} m_{\bar{s}}^2 = 1\cdot 10^{14} \gev^2\,.&
\end{eqnarray*}

Obviously, in the case of large $\lambda$, the additional loops can easy push the Higgs mass up by more than 5~GeV and the full calculation presented here is really necessary.  

\section{Fine Tuning and precision calculations}
\label{sec:FT}
\subsection{Fine Tuning measure}
In Ref.~\cite{Ellis:1986yg, Barbieri:1987fn} a quantitative estimate of the the fine tuning with respect to a set of independent parameters, $p$,  was 
introduced as
\begin{equation} 
\label{eq:measure}
\Delta \equiv \max {\text{Abs}}\big[\Delta _{p}\big],\qquad \Delta _{p}\equiv \frac{\partial \ln
  v^{2}}{\partial \ln p} = \frac{p}{v^2}\frac{\partial v^2}{\partial p} \;.
\end{equation}
The quantity $\Delta^{-1}$ gives a measure of the accuracy to which independent parameters must be tuned to get the correct electroweak breaking scale \cite{Ghilencea:2012qk}. The fine tuning of a model of course depends on what one takes to be the fundamental parameters and at which scale these are defined.
Given the success of gauge coupling unification in supersymmetric models it is natural to assume an underlying GUT structure and
to define the fundamental parameters at the GUT scale. We therefore assume in the following study that all sfermion soft terms unify at the GUT scale to $m_0$. 
In addition, the mass scale of the gaugino is set by the parameter $m_{1/2}$. However, we do not necessarily impose the unification of all gaugino masses (which can still be consistent with an underlying GUT) by allowing for 
$M_1 = a \cdot m_{1/2}$, $M_2 = b \cdot m_{1/2}$, $M_3 = m_{1/2}$ and studying the cases $a=b=1$ and deviations from it separately.
It has been observed that such a non-universality can reduce the hierarchy problem through the appearance of a new ``focus point''  that makes the Higgs mass less sensitive to the gaugino mass scale~\cite{Choi:2005hd,Choi:2006xb,Abe:2007kf,Lebedev:2005ge,Horton:2009ed,Asano:2012sv,Antusch:2012gv,Abe:2012xm,Badziak:2012yg,Gogoladze:2012yf,Yanagida:2013ah}.  
We assume that $a$ and $b$ are fixed by the underlying theory such that contributions to the fine tuning are absent.  As discussed in \cite{Kaminska:2013mya, Horton:2009ed} values of $a$ and $b$ in the low-focus-point region occur naturally in a variety of models. However, as discussed below, dropping this assumption does not greatly increase the minimum fine tuning. 

We calculate the fine tuning with respect to all independent parameters in the DiracNMSSM, defined at the GUT scale,
\begin{equation}
p \in \{m_0, m_{1/2},  A_0, \mu, b\mu,  \lambda, A_\lambda, M_s, b_s, t_s, t_{\bar{s}}, m_s^2, m_{\bar{s}}^2, m_{h_d}^2, m_{h_u}^2\}  \;.
\end{equation}

In many fine tuning analyses it has become customary to consider the fine tuning in terms of electroweak scale parameters only.
Specifically for the DiracNMSSM the following measure has been used in a previous study \cite{Lu:2013cta}
\begin{equation}
\Delta_h  \equiv \frac{2}{m_h^2} \max\{m_{h_d}^2, m_{h_u}^2, \beta^{(1)}_{m^2_{h_d}} L, \beta^{(1)}_{m^2_{h_u}} L, B_\text{eff}, \delta h \}
\end{equation}
with the one-loop $\beta$-functions for the Higgs soft terms, $L=\log\frac{M_{GUT}}{M_{SUSY}}\simeq 30$ and $\delta h = \frac{(\lambda M_s)^2}{(4\pi)^2}\log\frac{M_s^2 + m_s^2}{M^2}$. Here the factor $L$ is supposed to account for the fine tuning
from running, i.e.\ mimicking the effect of defining the parameters at the high scale. 

\begin{figure}[!h!]
 \centering
 \includegraphics[width=0.45\linewidth]{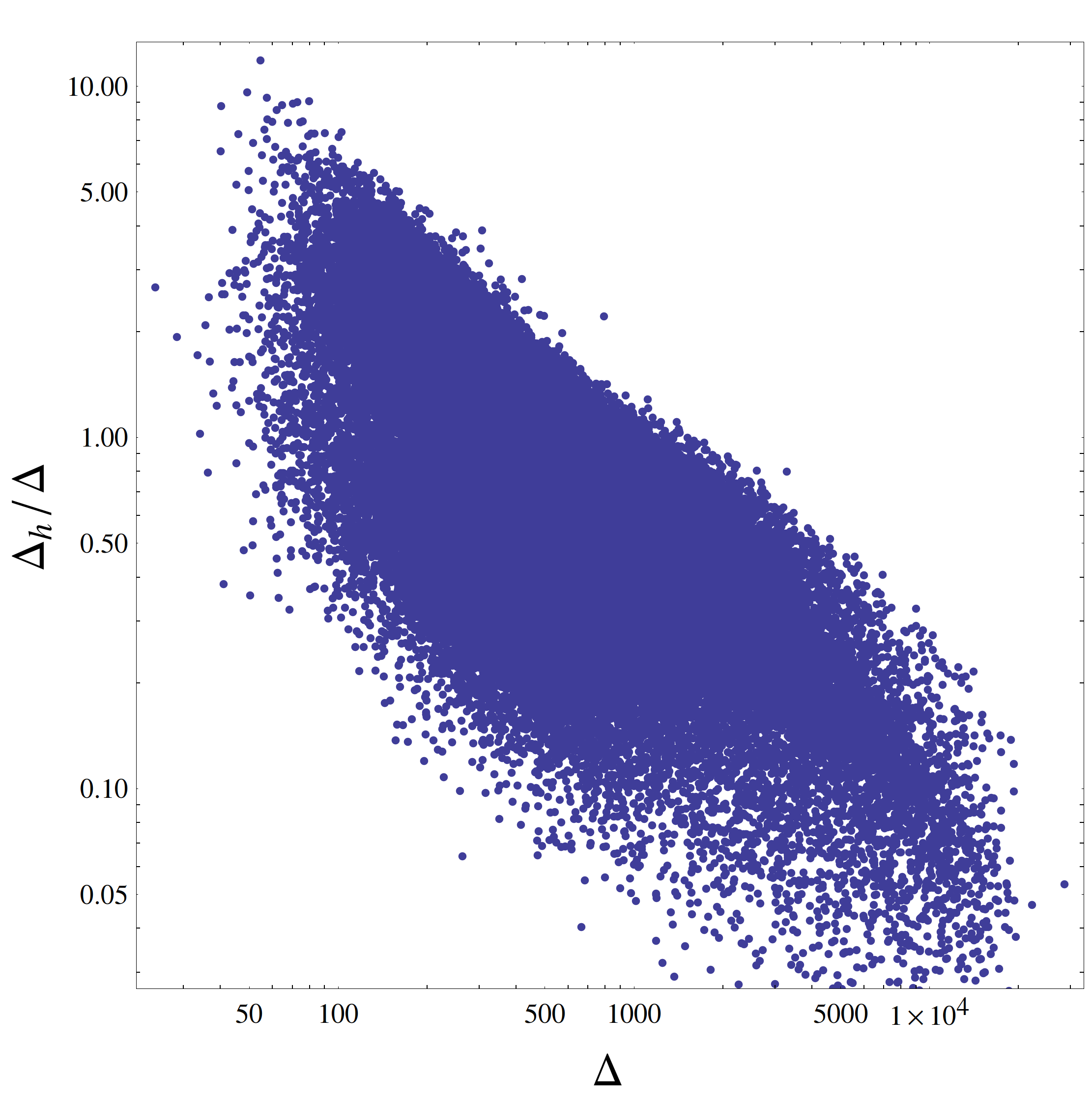} 
 \caption{Comparison of the two fine tuning measures.}
 \label{fig:FTmeasures}
 \end{figure}

We have compared both measures. The result is shown for a set of 100,000 points in Fig.~\ref{fig:FTmeasures}. Even if they usually predict a FT of the same order, the differences can be sizable, with a factor of more than an order of magnitude in both directions.
One feature that does not show up in the measure $\Delta_h$ is the appearance of focus point correlations, hence the fine tuning
can be overestimated. We believe however that it is interesting to study such correlations among parameters which
reduce the fine tuning as this might give valuable information about the desirable structure of the high energy theory.
On the other hand for large SUSY parameters which usually lead to very large $\Delta$, the FT in $\Delta_h$ appears to be much smaller. The reason is that for large SUSY parameters $\beta^{(1)}_{m^2_{h_{u,d}}} L$ is no longer a good approximation for the RGE dependence of the soft terms. Also the 'source' of fine tuning is less clear in the low scale
picture. For instance, the strong dependence on the gluino mass which just enters indirectly in the running of $m_{h_u}^2$ is completely missed \cite{Arvanitaki:2013yja}.  

\subsection{Fine tuning calculation}
Fine tuning studies have a long tradition. However, the precision of these studies has not necessarily improved 
as other theoretical predictions have. In particular for models which go beyond the MSSM 
the prediction of the fine tuning often suffers from large theoretical uncertainties which 
are often not mentioned:
\begin{itemize}
 \item Higgs mass prediction: in many BMSSM studies only the dominant radiative corrections from (s)tops 
 known from the MSSM are included. In addition, also the impact of external momenta is neglected. However,
 there are two main issues with this approach: (i) potentially large, additional loop corrections which don't
 exist in the MSSM are completely missed. The best example is the NMSSM with large $\lambda$. This has been 
 demonstrated for the NMSSM in Ref.~\cite{Degrassi:2009yq} and we also show an example for the DiracNMSSM 
 in sec.~\ref{sec:ModelHiggsLoop}.  (ii) the 
 effective potential approach corresponding to $p^2=0$ can differ significantly from a full one-loop correction 
 demanding $p^2=m_h^2$. This has been shown for an NMSSM extension in Ref.~\cite{Benakli:2012cy}.
 Both effects can be larger than the experimental uncertainty. 
 \item Dark matter prediction: in particular, in co-annihilation regions the relic density of a particle has to 
 be calculated numerically and \MO \cite{Belanger:2006is,Belanger:2007zz,Belanger:2010pz} is often used for that. 
 There exist nowadays several 
 possibilities to create model files for \MO for models which are not included in the public version \cite{Semenov:1998eb,Christensen:2008py,Staub:2009bi}. Using one of these codes should be strongly favored instead
 of modifying existing files by hand which is very error prone. 
  However, even if one uses the correct model files one must keep in mind that  the masses 
 used as input for \MO suffer from an uncertainty. It has been pointed out in Ref.~\cite{Staub:2010ty} 
 that for instance the stau co-annihilation band in the NMSSM gets a sizable shift when going from tree
  to the one-loop level. 
 \item RGE running: for phenomenological studies in the MSSM or NMSSM it has become standard that the full 2-loop 
 RGEs are solved numerically. Even if there exist now several public tools which can calculate the RGEs for a 
 given theory at the two-loop level \cite{Staub:2010jh,Fonseca:2011sy,Lyonnet:2013dna}, often one-loop approximations are used for FT studies. In particular for parameter
 regions which need cancellations between different contributions to the RGE running for specific parameters like the 
 Higgs mass terms, those approximations can fail badly. Moreover, it is well known that the GUT scale defined by $g_1=g_2$
 is shifted  at the 2-loop level quite a bit. Already this effect causes a non-negligible difference between the running 
 parameters at the SUSY scale and changes in particular the masses of particles which depend on the strong interaction \cite{Jack:2004ch}. 
 \item SUSY thresholds: the running parameters entering the RGEs have to be derived from the measured observables. To get an accurate 
 set of gauge and Yukawa couplings at the SUSY scale the SUSY thresholds have to be included \cite{Pierce:1996zz}.  
 In particular, $g_3^{\overline{\text{DR}}}$ can differ by several percent from $g_3^{\overline{\text{MS}}} = \sqrt{4\pi\alpha_s(M_Z)}$ \cite{Carena:1993ag}. Depending on the size of these threshold corrections the perturbativity limit of $\lambda$ is shifted 
 in singlet extensions of the NMSSM \cite{Ellwanger:2009dp}. 
 \end{itemize}
In the current study we take care of all of these issues by using the public codes \SARAH and \SPheno: we have implemented 
the DiracNMSSM in \SARAH \cite{Staub:2008uz,Staub:2009bi,Staub:2010jh,Staub:2012pb,Staub:2013tta} \footnote{The model files might become public with a future release of \SARAH. If you are interested in them beforehand, please contact the authors.}. \SARAH has then been used to create a \SPheno \cite{Porod:2011nf,Porod:2003um} module to get a full fledged spectrum generator for the DiracNMSSM which provides the following features:
\begin{itemize}
 \item All masses of SUSY and Higgs particles are calculated at the one-loop level including all contributions present in the DiracNMSSM 
 and including the impact of the external momenta in the loop functions. Details about the calculation in the Higgs sector are given in
 Appendix~\ref{app:HiggsLoop}.
 \item The dominant 2-loop corrections ($\mathcal{O}(\alpha_t^2), \mathcal{O}(\alpha_t \alpha_s), \mathcal{O}(\alpha_t \alpha_b)$) known from the MSSM are included \cite{Degrassi:2001yf,Brignole:2001jy,Brignole:2002bz,Dedes:2002dy,Dedes:2003km}.  By including these corrections, we have reduced the theoretical uncertainty of the Higgs mass. However, since two loop contributions involving $\lambda$ which are potentially important for large $\lambda$ are missing, the remaining uncertainty is still expected to be slightly larger in comparison to the MSSM with the Higgs mass calculated at the same level. 
 \item Prediction for precision observables like $b\to s\gamma$, $g-2$ or $B_{s,d}^0 \to \bar{\mu}\mu$ are derived at the one-loop level which can be used to further constrain models. Also all of these calculations are automatically adjusted to the present model by \SARAH as explained in Ref.~\cite{Dreiner:2012dh}.
 \item All SUSY thresholds are included when calculating the gauge and Yukawa couplings in the $\overline{\text{DR}}$ scheme from the measured values of SM fermion and gauge boson masses, $G_F$ and $\alpha_s(M_Z)$.
 \item The RGE running is performed at the 2-loop level without any approximation.  To calculate the fine tuning, we vary each of the independent parameters at the GUT scale, run all the parameters down to the weak scale and evaluate the shift in the Z mass by consistently solving all tadpole equations with respect to all vevs numerically. 
 The last step, which is necessary to get the precise fine tuning is not yet part of the public version of \SARAH but will be included in the next update. 
\end{itemize}
For the calculation of the relic density we have used \MO and created the corresponding model files for the DiracNMSSM also with \SARAH. To perform 
the parameter scans we made use of \SSP \cite{Staub:2011dp}. The exchange of parameter values between \SPheno modules written by \SARAH and \MO 
model files also written by \SARAH happens automatically by using the SLHAplus functionality  \cite{Belanger:2010st} of \CalcHep \cite{Pukhov:2004ca,Belyaev:2012qa}.

We have to mention that there is one issue which is still hard to include with the same precision as the other aspects: the question if a parameter point is ruled out by LHC searches or not. To be sure one would have to make a collider study for each parameter point what is far beyond the scope of this work here.\footnote{However, also this situation is expected to be improved significantly since several tools are currently developed to test SUSY points against LHC results \cite{Drees:2013wra,atom,Kraml:2013mwa}.} As discussed below, the non-universal gaugino mass case in the GNMSSM often leads to a compressed SUSY spectrum with small mass differences between gauginos and the LSP that makes SUSY discovery more difficult. To account for this in a manner consistent with the non-observation of superpartners at the LHC we implemented a cut on the gluino mass as a function of the gluino-LSP mass difference as presented in \cite{ATLAS-CONF-2013-047,CMS-PAS-SUS-13-008}. 
In Fig.~7 of \cite{ATLAS-CONF-2013-047} two bounds are shown, a weaker one for decoupled squarks and a stronger one for $m_\text{squark} \sim m_\text{gluino}$. Most parameter space points of interest to us are in the intermediate regime, but we will use the stronger bound here.\footnote{This of course assumes that the bound on $m_\text{gluino} \sim m_\text{squark}$ is at least as strong as say $m_\text{gluino} \sim 0.7 \cdot m_\text{squark}$. This is not quite clear as in the case of compressed spectra the gluino is still very close in mass to the lightest neutralino while some missing $E_T$ could be coming from the squarks. On the other hand the production cross section will be smaller in this case. As noted above a thorough study would need to examine every parameter point individually. Here we will content ourselves with
this approximate cut.}
We further require the chargino and slepton masses to be above $100\gev$.
We also require that the lightest supersymmetric particle (LSP) is a neutralino which is a good dark matter candidate and its
relic density does not exceed the $5\sigma$ PLANCK \cite{Ade:2013zuv} upper bound of $\Omega h^2 \le 0.1334$.

\section{Exploring the DiracNMSSM vs.\ the GNMSSM}
\label{sec:results}

In the following we will study the fine tuning of the DiracNMSSM and compare it with the GNMSSM. 
The superpotential of the GNMSSM is given by
\begin{equation}
 \mathcal{W} =  \mathcal{W}_\text{MSSM} +\lambda S H_u H_d + \xi_s S + \frac{1}{2} \mu_s S^2+ \frac{1}{3}\kappa S^3  \;.
\end{equation}
and the corresponding soft-breaking terms read
\begin{align}
 V_\text{soft}  &= V_\text{soft,MSSM} +  m_s^2 |s|^2 + \left(\frac{1}{3} \kappa A_\kappa s^3 + \lambda A_\lambda s h_u h_d +  b_s s^2  + t_s s + h.c.\right) \;.
\end{align}

\subsection{Universal gaugino masses}
Let us first concentrate on the fine tuning in the DiracNMSSM in the case that all gaugino masses unify at the GUT scale. We solve the four conditions for correct EWSB for $\{ \mu, b\mu, t_s, t_{\bar{s}}\}$. This leaves us with 16 input parameters
\begin{equation}
m_0, m_{1/2}, A_0, \tan\beta, m_{h_u}^2,  m_{h_d}^2, \lambda, A_\lambda, v_s, v_{\bar{s}}, M_s, b_s, m_s^2, m_{\bar{s}}^2, \xi_s, \xi_{\bar{s}}
\end{equation}
For the case of universal gaugino masses the lightest gaugino will always be a bino. 
This potentially leaves a mixture of bino-, higgsino, and singlino-like neutralinos as LSP candidates.
For the bino it is difficult to achieve a small enough relic abundance except in the stau coannihilation or Higgs funnel regions 
while the singlino turns out to be
typically rather heavy in the region of interest. Accordingly most of the viable points we find have a LSP with a sizable
higgsino fraction.

\begin{figure}[!h!]
 \centering
 \includegraphics[width=0.45\linewidth]{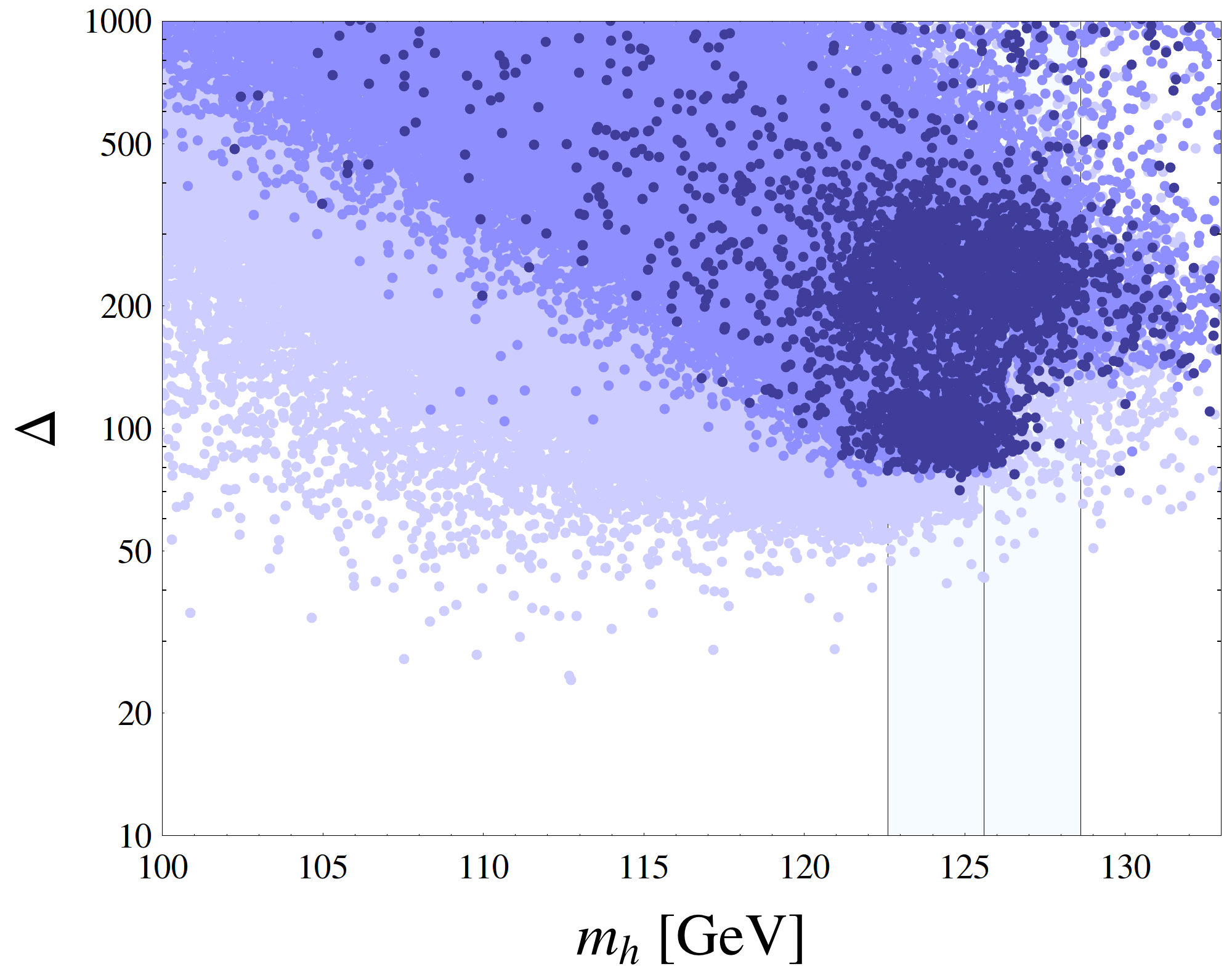} 
 \includegraphics[width=0.45\linewidth]{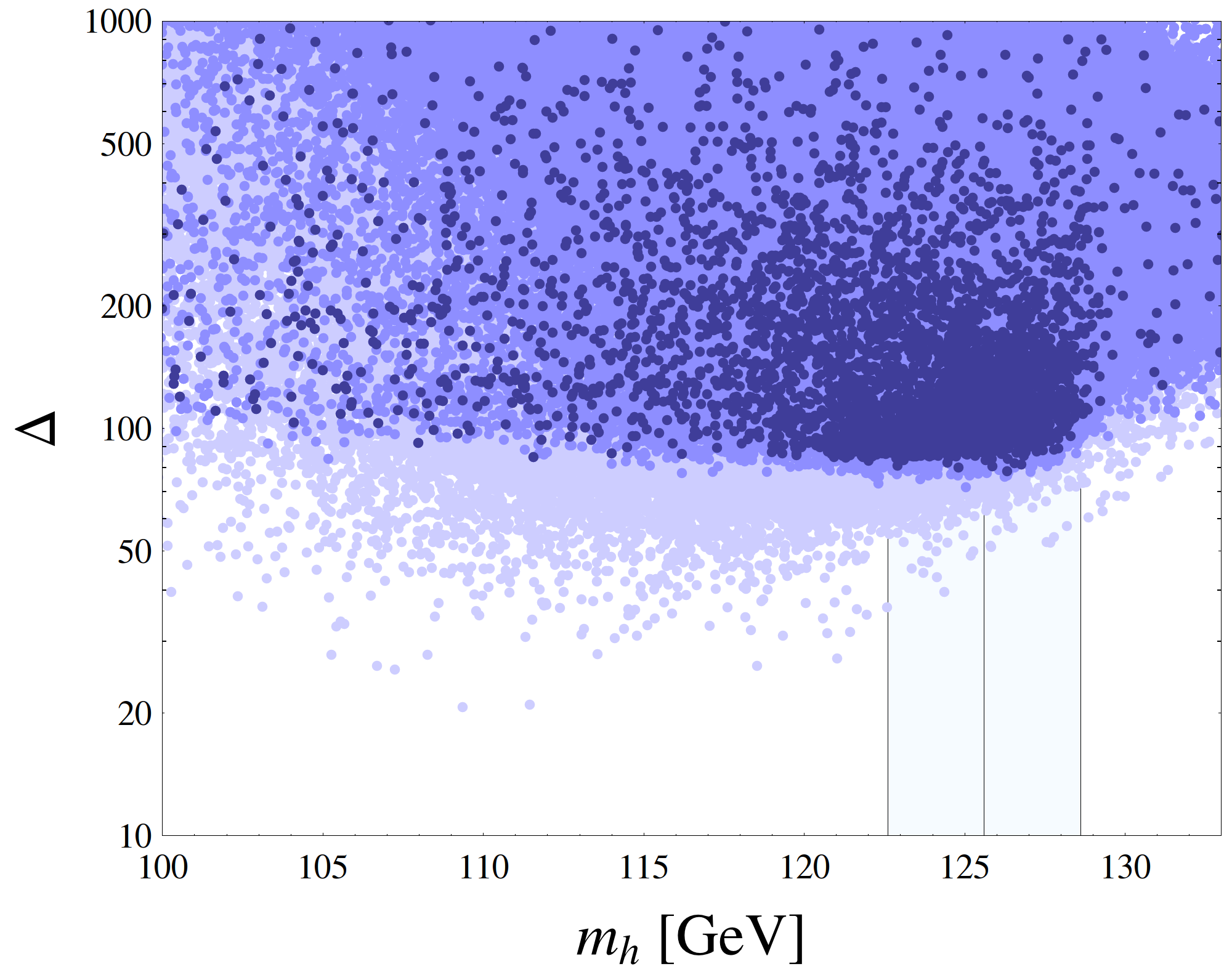}
 \caption{The fine tuning as a function of the SM like Higgs mass for the DiracNMSSM (left) and the GNMSSM (right). 
 The light blue points are before any cuts.
 For the dark blue points we use appropriate SUSY and dark matter cuts. }
 \label{fig:DiracNMSSMnuhm}
 \end{figure}
 
The overall result is summarized in Fig.~\ref{fig:DiracNMSSMnuhm}. This plot shows the fine tuning as function of the SM-like Higgs mass for the DiracNMSSM (left) and the GNMSSM (right). Several different cuts are indicated: (i) light blue points correspond to no cuts, (ii) medium blue points include LHC limits on SUSY masses, and (iii) dark blue points include in addition the upper limit on the dark matter abundance. Note that these plots are combinations of several scans of different regions in the parameter space of the models. Since we have zoomed into interesting regions with a small or moderate fine tuning the density of points can't be interpreted as some probability measure. 
Nevertheless, it is interesting that there is no big difference between the best fine tuning after the LHC and after the LHC and dark matter cut for the DiracNMSSM. The reason is that the lightest neutralino in the DiracNMSSM turns out to have easily a large Higgsino fraction which is sufficient to reduce the dark matter abundance to the allowed level. One can see that the best fine tuning consistent with all observables in the DiracNMSSM with unified gaugino masses is about 70. This is not improved if we dropped the upper limit on the neutralino relic density. This value is, of course, significantly better than in the MSSM where one expects $\Delta \gtrsim 300$ and also improves the situation compared to the NMSSM. However, it is of the same size as the fine tuning reported for the GNMSSM in a similar setup, see the right plot in Fig.~\ref{fig:DiracNMSSMnuhm}. Here we find as best fine tuning $\Delta \simeq 78$ including the DM cut and $\Delta \simeq 71$ without the DM cut.

To compare the dependence of the fine tuning in both models we have randomly picked 200,000 valid points for each model. The parameter values for all points are in the ranges
\begin{eqnarray*}
&m_0 \in [0,1]~\text{TeV}\,,\hspace{1cm} m_{1/2} \in [0,1]~\text{TeV}\,,\hspace{1cm} \tan\beta \in [1.5,4.0]&\\
&\lambda \in [1.0,1.7]\,,\hspace{1cm} A_\lambda \in [-1.5,1.5]~\text{TeV}\,,\hspace{1cm} A_0 \in [-2.5,2.5]~\text{TeV}&\\
&M_s,\mu_s \in [-10,10]~\text{TeV}\,,\hspace{1cm} b_s \in [-10,10]~\text{TeV}^2\,,\hspace{1cm} v_s \in [-2,2]~\text{TeV}& \\
&m_{h_d}^2 \in [-5,5]~\text{TeV}^2\,,\hspace{1cm} m_{h_u}^2 \in [-5,5]~\text{TeV}^2\,,\hspace{1cm} m_s^2 \in [-5,5]~\text{TeV}^2&
\end{eqnarray*}
in addition, the specific parameters for the DiracNMSSM have been chosen to be
\begin{eqnarray*}
&v_{\bar{s}} \in [-2,2]~\text{GeV} \,,\hspace{1cm} m_{\bar{s}}^2 \in [0,10^4]~\text{TeV}^2 &
\end{eqnarray*}
and those for the GNMSSM
\begin{eqnarray*}
\kappa \in [-1,1]\,,\hspace{1cm} A_\kappa \in [-1.5,1.5]~\text{TeV} 
\end{eqnarray*}
Note, since the points shown in the following are chosen randomly from the full data-set they don't include the points with the best fine tuning which have been found for both models. They are just meant to demonstrate the general similarities and differences in the fine tuning of both models.

The effect of the different cuts can be summarized as follows: 
\begin{center}
\begin{tabular}{|c|c|c|}
\hline 
   & DiracNMSSM & GNMSSM \\
\hline
\hline
Valid points & 200,000 & 200,000 \\
After SUSY cuts & 52,623 & 64,514 \\
After Higgs cut & 5,527 & 23,783 \\
After DM cut & 943 & 204 \\
\hline
\end{tabular}
\end{center}
In both models the SUSY cuts rule out about 2/3 to 3/4 of the valid points. That's not surprising since the changes in the Higgs sector are expected to have only a small impact on the squark and gluino masses and for $m_{1/2} \lesssim 750\gev$ the gluino mass turns out to be too light. However, there are two obvious differences: while it seems to be much easier to accommodate a Higgs mass in the expected range in the GNMSSM than in the DiracNMSSM, the DiracNMSSM is doing better in satisfying the upper limit of the dark matter abundance. The reason for this is, as mentioned already above, that the lightest neutralino in the DiracNMSSM is much more often a Higgsino which annihilates very efficiently. 

\begin{figure}[tb]
\includegraphics[width=0.3\linewidth]{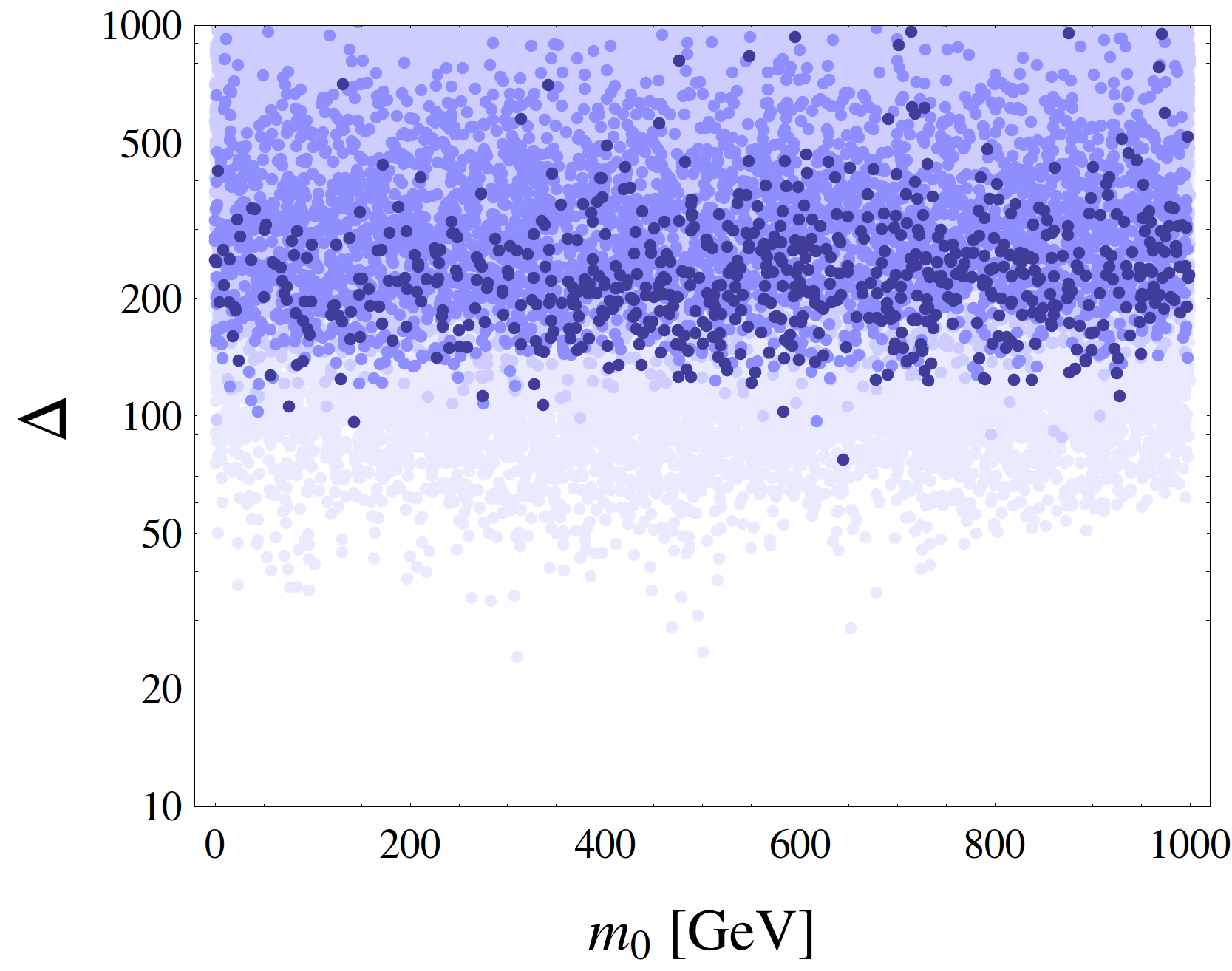}  
\hfill 
\includegraphics[width=0.3\linewidth]{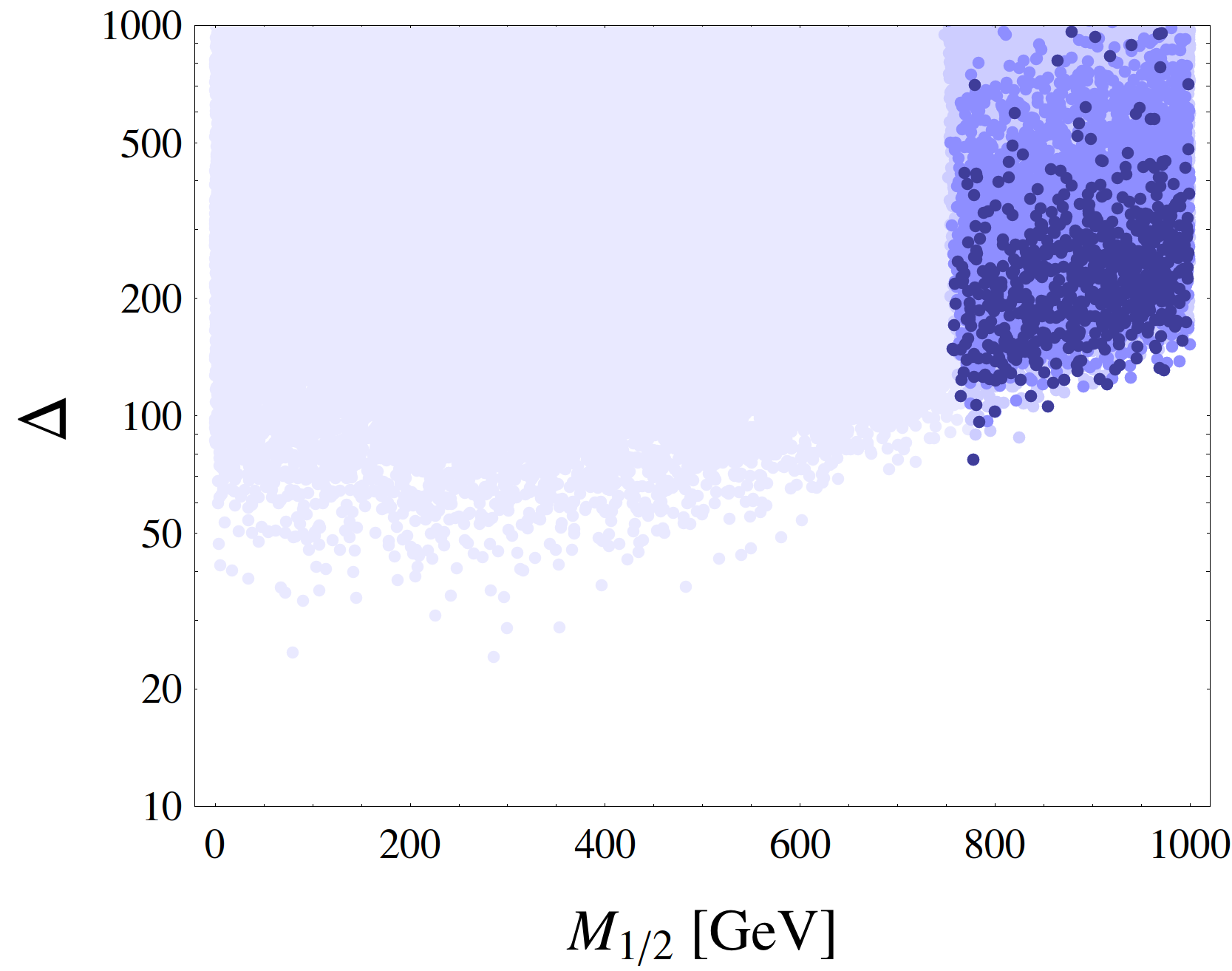}  
\hfill
\includegraphics[width=0.3\linewidth]{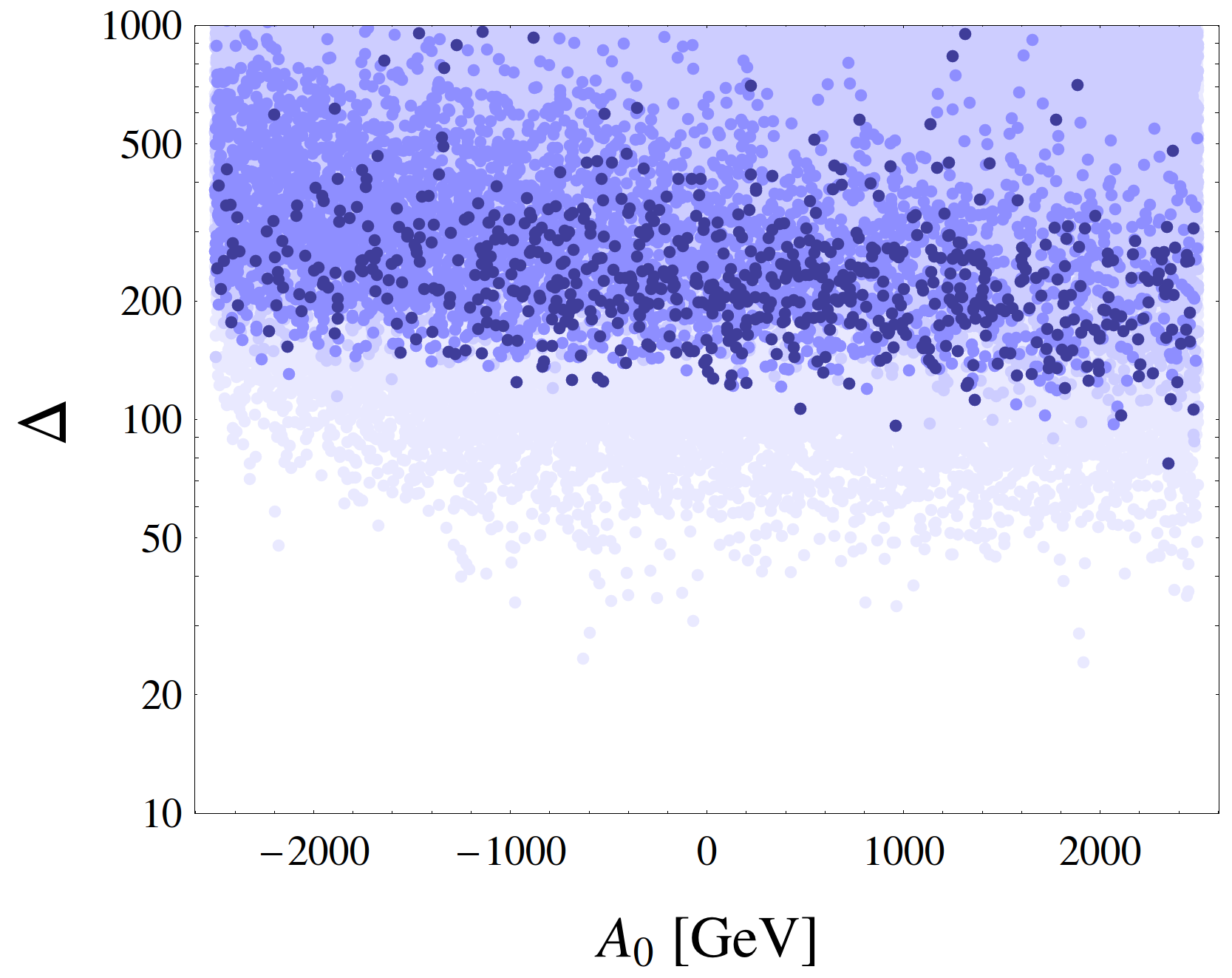}  \\
\includegraphics[width=0.3\linewidth]{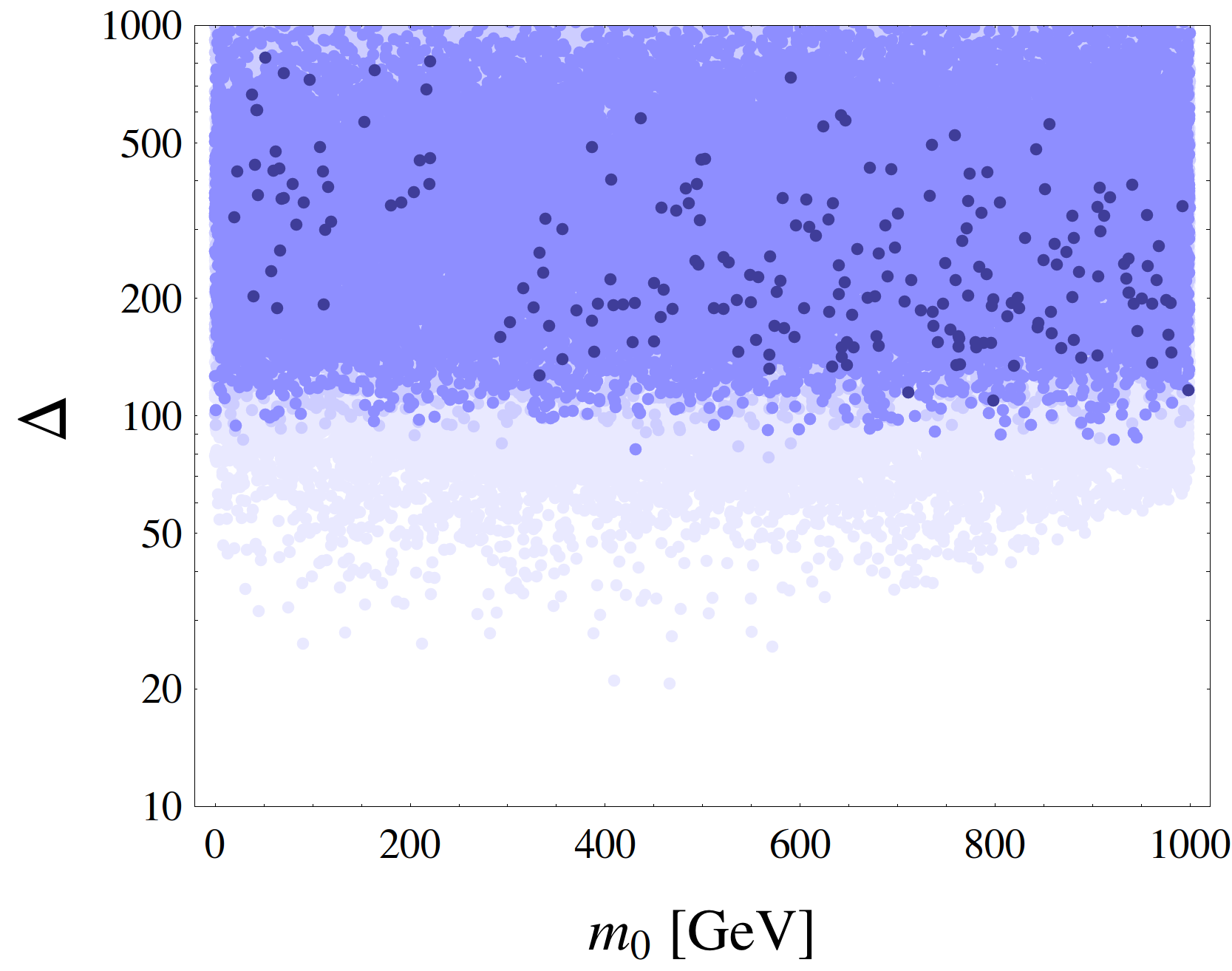}  
\hfill 
\includegraphics[width=0.3\linewidth]{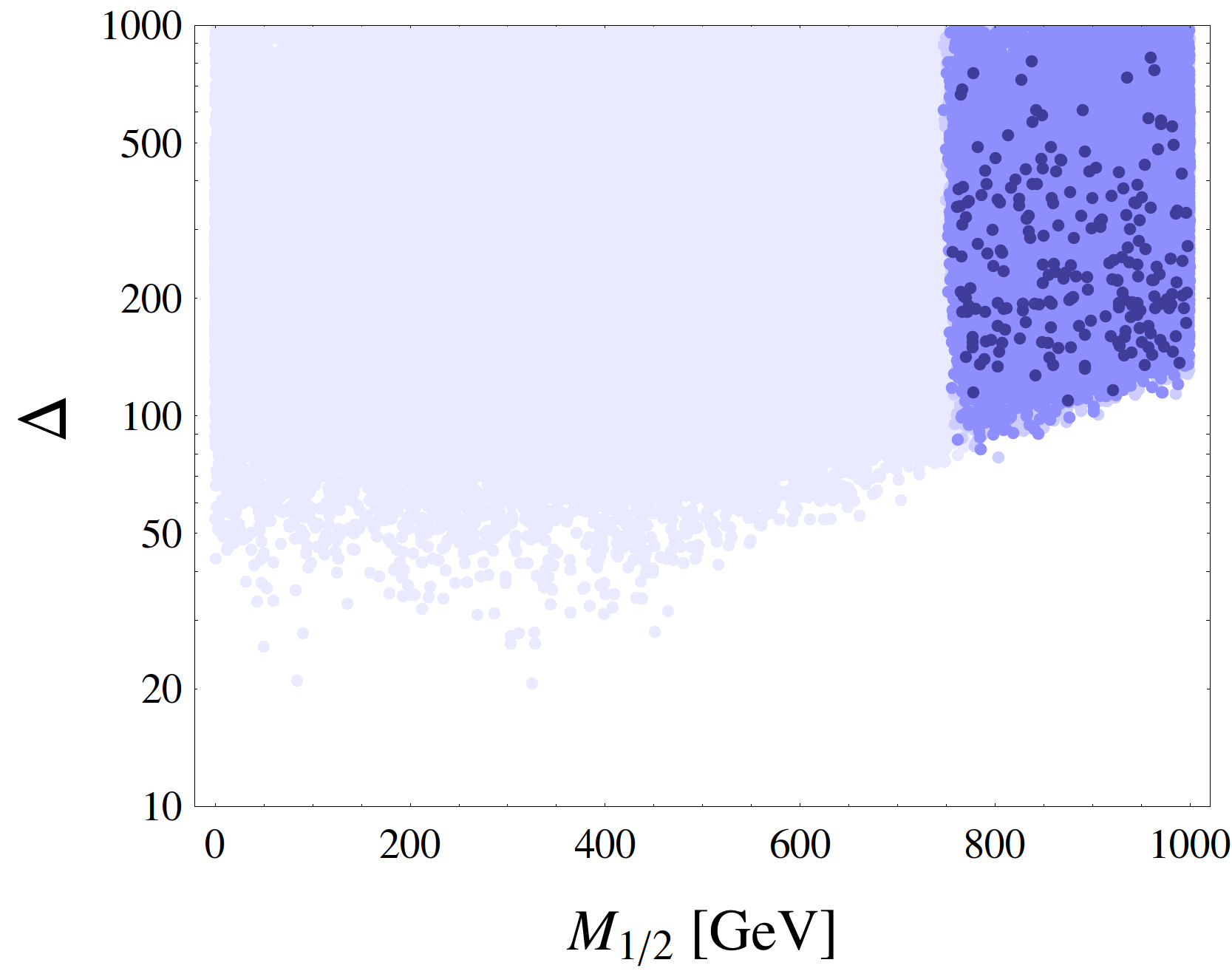} 
\hfill 
\includegraphics[width=0.3\linewidth]{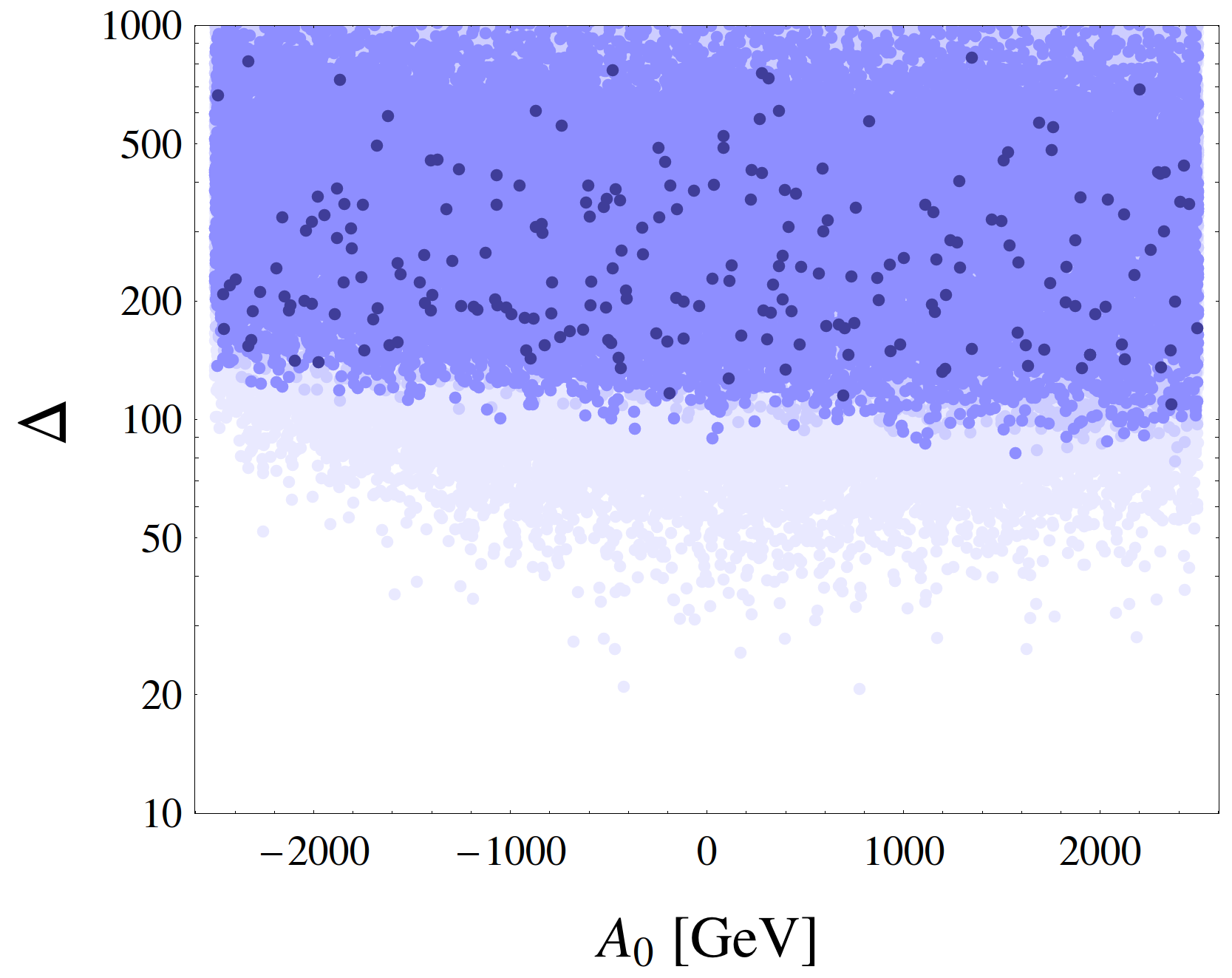} 
\caption{Fine tuning vs. $m_0$, $m_{1/2}$ and $A_0$ in the DiracNMSSM (first line) 
and GNMSSM (second line). With increasing saturation the following cuts are applied:
(i) no cut, (ii) cut on SUSY masses, (iii) cut on the SM like Higgs mass, (iv) cut 
on the upper limit of the dark matter abundance. }
\label{fig:compFT1}
\end{figure}
\begin{figure}[tb]
\includegraphics[width=0.3\linewidth]{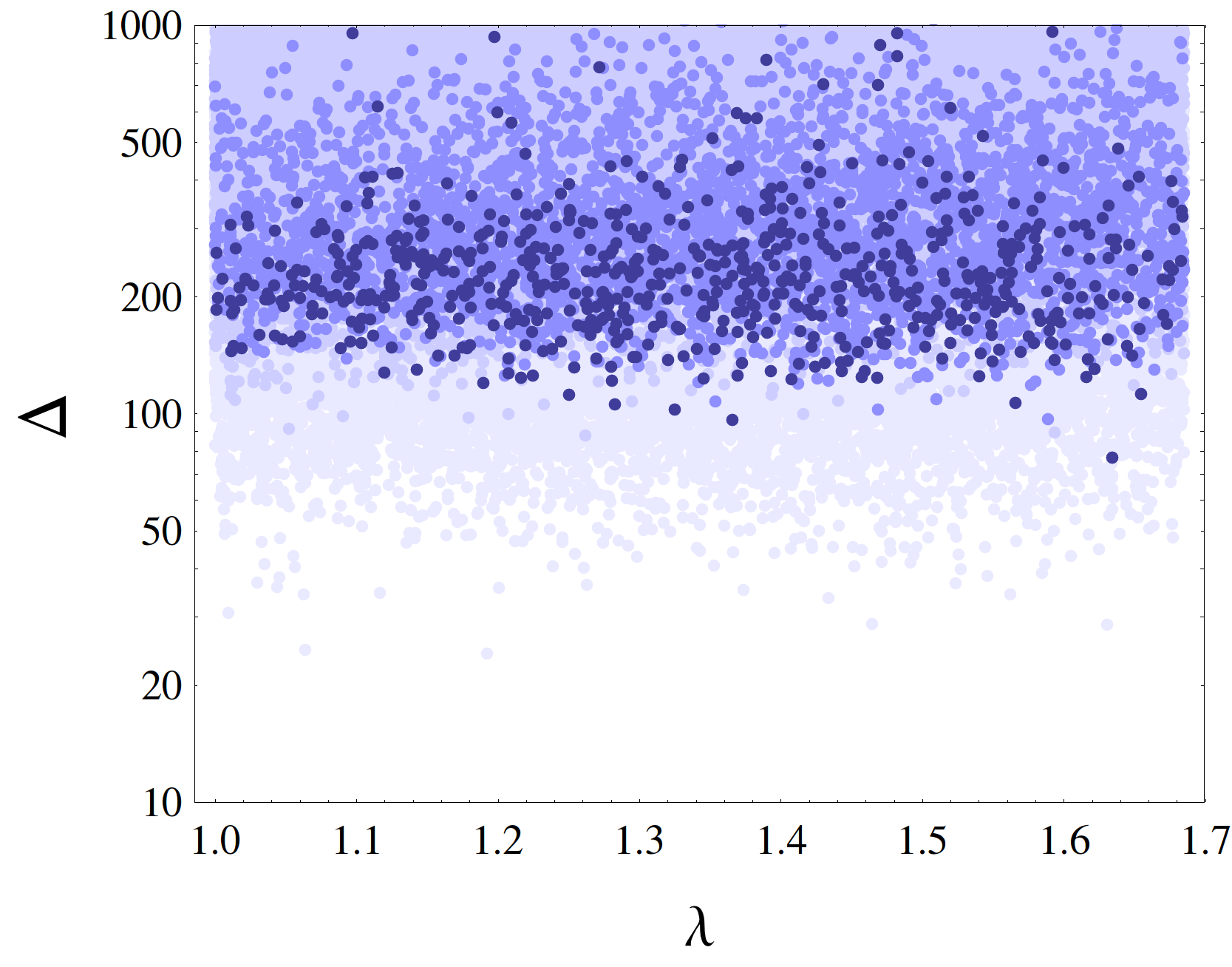}  
\hfill
\includegraphics[width=0.3\linewidth]{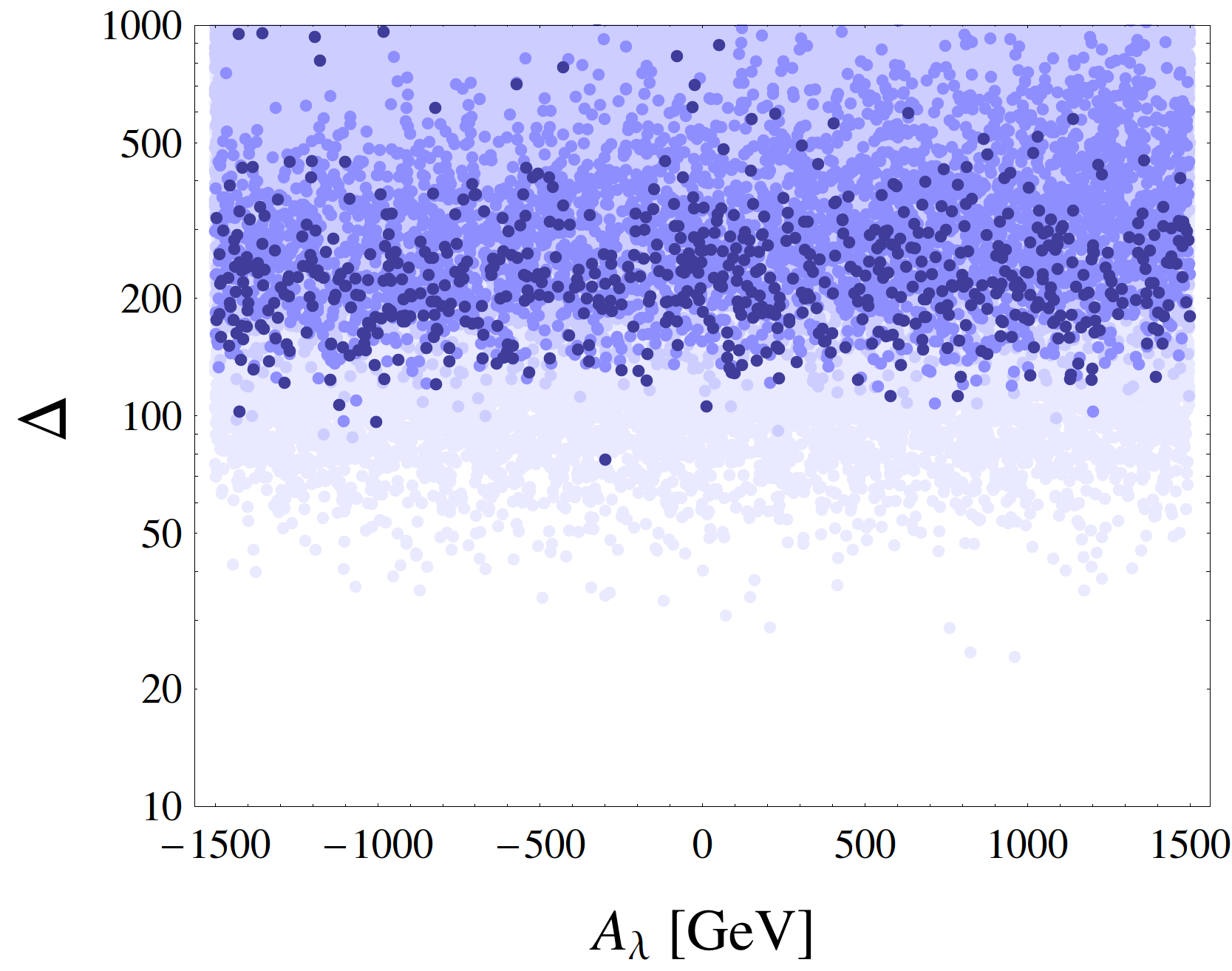}  
\hfill 
\includegraphics[width=0.3\linewidth]{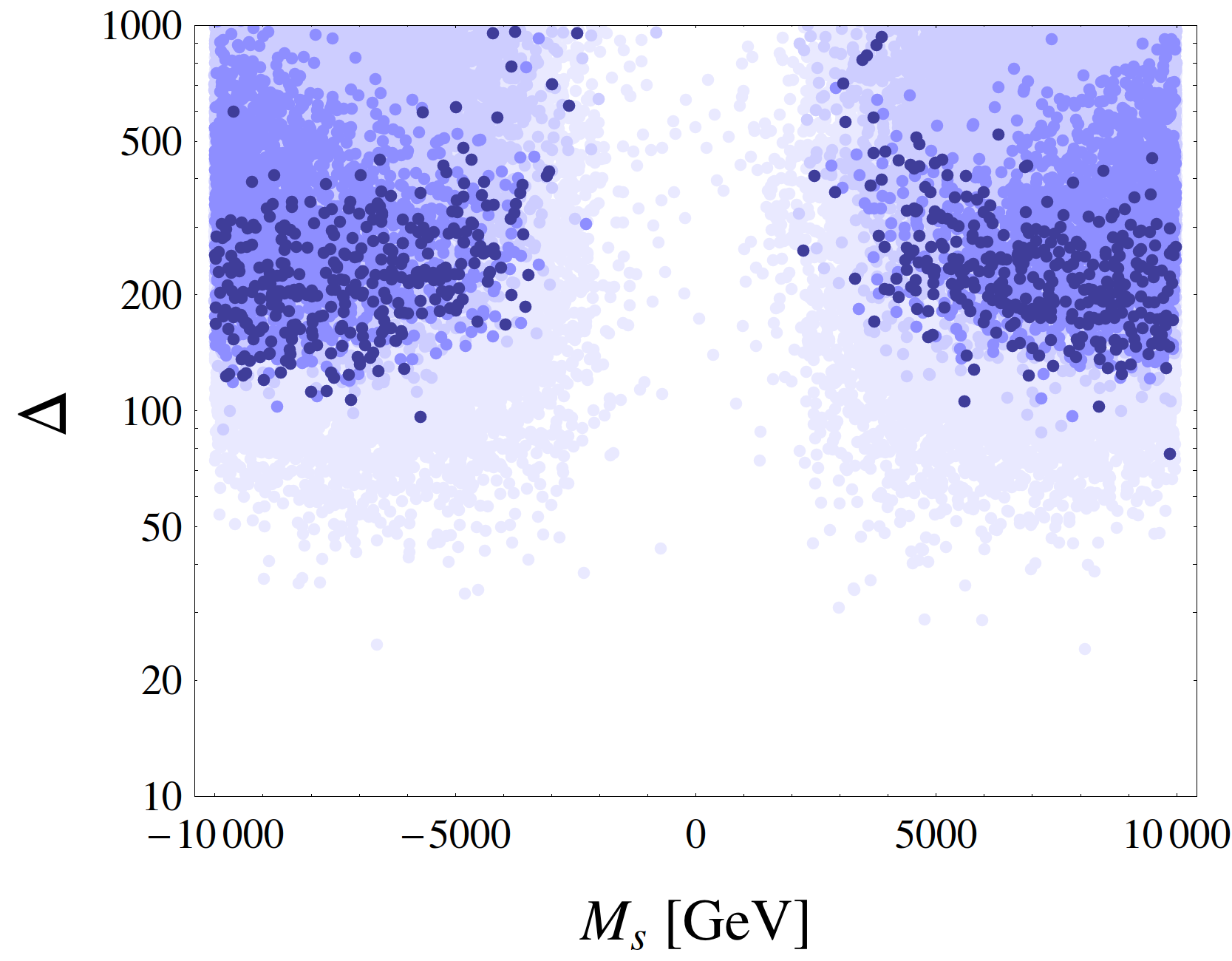}  \\
\includegraphics[width=0.3\linewidth]{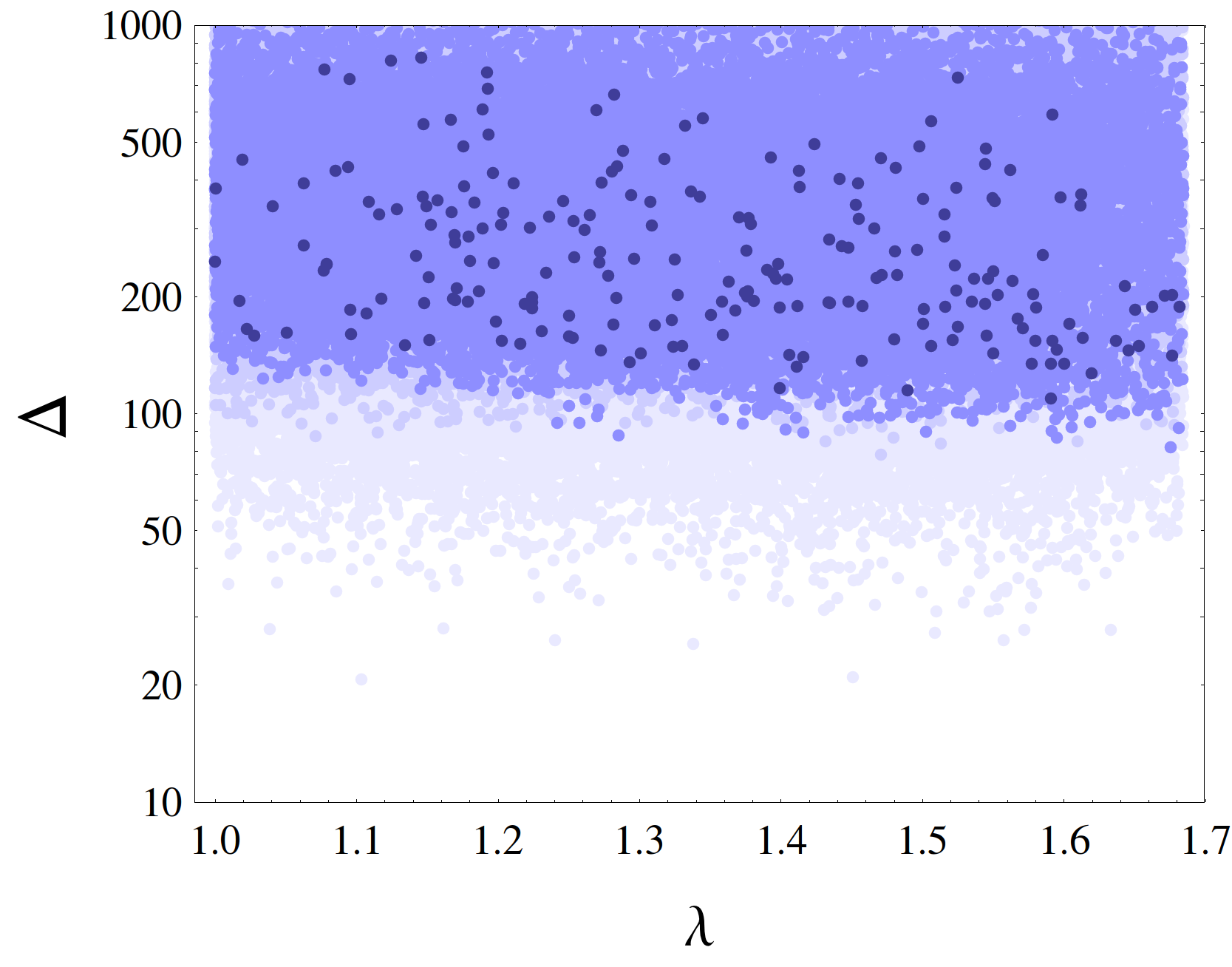}  
\hfill 
\includegraphics[width=0.3\linewidth]{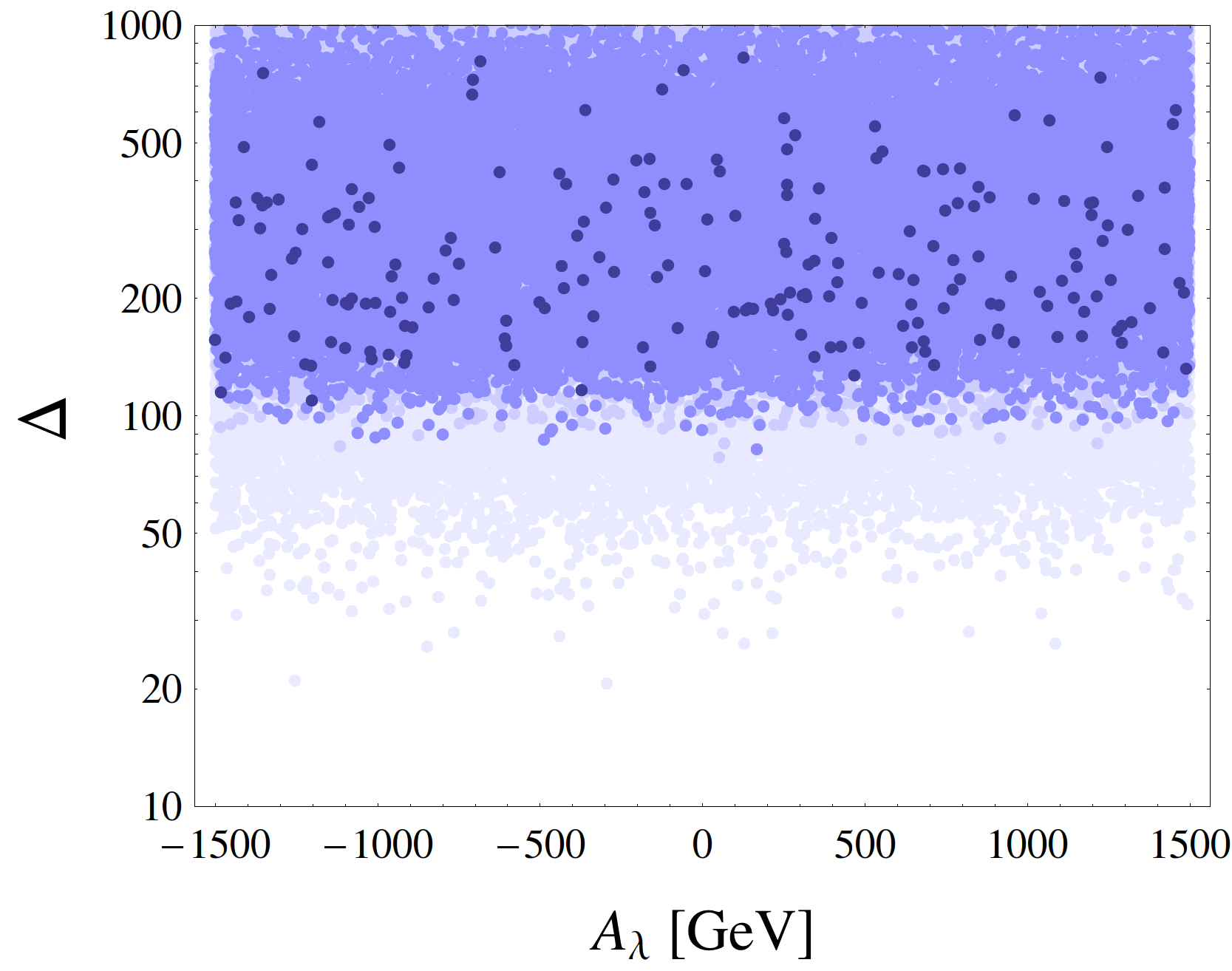}  
\hfill 
\includegraphics[width=0.3\linewidth]{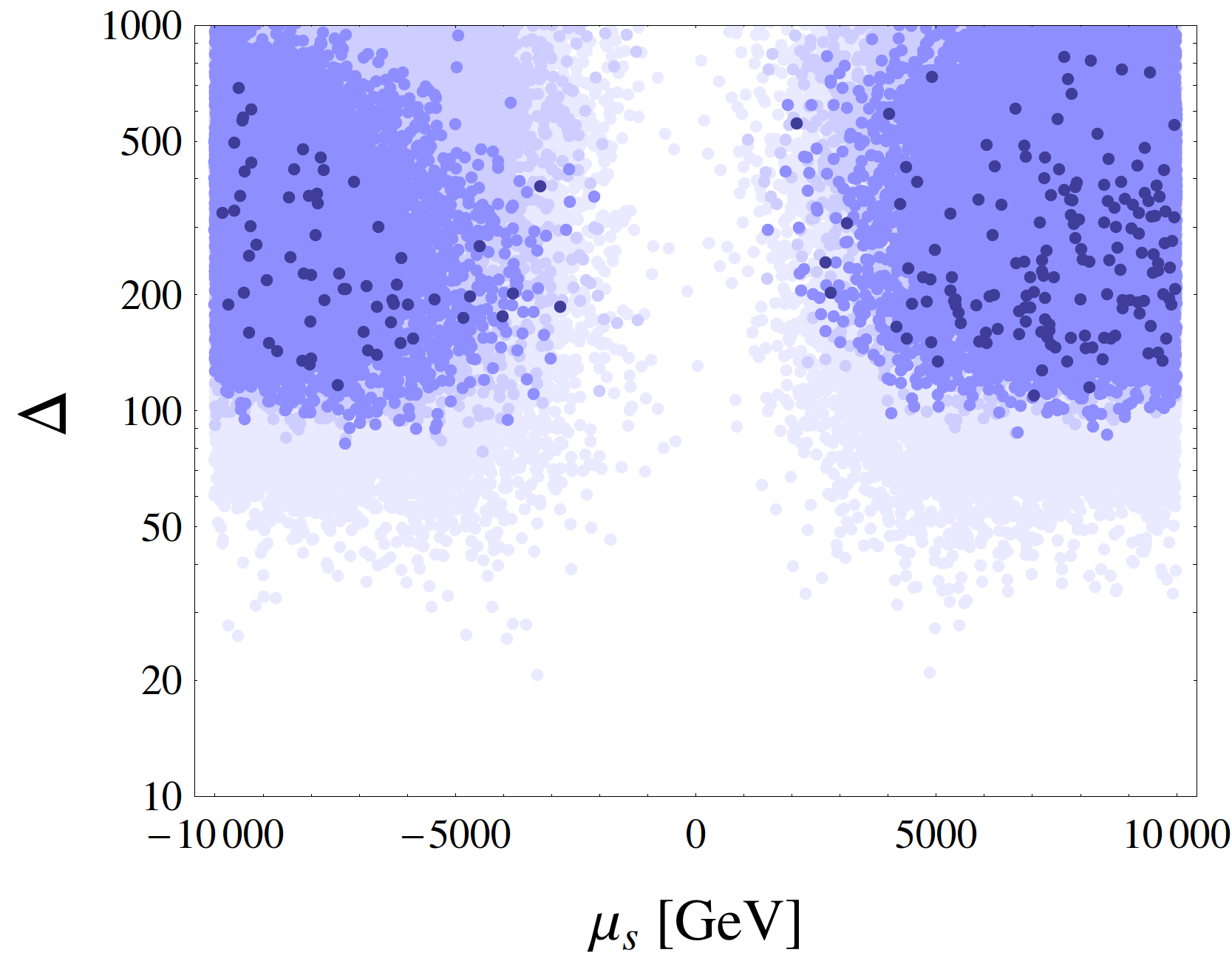}  
\caption{Fine tuning vs. $\lambda$, $A_\lambda$ and $M_s$ in the DiracNMSSM (first line) 
and GNMSSM (second line). The color code is the same as in Fig.~\ref{fig:compFT1}.}
\label{fig:compFT2}
\end{figure}
\begin{figure}[tb]
\includegraphics[width=0.3\linewidth]{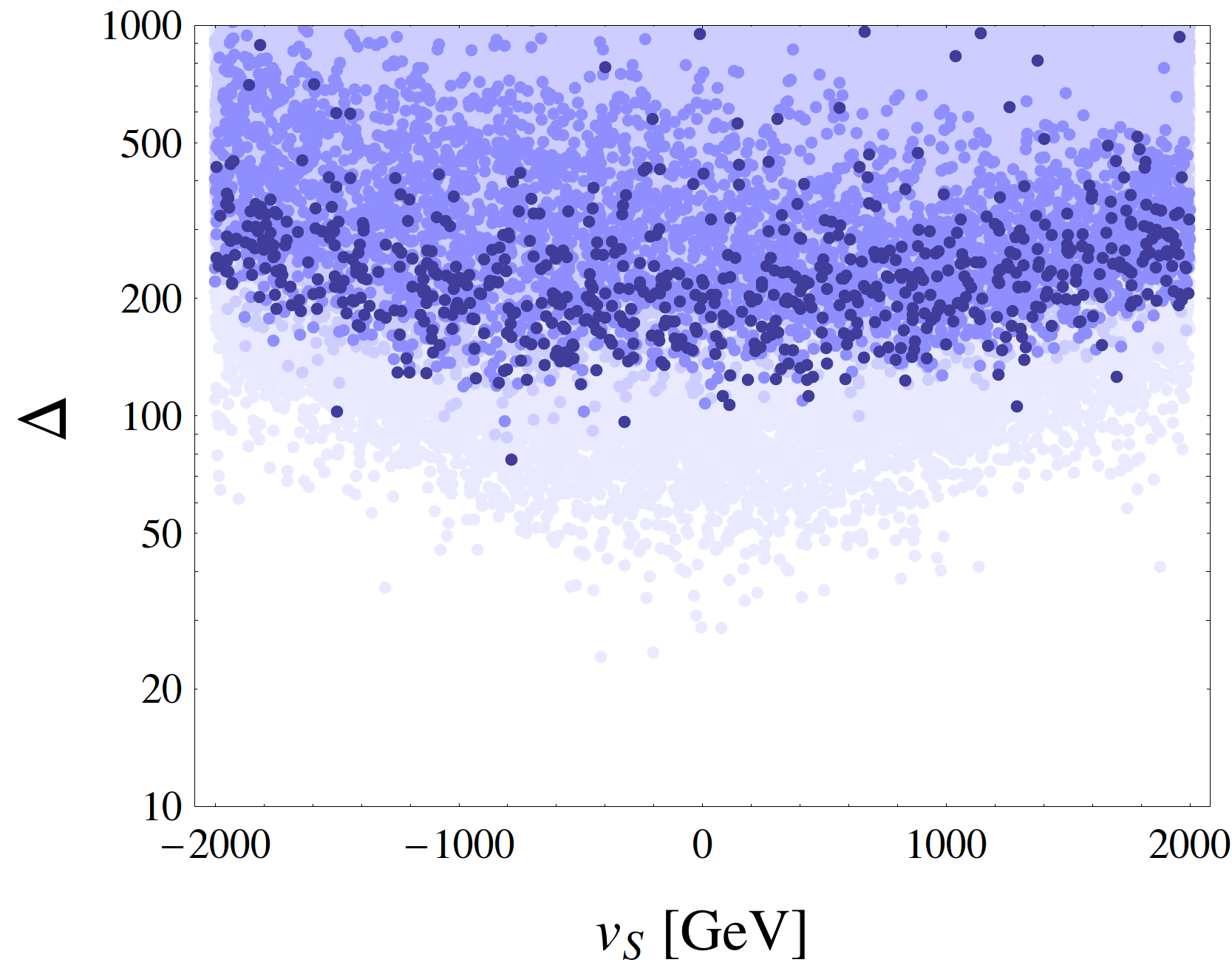}  
\hfill
\includegraphics[width=0.3\linewidth]{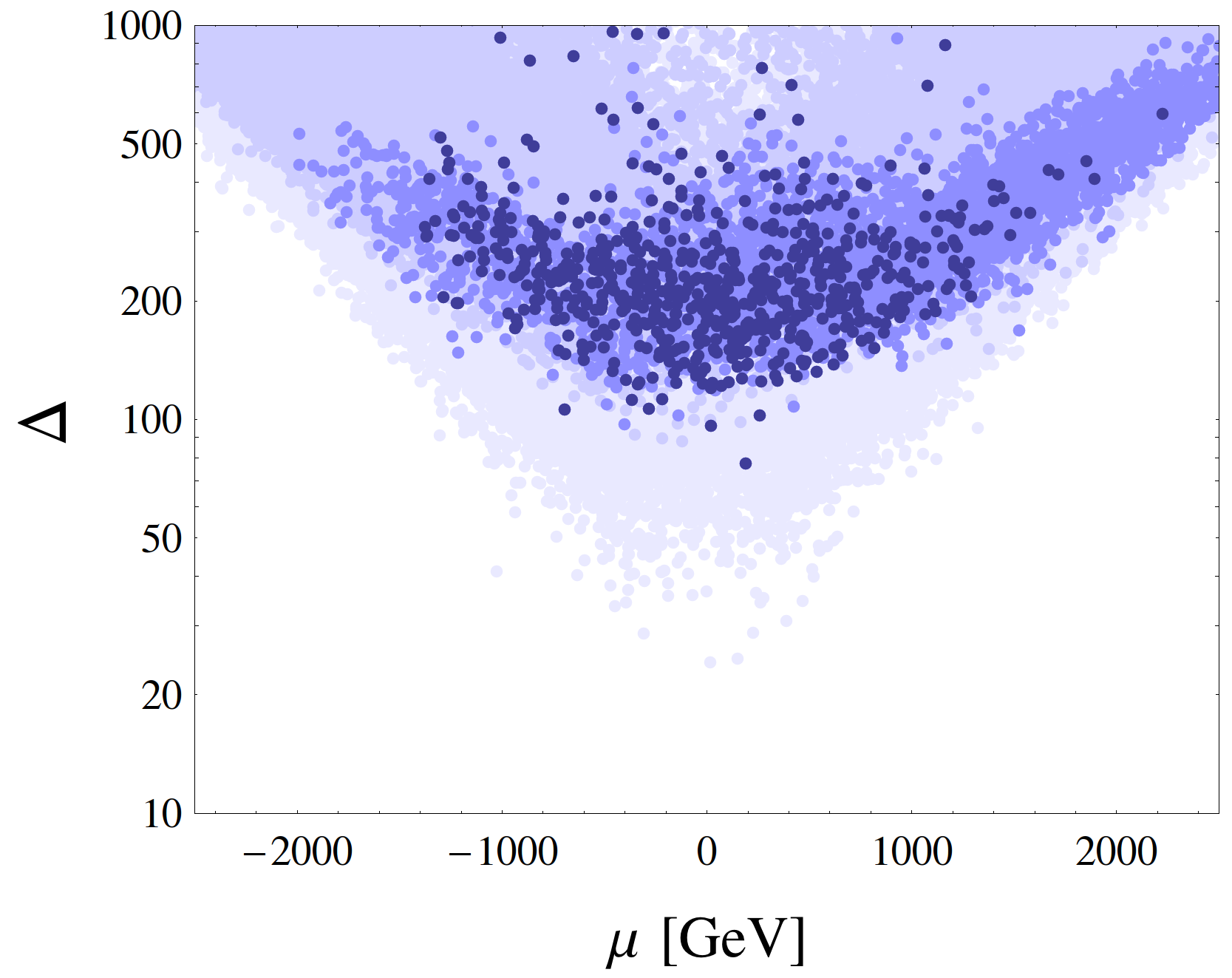}  
\hfill 
\includegraphics[width=0.3\linewidth]{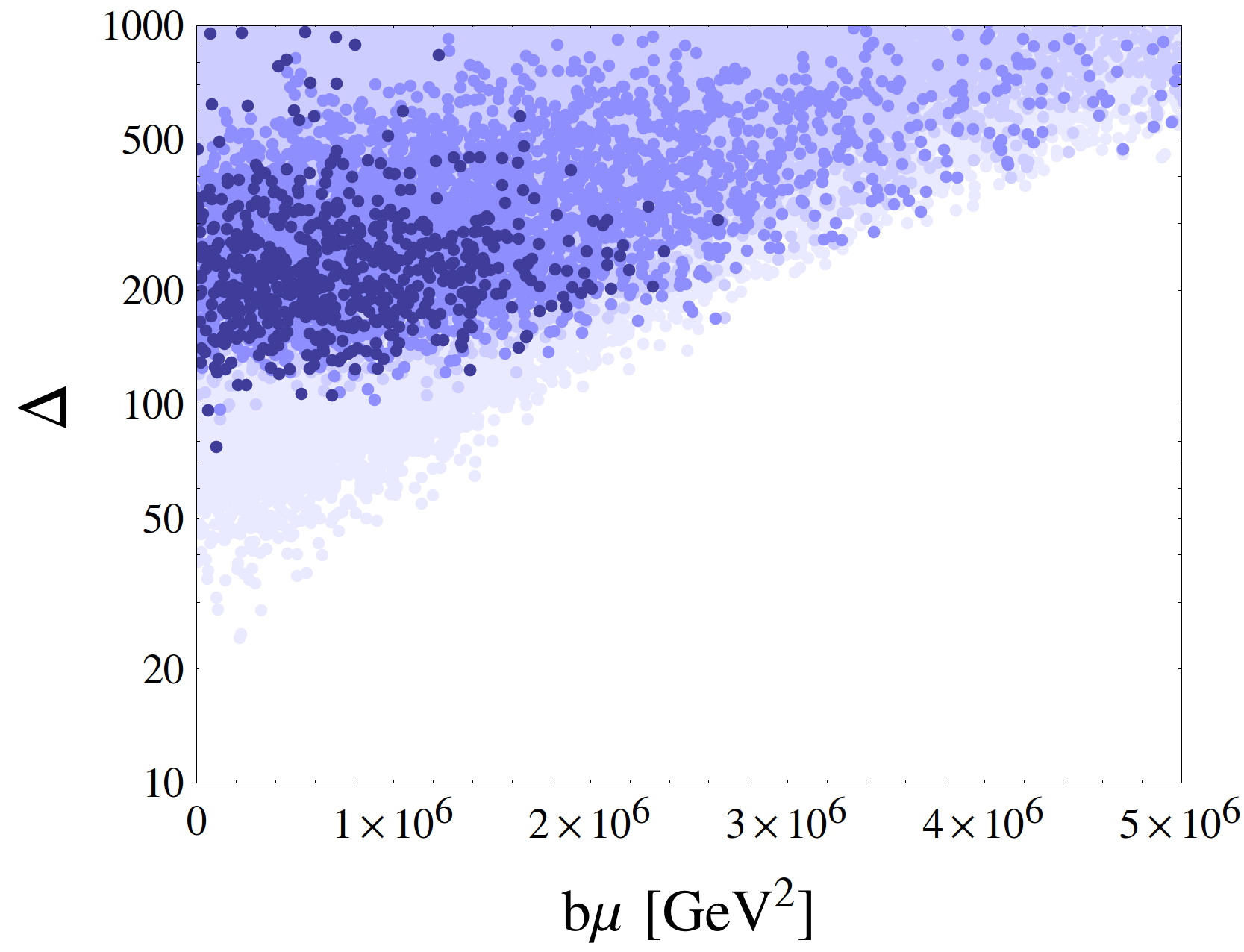}  \\
\includegraphics[width=0.3\linewidth]{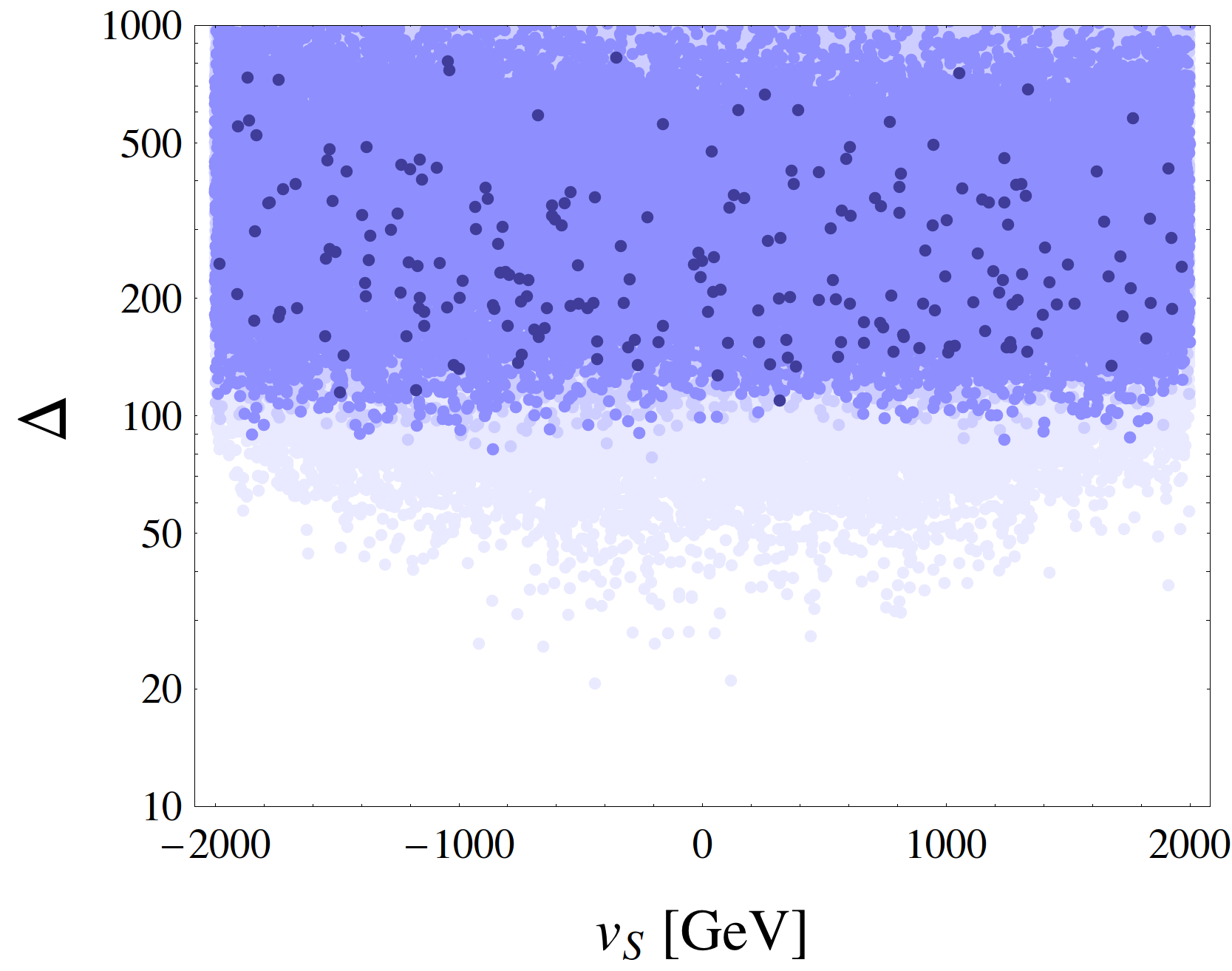}  
\hfill 
\includegraphics[width=0.3\linewidth]{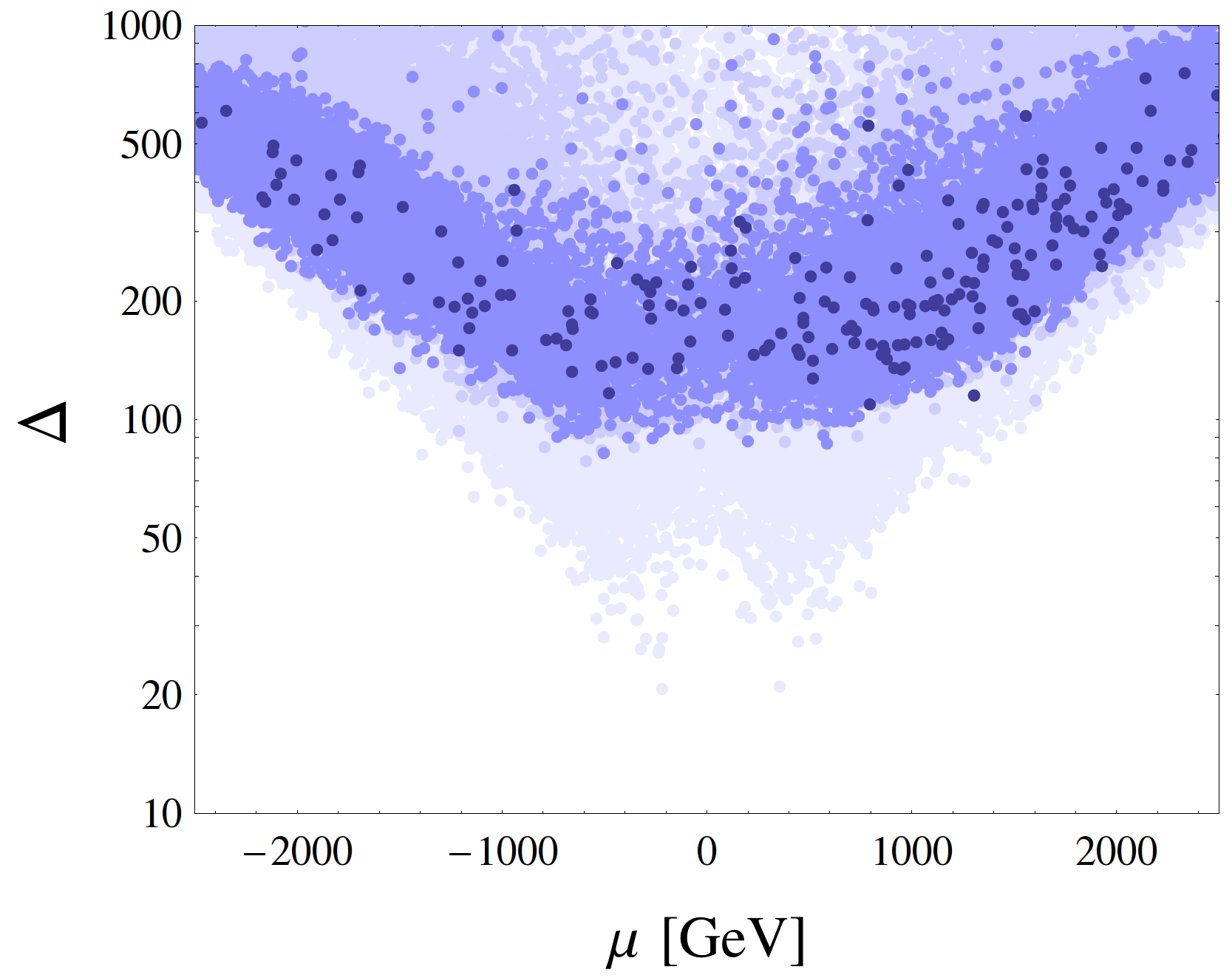}  
\hfill 
\includegraphics[width=0.3\linewidth]{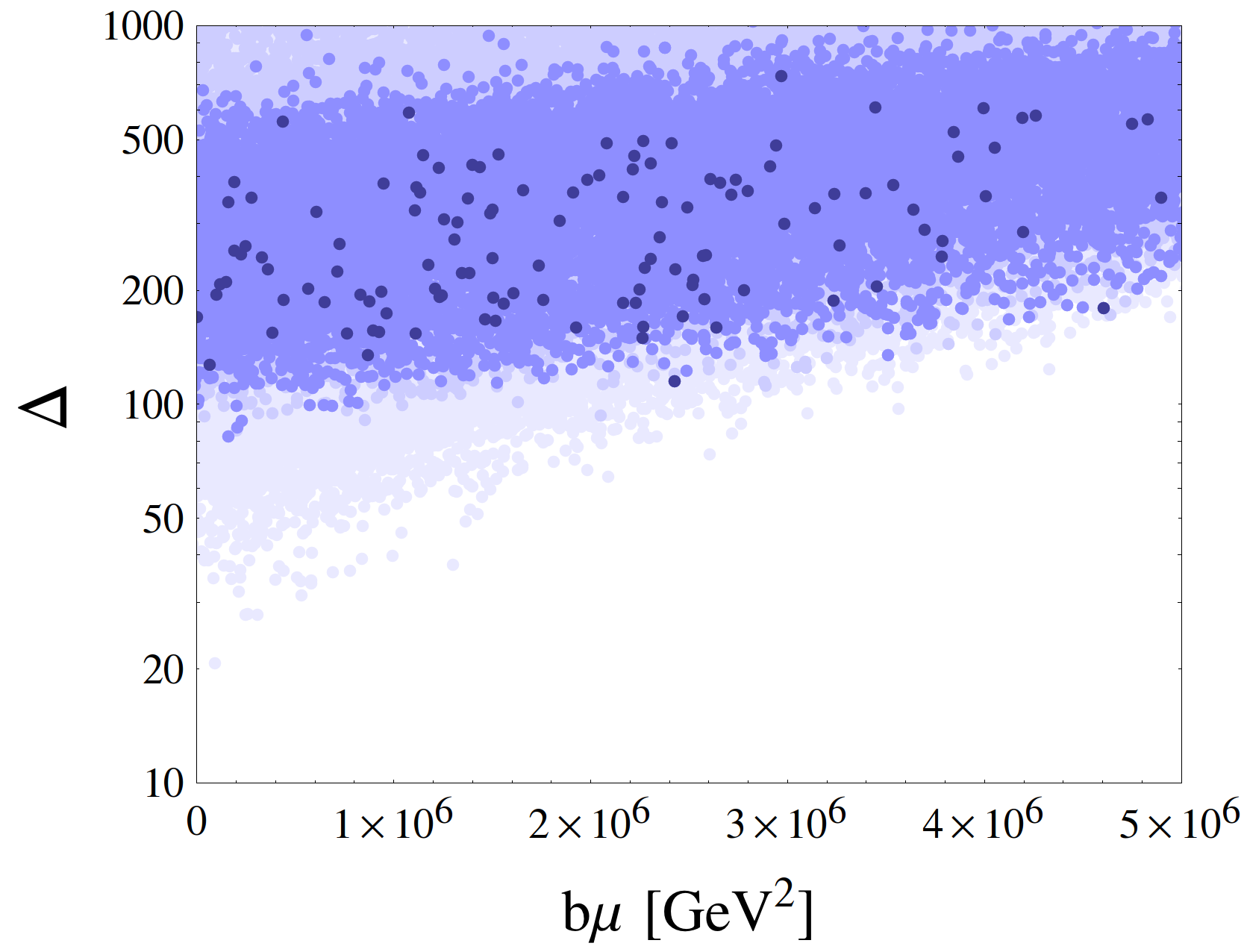}  
\caption{Fine tuning vs.\ $v_s$, $\mu$ and $b\mu$ in the DiracNMSSM (first line) 
and GNMSSM (second line). Here $\mu$ and $b\mu$ are given at the SUSY scale. The color code is the same as in Fig.~\ref{fig:compFT1}.}
\label{fig:compFT4}
\end{figure}
\begin{figure}[tb]
\includegraphics[width=0.3\linewidth]{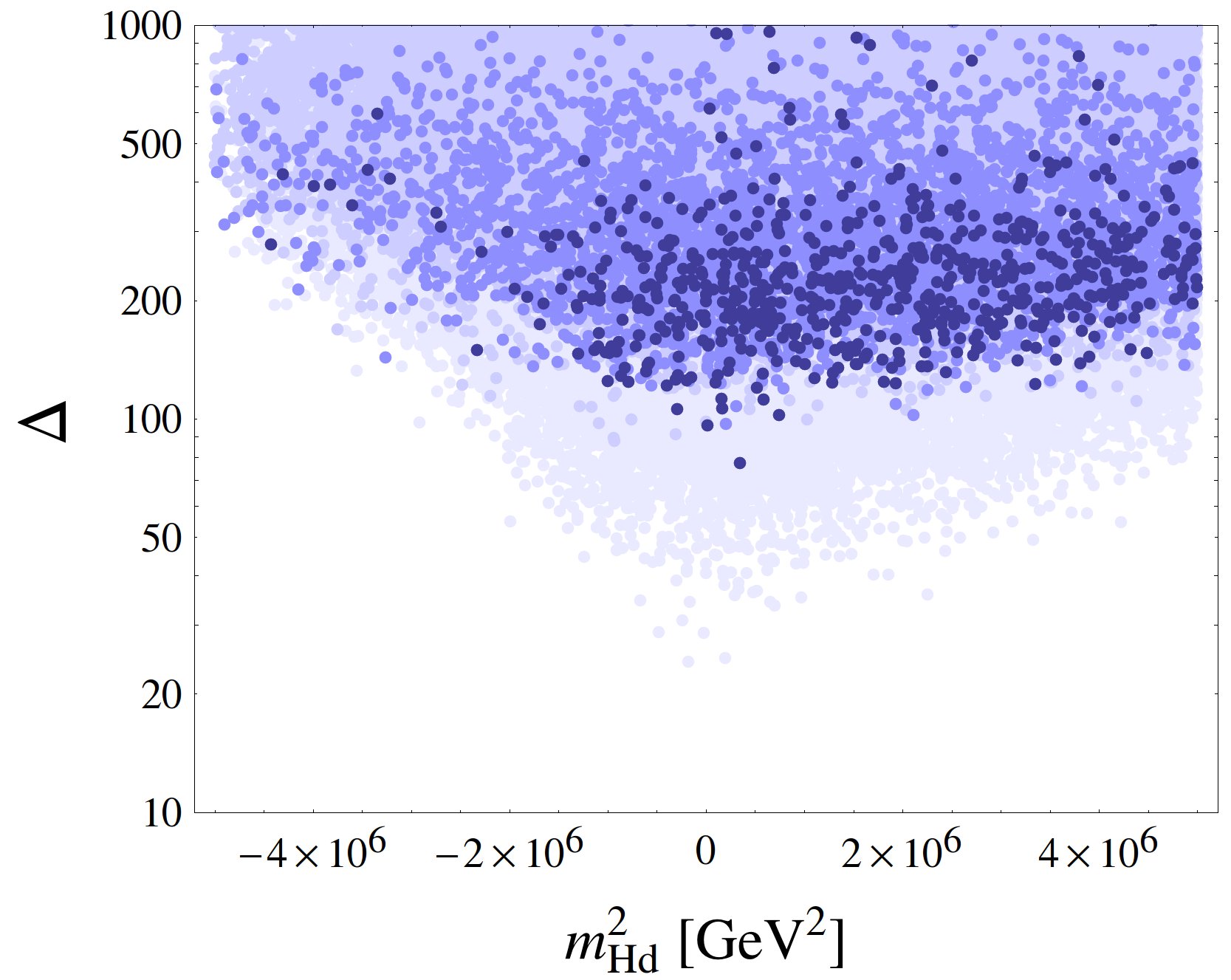}  
\hfill 
\includegraphics[width=0.3\linewidth]{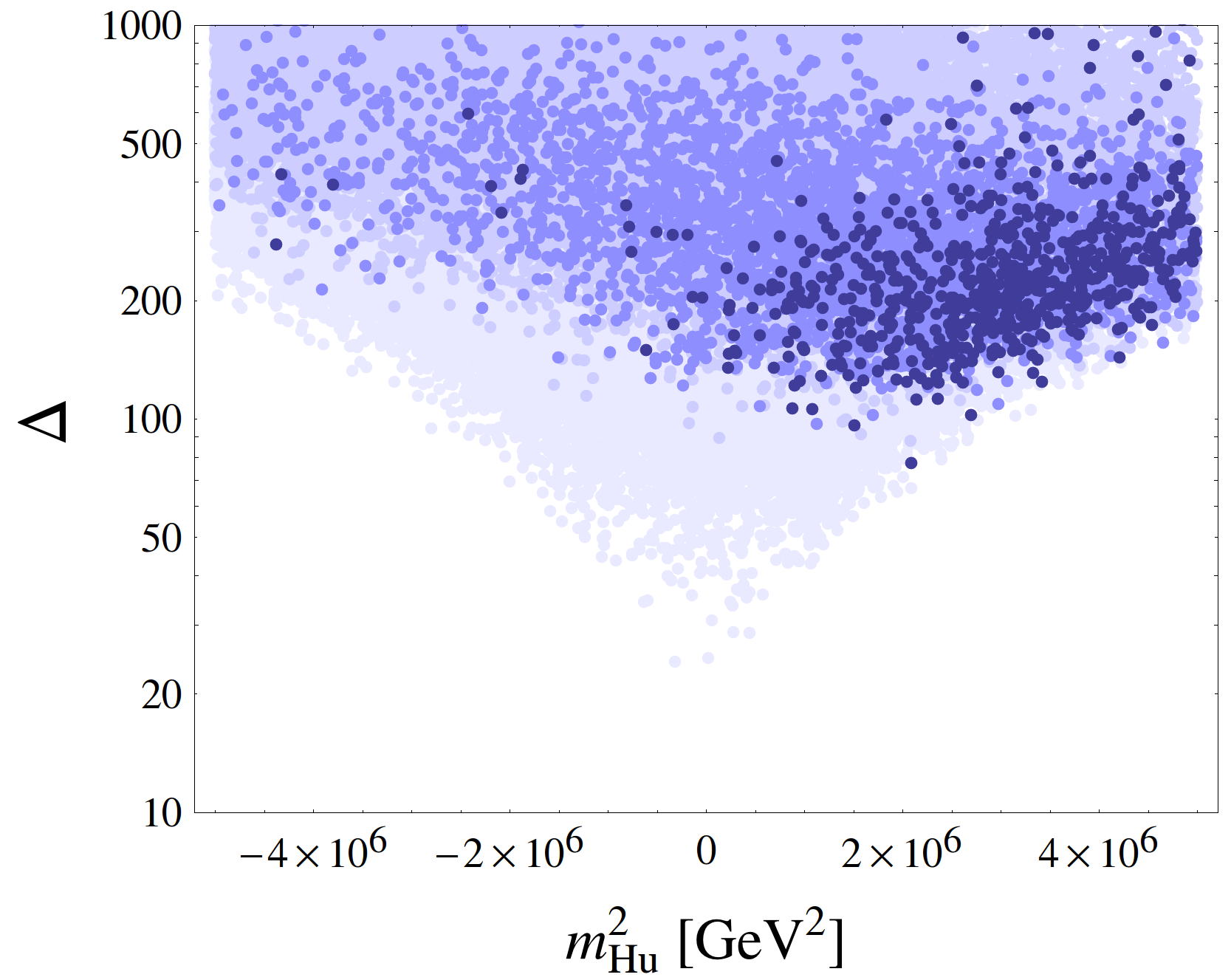}  
\hfill 
\includegraphics[width=0.3\linewidth]{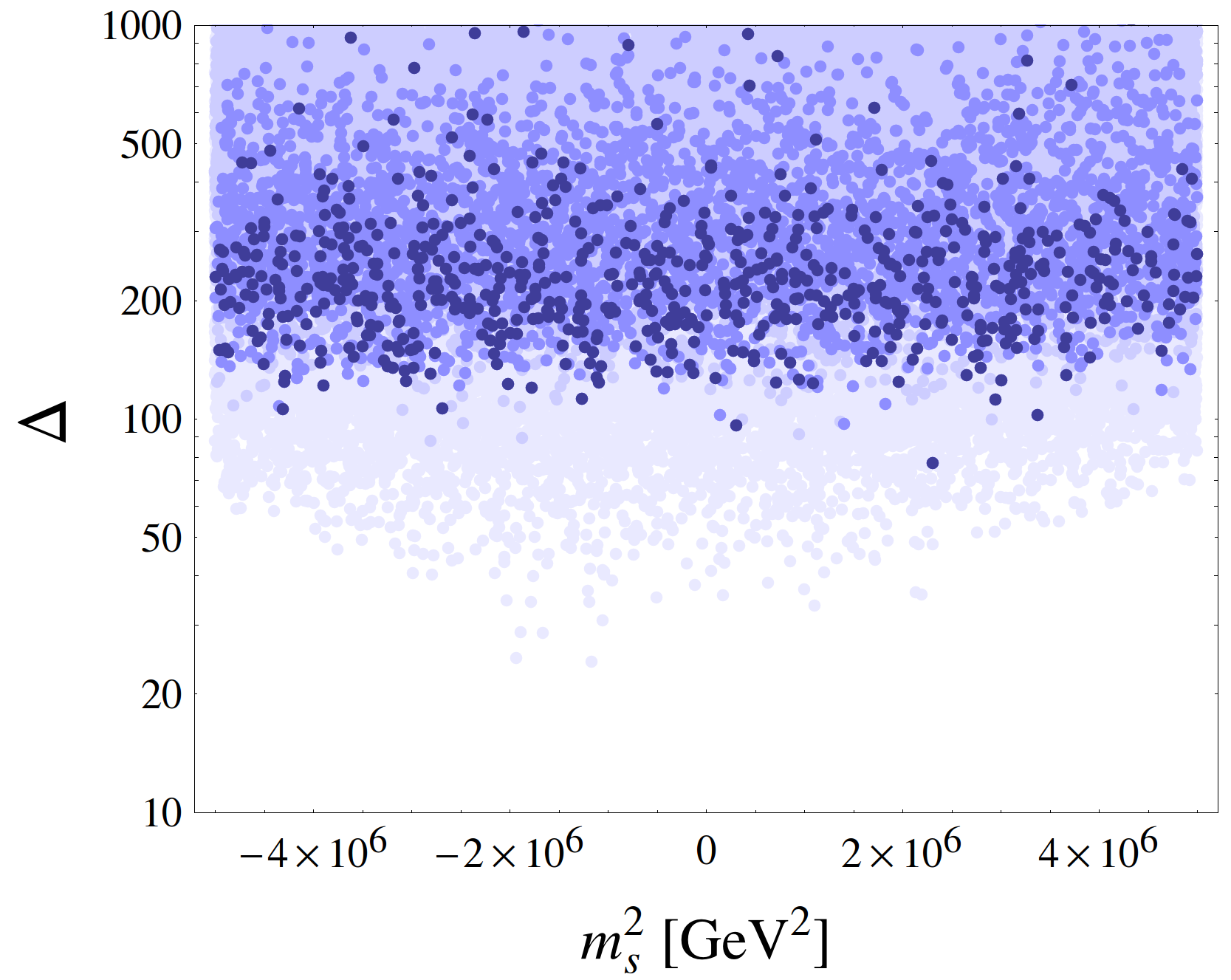}   \\
\includegraphics[width=0.3\linewidth]{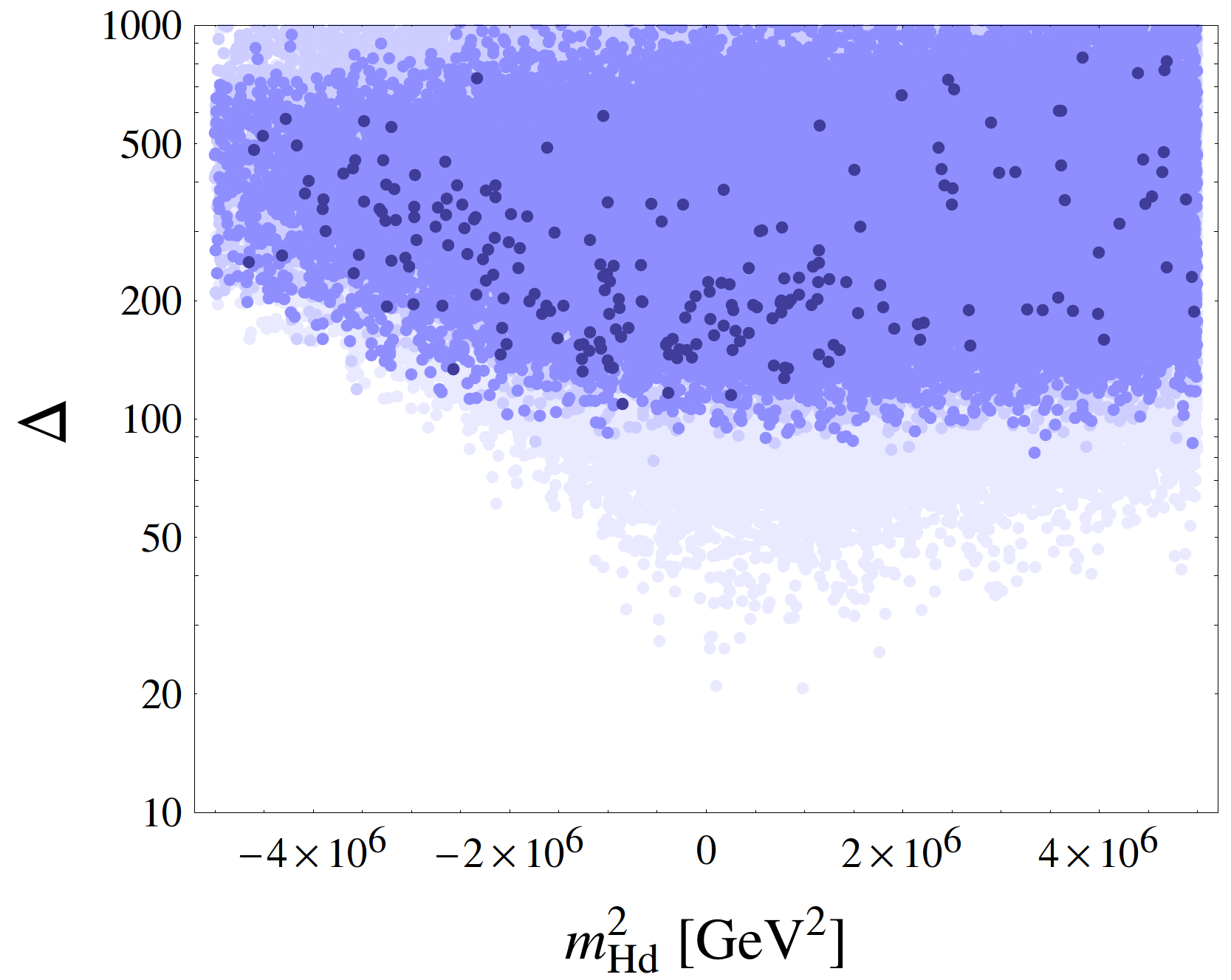}  
\hfill 
\includegraphics[width=0.3\linewidth]{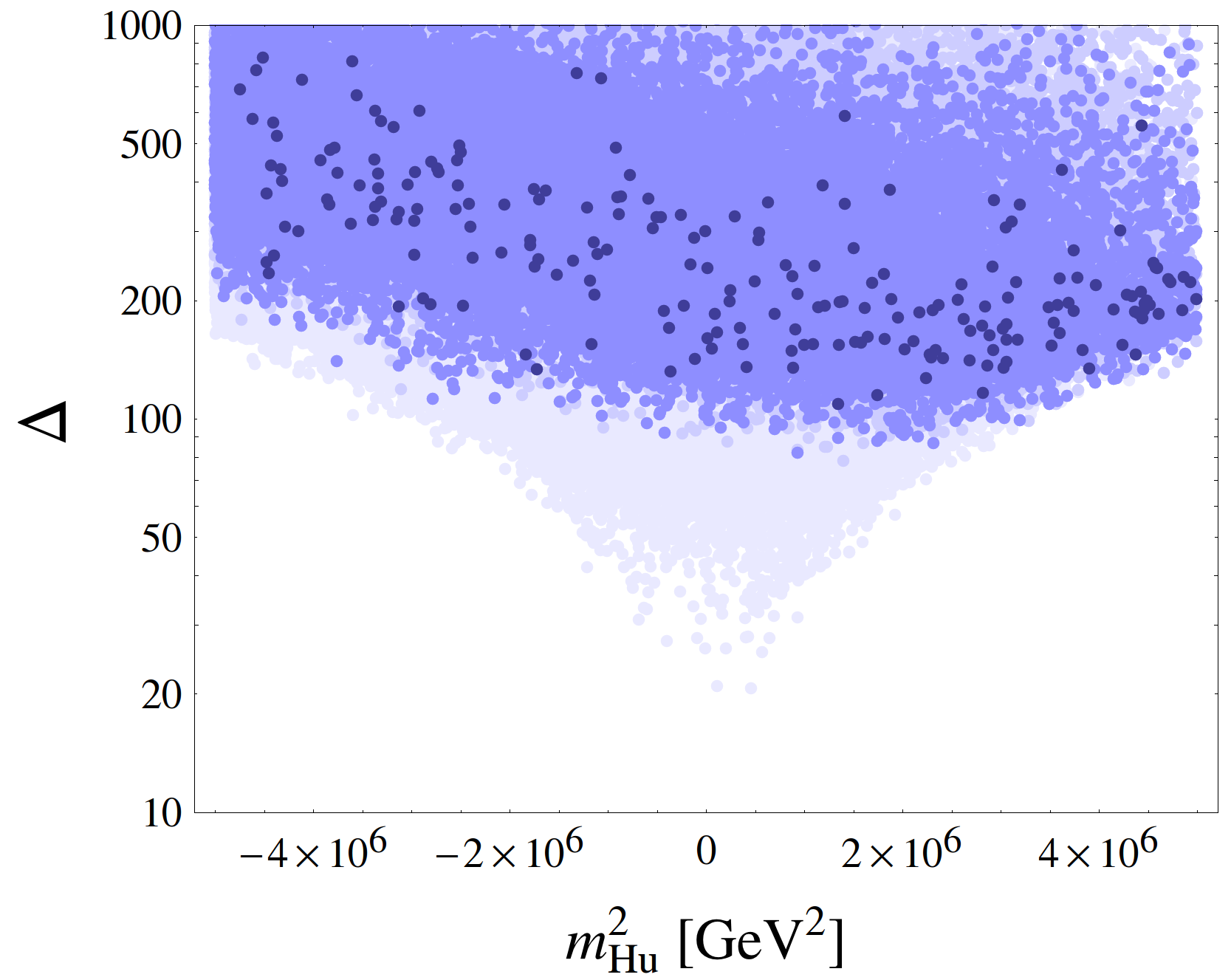}  
\hfill 
\includegraphics[width=0.3\linewidth]{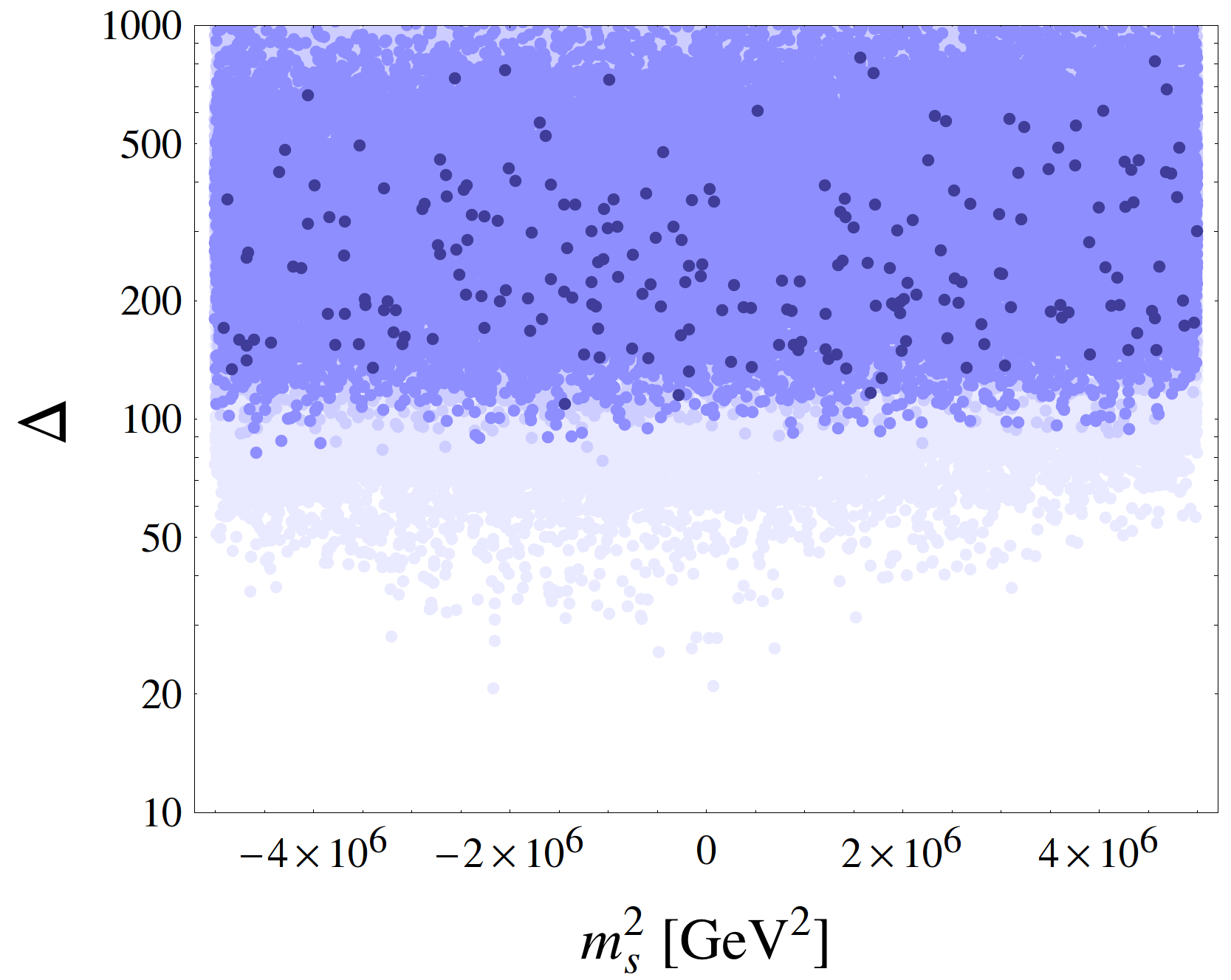}  
\caption{Fine tuning vs. $m_{h_d}^2$, $m_{h_u}^2$ and $m_s^2$ in the DiracNMSSM (first line) 
and GNMSSM (second line). The color code is the same as in Fig.~\ref{fig:compFT2}.}
\label{fig:compFT3}
\end{figure}

We turn now to the dependence of the fine tuning on the different parameters. For this purpose the fine tuning versus $m_0$, $m_{1/2}$ and $A_0$ is shown in Fig.~\ref{fig:compFT1}, versus $\lambda$, $A_\lambda$ and $M_s / \mu_s$ in Fig.~\ref{fig:compFT2}, versus $v_s$, $\mu$ and $b\mu$ in Fig.~\ref{fig:compFT4}, and versus $m_{h_d}^2$, $m_{h_u}^2$ and $m_s^2$ in Fig.~\ref{fig:compFT3}. For the parameters not shown here ($b_s$, $m_{\bar{s}}^2$, $v_{\bar{s}}$, $\kappa$, $A_{\kappa}$) there is no visible dependence of the fine tuning on those parameters.

In Fig.~\ref{fig:compFT1} there is a very strong dependence of the fine tuning in both models on the gaugino mass parameters $m_{1/2}$. Small values for $m_{1/2}$ 
are usually forbidden by the gluino searches at the LHC. That's one of the main reasons which pushes the fine tuning to larger values. The impact of $m_0$ is rather moderate in both models, as long as it is not too large (as we
don't assume the sfermion masses to unify with the soft Higgs masses, the focus point solution for large $m_0$ won't work here). Further the DiracNMSSM and the GNMSSM seem to slightly favour positive $A_0$. 
The fine tuning as function of the additional superpotential parameters $\lambda$ and $M_s$ or $\mu_s$ is shown in Fig.~\ref{fig:compFT2}. While $\lambda$ plays an important role in lifting the Higgs mass to the desired range, only a mild preference towards very large $\lambda$ in terms of the fine tuning is seen. On the other hand the dependence on $M_s$ is much more pronounced. First, there are hardly any valid points for $|M_s| < 1$~TeV. The reason for this in the DiracNMSSM has already been discussed in sec.~\ref{sec:HiggsTree}: usually, large values of $M_s^2$ are needed to get a large enough Higgs mass or even three positive eigenvalues of the Higgs mass matrix squared. Similarly, it is very difficult to find valid parameter points in the GNMSSM which don't suffer from tachyonic states in the Higgs sector\footnote{This seems at odds with a number of NMSSM studies, which of course have $\mu_s=0$. One should note however that the density of valid points in the NMSSM is significantly smaller than in the 
MSSM, which 
indirectly shows up here.}. Furthermore, even for $|M_s| > 4$~TeV there is a strong correlation between the fine tuning and the value of $M_s$ in the DiracNMSSM after the Higgs cut is applied. Usually very large values of $M_s$ are needed to reduce the fine tuning. The reason is that the tree-level mass of the SM like Higgs increases with $M_s$ and therefore for large values of $M_s$ the necessity of large loop corrections due to heavy stops is reduced. 
That's different to the GNMSSM where the relation between the fine tuning and $\mu_s$ is roughly flat in this range even after the Higgs mass cut. 
In Fig.~\ref{fig:compFT4} we show the fine tuning as function of $v_s$. While the fine tuning in the GNMSSM shows hardly any dependence on the singlet VEV after all cuts, the fine tuning in the DiracNMSSM increases with increasing $|v_s|$. Thus, singlet VEVs below 1~TeV are preferred in the DiracNMSSM. One might be surprised that the fine tuning depends on $v_s$ at all since this parameter does not enter eq.~(\ref{eq:measure}). However, one can see from eq.~(\ref{eq:tadMu}) that large $|v_s|$ also leads in general to larger $|\mu|$. That the fine tuning in the MSSM strongly depends on $\mu$ is well known. As we can see in the middle of Fig.~\ref{fig:compFT4} this is also the case for the DiracNMSSM and the GNMSSM. $|\mu|$ should be not larger than a few hundred GeV to have good fine tuning. 

From the model building point of view, the main difference between the GNMSSM and DiracNMSSM is the extended singlet sector of the DiracNMSSM. A second singlet was introduced to allow for smaller values of $m_s^2$ without decoupling of the tree-level contribution to the Higgs mass for large singlet masses. This in turn was supposed to reduce the fine tuning in the DiracNMSSM. However, the plots in the last row in Fig.~\ref{fig:compFT3} show that the fine tuning with respect to $m_s^2$ is rather mild and also in the GNMSSM there are points after all cuts with $m_s^2=0$. This might be the main reason why there is not a significant improvement in the fine tuning in the DiracNMSSM compared to the GNMSSM. Actually, it turns out that the fine tuning in both models is roughly the same for universal gaugino masses as we have seen. 

\subsection{Non-universal gaugino masses}
\label{sec:nonuniversal}
As we have mentioned before, non-universal gaugino masses tend to improve the fine tuning through the appearance of a new ``focus point''  that makes the Higgs mass less sensitive to the gaugino mass scale~\cite{Choi:2005hd,Choi:2006xb,Abe:2007kf,Lebedev:2005ge,Horton:2009ed,Asano:2012sv,Antusch:2012gv,Abe:2012xm,Badziak:2012yg,Gogoladze:2012yf,Yanagida:2013ah}.  
We assume that $a$ and $b$ are fixed by the underlying theory such that their contributions to the fine tuning are absent. As discussed in \cite{Kaminska:2013mya, Horton:2009ed} values of $a$ and $b$ in the low-focus-point region occur naturally in a variety of models.
 
\begin{figure}[!h!] 
 \centering
 \includegraphics[width=0.45\linewidth]{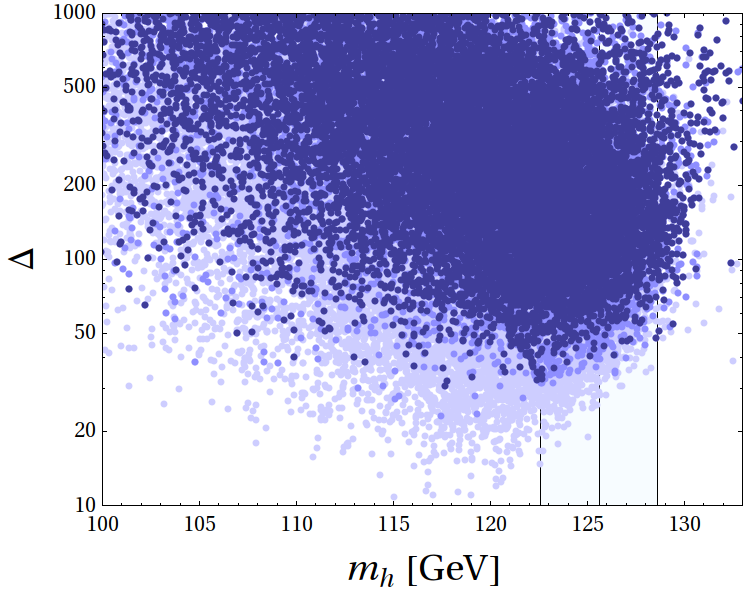}  
 \includegraphics[width=0.45\linewidth]{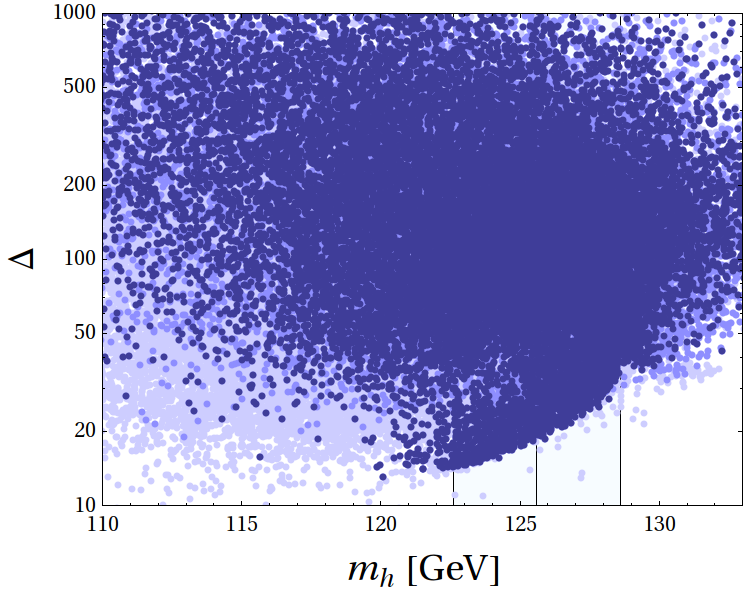}
 \caption{The fine tuning as a function of the SM like Higgs mass allowing for non-universal gaugino masses
 for the DiracNMSSM (left) and the GNMSSM (right). 
 The light blue points are before any cuts.
 For the dark blue points we use appropriate SUSY and dark matter cuts. The minimal FT we find in the DiracNMSSM is $\Delta=32$, in the GNMSSM it is $\Delta=14$.}
 \label{fig:4}
 \end{figure}  
 
We show in Fig.~\ref{fig:4} the overall fine tuning vs.\ the mass of the SM-like Higgs for this case. We see that in both models it improves significantly. The best fine tuning we find now for the DiracNMSSM fulfilling all experimental constraints and satisfying the upper limit of the neutralino abundance is $\Delta \simeq 32$, for the GNMSSM we even find points with $\Delta \simeq 14$ \footnote{However, a strict ratio of the wino to gluino masses is not necessary as long as both are comparable at the weak scale; in the case of the GNMSSM, including the fine tuning with respect to a and b, the minimum fine tuning is still only 35.  Hence, dropping the assumption of definite a and b ratios does not greatly increase the fine tuning.}. The plot also shows that before SUSY cuts the fine tuning in both models is very similar.
We find the difference between the two models in the case of nonuniversal gaugino masses to be the following: The mass of the higgsino-like neutralino is set
by $\mu_\text{eff}=\mu + \tfrac{1}{\sqrt{2}} \lambda_{EW} v_s$, while the fine tuning is mainly dominated by $\mu$. In the GNMSSM it seems to be easier to have sizable $\mu_\text{eff} \gtrsim 700 \gev$ which would allow for a compressed spectrum and hence lighter gluinos. In the DiracNMSSM on the other hand $\mu_\text{eff}$ is typically closer to $\mu$.
This means that for very compressed spectra with $m_\text{higgsino} \sim m_\text{gluino} \sim 700 \gev$ the $\mu$ term is sizable, implying moderate fine tuning.
On the other hand for very small values of $\mu$ we typically have a rather light higgsino-like LSP. In this case a compressed spectrum is not possible and the gluino has to be 
correspondingly heavier, implying larger fine tuning. 

\begin{figure}[!h!]
\begin{center}
\includegraphics[width=0.49\linewidth]{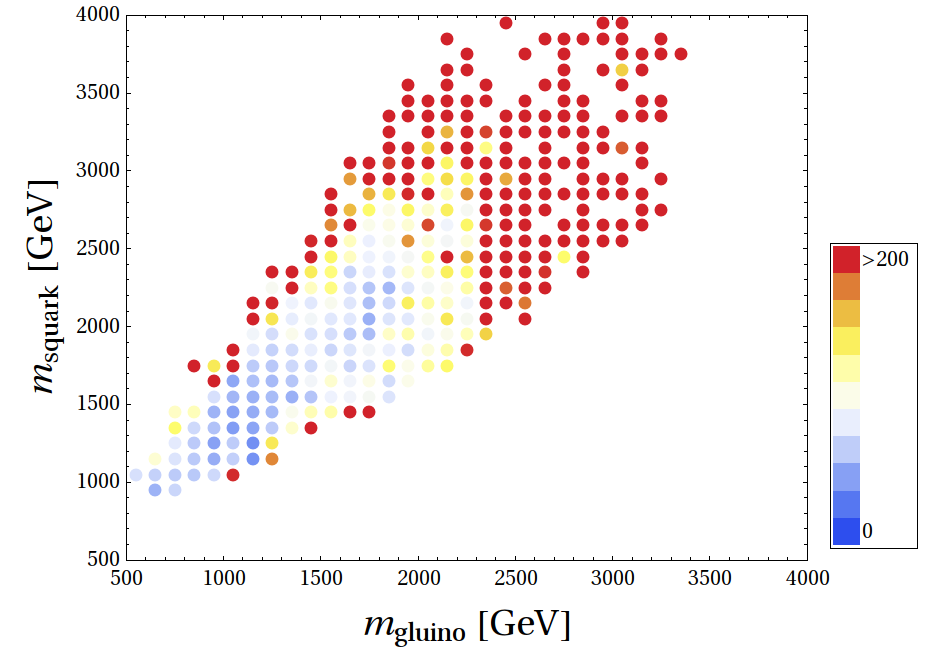}
\includegraphics[width=0.49\linewidth]{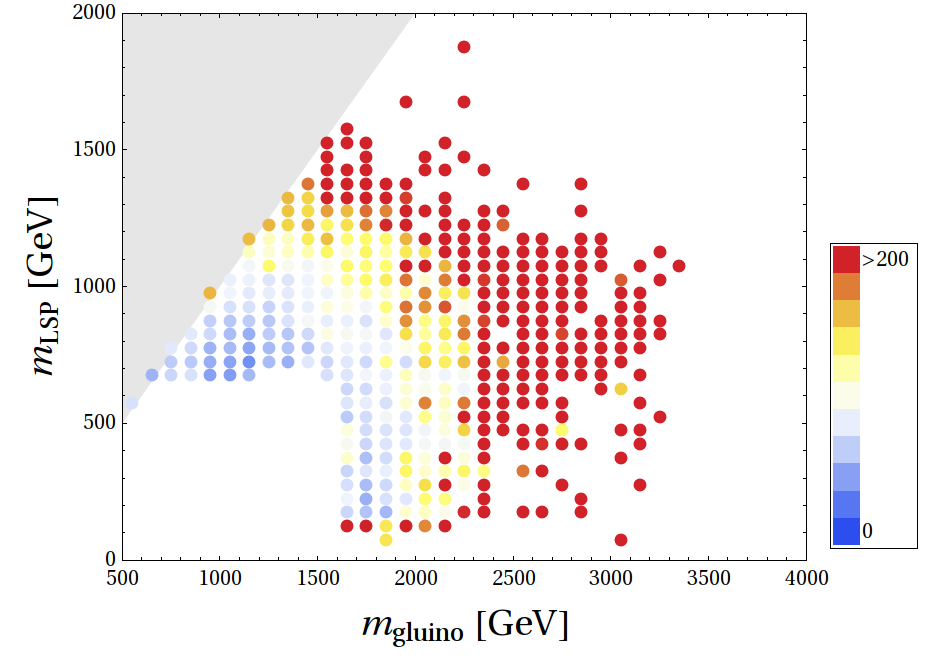}
\includegraphics[width=0.49\linewidth]{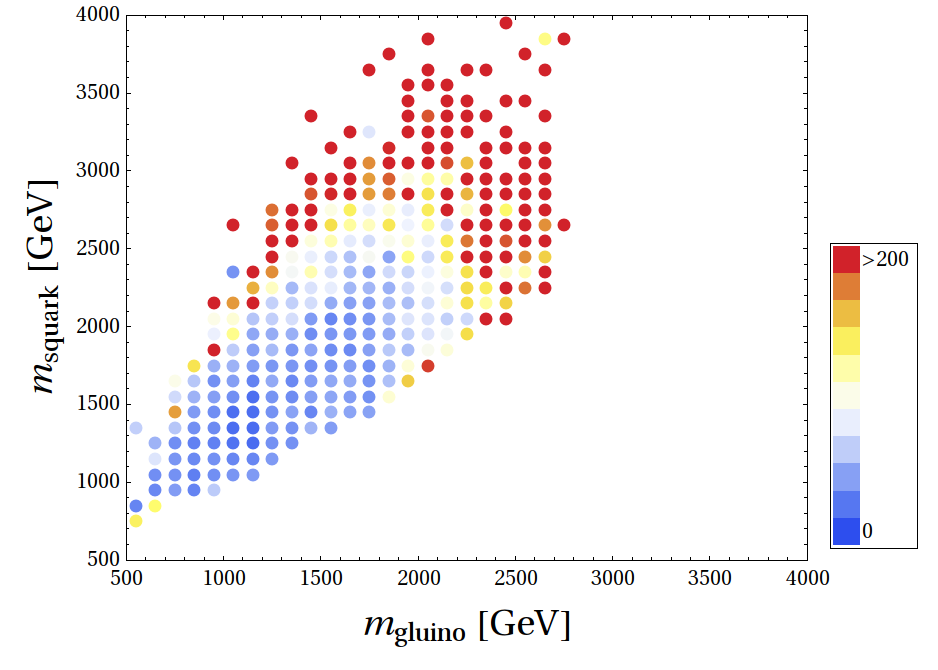}
\includegraphics[width=0.49\linewidth]{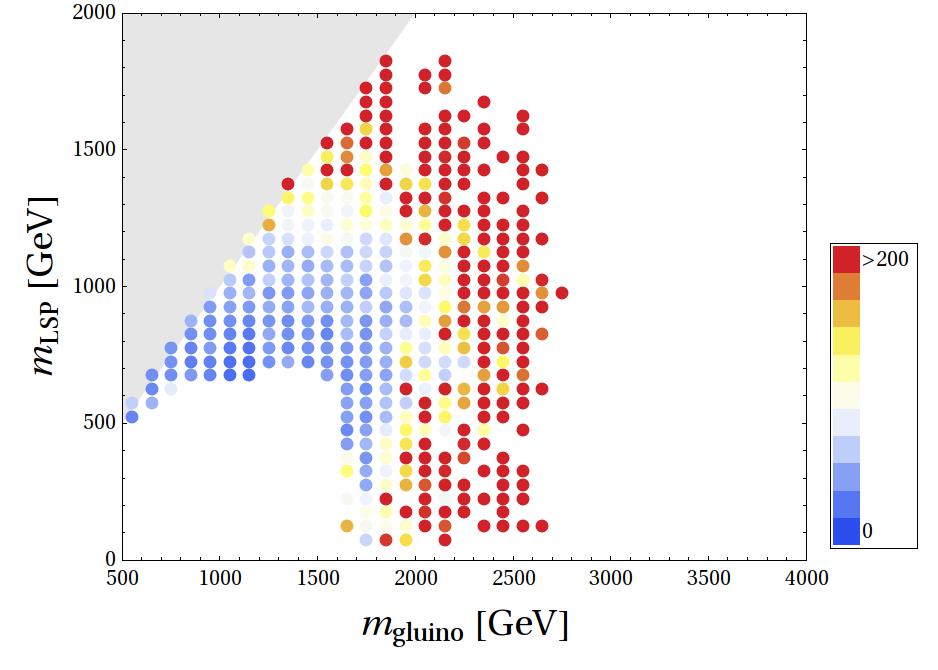}
\caption{Smallest fine tuning in the gluino--squark and gluino--LSP plane for the DiracNMSSM (top)
and the GNMSSM (bottom).
It can be seen that low fine tuning often corresponds to compressed spectra.}
\label{GNMSSMfinetuningab}
\end{center}
\end{figure}

In Fig.~\ref{GNMSSMfinetuningab} we present the best fine tuning we find in the $(m_{\rm gluino},m_{\rm squark})$ and in the  $(m_{\rm gluino}, m_{\rm LSP})$ plane. We see that a very good fine tuning is not only possible close to the parameter regions where the gluino is nearly degenerate with the LSP but also for heavy gluinos above 2~TeV. Also note that our scans were optimized to find the smallest fine tuning, so it is conceivable that the fine-tuning for large masses is overestimated. The heavy gluino case will be hard to exclude by the next LHC runs \cite{Ulmer:2013csa}. Thus, despite the excellent performance of the LHC experiments there is still the possibility of SUSY with a fine tuning less than 100 which can't be tested in the near future. However, to find models with this small fine tuning for gluinos above 2~TeV one has to give up the most constrained models with universal soft parameters for all scalars and all gauginos. 

\section{Summary and Conclusions}
\label{sec:conclusion}
We have performed a careful evaluation of the level of fine tuning in the DiracNMSSM. For this purpose we have implemented this model in public computer tools to get a precise prediction for the Higgs mass, the dark matter relic abundance, and the SUSY mass spectrum. We have considered rather general high scale boundary conditions -- in particular we assumed that the Higgs and singlet soft terms are independent of $m_0$ and of each other. Also all the A-terms were taken to be independent. If we force the gaugino mass terms to unify at the GUT scale the minimal fine tuning allowed by all experimental constraints is about 70 for both the DiracNMSSM and the GNMSSM. If we relax the unification conditions in the gaugino sector the minimal fine tuning gets improved to 32 in the DiracNMSSM and 14 in the GNMSSM.
Hence, both models significantly improve the fine tuning situation compared to the MSSM and the level of fine tuning is comparable in both models, albeit slightly lower in the GNMSSM.

\section*{Acknowledgements}
The research presented here was partially supported by the EU ITN grant UNILHC 237920 (Unification in the LHC era).
FS is supported by the BMBF
PT DESY Verbundprojekt 05H2013-THEORIE 'Vergleich von LHC-Daten mit supersymmetrischen Modellen'.

\begin{appendix}

\section{Renormalisation Group Equations}
\label{app:RGEs}
For those parameters present in the MSSM, we give only the difference to the RGEs in comparison to the MSSM. The convention for the $\beta$ function and
anomalous dimensions are
\begin{equation}
\beta_x = \frac{1}{16\pi^2} \beta^{(1)}_x+\frac{1}{(16\pi^2)^2} \beta^{(2)}_x\,,\hspace{1cm}
\gamma_x = \frac{1}{16\pi^2} \gamma^{(1)}_x+\frac{1}{(16\pi^2)^2} \gamma^{(2)}_x
\end{equation}
The calculation of the $\beta$-functions is performed by \SARAH. The calculation of $\beta$-functions for SUSY models
up to two loop with \SARAH are based on the generic results given in Refs.~\cite{Martin:1993zk,Yamada:1994id,Fonseca:2011vn,Goodsell:2012fm,Sperling:2013eva,Sperling:2013xqa}. 

\subsection{Anomalous Dimensions}
{\allowdisplaybreaks \begin{align} 
\Delta \gamma_{\hat{q}}^{(2)} & =  
- |\lambda|^2 \Big({Y_{d}^{\dagger}  Y_d} + {Y_{u}^{\dagger}  Y_u}\Big)\\ 
\Delta  \gamma_{\hat{l}}^{(2)} & =  
- |\lambda|^2 {Y_{e}^{\dagger}  Y_e} \\ 
\Delta \gamma_{\hat{H}_d}^{(1)} & =  
|\lambda|^2\\ 
\Delta \gamma_{\hat{H}_d}^{(2)} & =  
-3 |\lambda|^2 \Big(|\lambda|^2  + \mbox{Tr}\Big({Y_u  Y_{u}^{\dagger}}\Big)\Big)\\ 
\Delta \gamma_{\hat{H}_u}^{(1)} & =  
|\lambda|^2\\ 
\Delta \gamma_{\hat{H}_u}^{(2)} & =  
- |\lambda|^2 \Big(3 |\lambda|^2  + 3 \mbox{Tr}\Big({Y_d  Y_{d}^{\dagger}}\Big)  + \mbox{Tr}\Big({Y_e  Y_{e}^{\dagger}}\Big)\Big)\\ 
\Delta \gamma_{\hat{d}}^{(2)} & =  
-2 |\lambda|^2 {Y_d^*  Y_{d}^{T}} \\ 
\Delta \gamma_{\hat{u}}^{(2)} & =  
-2 |\lambda|^2 {Y_u^*  Y_{u}^{T}} \\ 
\Delta \gamma_{\hat{e}}^{(2)} & =  
-2 |\lambda|^2 {Y_e^*  Y_{e}^{T}} \\ 
\gamma_{\hat{s}}^{(1)} & =  
2 |\lambda|^2 \\ 
\gamma_{\hat{s}}^{(2)} & =  
-\frac{2}{5} |\lambda|^2 \Big(10 |\lambda|^2  -15 g_{2}^{2}  + 15 \mbox{Tr}\Big({Y_d  Y_{d}^{\dagger}}\Big)  + 15 \mbox{Tr}\Big({Y_u  Y_{u}^{\dagger}}\Big)  -3 g_{1}^{2}  + 5 \mbox{Tr}\Big({Y_e  Y_{e}^{\dagger}}\Big) \Big)\\
\gamma_{\hat{\bar{s}}}^{(1)} & =  0 \\
\gamma_{\hat{\bar{s}}}^{(2)} & =  0
\end{align} } 
\subsection{Gauge Couplings}
{\allowdisplaybreaks  \begin{align} 
\Delta  \beta_{g_1}^{(2)} & =  
-\frac{6}{5} g_{1}^{3} |\lambda|^2 \\ 
\Delta  \beta_{g_2}^{(2)} & =  
-2 g_{2}^{3} |\lambda|^2 
\end{align}} 
\subsection{Gaugino Mass Parameters}
{\allowdisplaybreaks  \begin{align} 
\Delta  \beta_{M_1}^{(2)} & =  
-\frac{12}{5} g_{1}^{2} \lambda^* \Big(M_1 \lambda  - T_{\lambda} \Big)\\ 
\Delta  \beta_{M_2}^{(2)} & =  
4 g_{2}^{2} \lambda^* \Big(- M_2 \lambda  + T_{\lambda}\Big)
\end{align}} 
\subsection{Trilinear Superpotential Parameters}
{\allowdisplaybreaks  \begin{align} 
\Delta \beta_{Y_d}^{(1)} & =  
Y_d |\lambda|^2 \\ 
\Delta \beta_{Y_d}^{(2)} & =  
- |\lambda|^2 \Big(3 Y_d |\lambda|^2  + 3 Y_d \mbox{Tr}\Big({Y_u  Y_{u}^{\dagger}}\Big)  + 3 {Y_d  Y_{d}^{\dagger}  Y_d}  + {Y_d  Y_{u}^{\dagger}  Y_u}\Big)\\ 
\Delta \beta_{Y_e}^{(1)} & =  
Y_e |\lambda|^2 \\ 
\Delta \beta_{Y_e}^{(2)} & =  
-3 |\lambda|^2 \Big(Y_e |\lambda|^2  + Y_e \mbox{Tr}\Big({Y_u  Y_{u}^{\dagger}}\Big)  + {Y_e  Y_{e}^{\dagger}  Y_e}\Big)\\ 
\Delta \beta_{Y_u}^{(1)} & =  
Y_u |\lambda|^2 \\ 
\Delta \beta_{Y_u}^{(2)} & =  
- |\lambda|^2 \Big(3 Y_u |\lambda|^2  + 3 Y_u \mbox{Tr}\Big({Y_d  Y_{d}^{\dagger}}\Big)  + 3 {Y_u  Y_{u}^{\dagger}  Y_u}  + Y_u \mbox{Tr}\Big({Y_e  Y_{e}^{\dagger}}\Big)  + {Y_u  Y_{d}^{\dagger}  Y_d}\Big)\\
\beta_{\lambda}^{(1)} & =  
-3 g_{2}^{2} \lambda  + 3 \lambda \mbox{Tr}\Big({Y_d  Y_{d}^{\dagger}}\Big)  + 3 \lambda \mbox{Tr}\Big({Y_u  Y_{u}^{\dagger}}\Big)  + 4 \lambda^{2} \lambda^*  -\frac{3}{5} g_{1}^{2} \lambda  + \lambda \mbox{Tr}\Big({Y_e  Y_{e}^{\dagger}}\Big) \\ 
\beta_{\lambda}^{(2)} & =  
-\frac{1}{50} \lambda \Big(500 |\lambda|^4-207 g_{1}^{4} -90 g_{1}^{2} g_{2}^{2} -375 g_{2}^{4} +20 \Big(g_{1}^{2}-40 g_{3}^{2} \Big)\mbox{Tr}\Big({Y_d  Y_{d}^{\dagger}}\Big) \nonumber \\ 
 &-20 g_{1}^{2} \mbox{Tr}\Big(3{Y_e  Y_{e}^{\dagger}}+2{Y_u  Y_{u}^{\dagger}}\Big) -30 |\lambda|^2 \Big(10 g_{2}^{2}  -5\mbox{Tr}\Big(3{Y_d  Y_{d}^{\dagger}}+3{Y_u  Y_{u}^{\dagger}}+{Y_e  Y_{e}^{\dagger}}\Big)  + 2 g_{1}^{2} \Big) \nonumber \\ 
 &-800 g_{3}^{2} \mbox{Tr}\Big({Y_u  Y_{u}^{\dagger}}\Big) +15 \mbox{Tr}\Big(3{Y_d  Y_{d}^{\dagger}  Y_d  Y_{d}^{\dagger}}+2{Y_d  Y_{u}^{\dagger}  Y_u  Y_{d}^{\dagger}}+{Y_e  Y_{e}^{\dagger}  Y_e  Y_{e}^{\dagger}}+3{Y_u  Y_{u}^{\dagger}  Y_u  Y_{u}^{\dagger}}\Big) \Big) 
\end{align}} 
\subsection{Bilinear Superpotential Parameters}
{\allowdisplaybreaks  \begin{align} 
\Delta \beta_{\mu}^{(1)} & =  
2 \mu |\lambda|^2 \\ 
\Delta \beta_{\mu}^{(2)} & =  
- \mu |\lambda|^2 \Big(3 \mbox{Tr}\Big({Y_d  Y_{d}^{\dagger}}\Big)  + 3 \mbox{Tr}\Big({Y_u  Y_{u}^{\dagger}}\Big)  + 6 |\lambda|^2  + \mbox{Tr}\Big({Y_e  Y_{e}^{\dagger}}\Big)\Big)\\ 
\beta_{M_s}^{(1)} & =  
2 M_s |\lambda|^2 \\ 
\beta_{M_s}^{(2)} & =  
-\frac{2}{5} M_s |\lambda|^2 \Big(10 |\lambda|^2  -15 g_{2}^{2}  + 15 \mbox{Tr}\Big({Y_d  Y_{d}^{\dagger}}\Big)  + 15 \mbox{Tr}\Big({Y_u  Y_{u}^{\dagger}}\Big)  -3 g_{1}^{2}  + 5 \mbox{Tr}\Big({Y_e  Y_{e}^{\dagger}}\Big) \Big)
\end{align}} 
\subsection{Trilinear Soft-Breaking Parameters}
{\allowdisplaybreaks  \begin{align} 
\Delta \beta_{T_d}^{(1)} & =  
\lambda^* \Big(2 Y_d T_{\lambda}  + \lambda T_d \Big)\\ 
\Delta \beta_{T_d}^{(2)} & =  
- \lambda^* \Big(3 |\lambda|^2 \Big(4 Y_d T_{\lambda}  + \lambda T_d \Big)+2 T_{\lambda} \Big(3 Y_d \mbox{Tr}\Big({Y_u  Y_{u}^{\dagger}}\Big)  + 3 {Y_d  Y_{d}^{\dagger}  Y_d}  + {Y_d  Y_{u}^{\dagger}  Y_u}\Big)\nonumber \\ 
 &+\lambda \Big(2 {Y_d  Y_{u}^{\dagger}  T_u}  + 3 T_d \mbox{Tr}\Big({Y_u  Y_{u}^{\dagger}}\Big)  + 4 {Y_d  Y_{d}^{\dagger}  T_d}  + 5 {T_d  Y_{d}^{\dagger}  Y_d}  + 6 Y_d \mbox{Tr}\Big({Y_{u}^{\dagger}  T_u}\Big)  + {T_d  Y_{u}^{\dagger}  Y_u}\Big)\Big)\\ 
\Delta \beta_{T_e}^{(1)} & =  
\lambda^* \Big(2 Y_e T_{\lambda}  + \lambda T_e \Big)\\ 
\Delta \beta_{T_e}^{(2)} & =  
- \lambda^* \Big(3 |\lambda|^2 \Big(4 Y_e T_{\lambda}  + \lambda T_e \Big)+6 T_{\lambda} \Big(Y_e \mbox{Tr}\Big({Y_u  Y_{u}^{\dagger}}\Big)  + {Y_e  Y_{e}^{\dagger}  Y_e}\Big)\nonumber \\ 
 &+\lambda \Big(3 T_e \mbox{Tr}\Big({Y_u  Y_{u}^{\dagger}}\Big)  + 4 {Y_e  Y_{e}^{\dagger}  T_e}  + 5 {T_e  Y_{e}^{\dagger}  Y_e}  + 6 Y_e \mbox{Tr}\Big({Y_{u}^{\dagger}  T_u}\Big) \Big)\Big)\\ 
\Delta \beta_{T_u}^{(1)} & =  
\lambda^* \Big(2 Y_u T_{\lambda}  + \lambda T_u \Big)\\ 
\Delta \beta_{T_u}^{(2)} & =  
- \lambda^* \Big(3 |\lambda|^2 \Big(4 Y_u T_{\lambda}  + \lambda T_u \Big)+2 T_{\lambda} \Big(Y_u \mbox{Tr}\Big(3{Y_d  Y_{d}^{\dagger}}+{Y_e  Y_{e}^{\dagger}}\Big)  + 3 {Y_u  Y_{u}^{\dagger}  Y_u}  + {Y_u  Y_{d}^{\dagger}  Y_d}\Big)\nonumber \\ 
 &+\lambda \Big(2 {Y_u  Y_{d}^{\dagger}  T_d} +4 {Y_u  Y_{u}^{\dagger}  T_u} +{T_u  Y_{d}^{\dagger}  Y_d}+5 {T_u  Y_{u}^{\dagger}  Y_u} +3 T_u \mbox{Tr}\Big({Y_d  Y_{d}^{\dagger}}\Big) +T_u \mbox{Tr}\Big({Y_e  Y_{e}^{\dagger}}\Big) \nonumber \\ 
 &+6 Y_u \mbox{Tr}\Big({Y_{d}^{\dagger}  T_d}\Big) +2 Y_u \mbox{Tr}\Big({Y_{e}^{\dagger}  T_e}\Big) \Big)\Big)\\
\beta_{T_{\lambda}}^{(1)} & =  
+\frac{6}{5} g_{1}^{2} M_1 \lambda +6 g_{2}^{2} M_2 \lambda +T_{\lambda} \Big(12 |\lambda|^2  -3 g_{2}^{2}  + \mbox{Tr}\Big(3{Y_d  Y_{d}^{\dagger}}+{Y_e  Y_{e}^{\dagger}}+ 3 {Y_u  Y_{u}^{\dagger}}\Big)  -\frac{3}{5} g_{1}^{2}\Big)\nonumber \\ 
 &+6 \lambda \mbox{Tr}\Big({Y_{d}^{\dagger}  T_d}\Big) +2 \lambda \mbox{Tr}\Big({Y_{e}^{\dagger}  T_e}\Big) +6 \lambda \mbox{Tr}\Big({Y_{u}^{\dagger}  T_u}\Big) \\ 
\beta_{T_{\lambda}}^{(2)} & =  
-50 |\lambda|^4 T_{\lambda} -\frac{3}{5} |\lambda|^2 \Big(T_{\lambda} \Big(15 \mbox{Tr}\Big({Y_e  Y_{e}^{\dagger}}\Big)  -30 g_{2}^{2}  + 45 \mbox{Tr}\Big({Y_d  Y_{d}^{\dagger}}+{Y_u  Y_{u}^{\dagger}}\Big)  -6 g_{1}^{2} \Big)\nonumber \\ 
 &+2 \lambda \Big(10 g_{2}^{2} M_2  + 15 \mbox{Tr}\Big({Y_{d}^{\dagger}  T_d}\Big)  + 15 \mbox{Tr}\Big({Y_{u}^{\dagger}  T_u}\Big)  + 2 g_{1}^{2} M_1  + 5 \mbox{Tr}\Big({Y_{e}^{\dagger}  T_e}\Big) \Big)\Big)\nonumber \\ 
 &+T_{\lambda} \Big(\frac{207}{50} g_{1}^{4} +\frac{9}{5} g_{1}^{2} g_{2}^{2} +\frac{15}{2} g_{2}^{4} -\frac{2}{5} \Big(g_{1}^{2}-40 g_{3}^{2}\Big)\mbox{Tr}\Big({Y_d  Y_{d}^{\dagger}}\Big) +\frac{2}{5} g_{1}^{2} \mbox{Tr}\Big(3{Y_e  Y_{e}^{\dagger}}+2{Y_u  Y_{u}^{\dagger}}\Big)  \nonumber \\ 
 &+16 g_{3}^{2} \mbox{Tr}\Big({Y_u  Y_{u}^{\dagger}}\Big) -3\mbox{Tr}\Big(3{Y_d  Y_{d}^{\dagger}  Y_d  Y_{d}^{\dagger}} +2 {Y_d  Y_{u}^{\dagger}  Y_u  Y_{d}^{\dagger}}+{Y_e  Y_{e}^{\dagger}  Y_e  Y_{e}^{\dagger}}+3{Y_u  Y_{u}^{\dagger}  Y_u  Y_{u}^{\dagger}}\Big) \Big)\nonumber \\ 
 &-\frac{2}{25} \lambda \Big(207 g_{1}^{4} M_1 +45 g_{1}^{2} g_{2}^{2} (M_1 +M_2) +375 g_{2}^{4} M_2 -10 \Big( g_{1}^{2} M_1 -40 g_{3}^{2} M_3\Big)\mbox{Tr}\Big({Y_d  Y_{d}^{\dagger}}\Big) \nonumber \\ 
 &+30 g_{1}^{2} M_1 \mbox{Tr}\Big({Y_e  Y_{e}^{\dagger}}\Big) +20 g_{1}^{2} M_1 \mbox{Tr}\Big({Y_u  Y_{u}^{\dagger}}\Big) +400 g_{3}^{2} M_3 \mbox{Tr}\Big({Y_u  Y_{u}^{\dagger}}\Big) +10 g_{1}^{2} \mbox{Tr}\Big({Y_{d}^{\dagger}  T_d}\Big) \nonumber \\ 
 &-400 g_{3}^{2} \mbox{Tr}\Big({Y_{d}^{\dagger}  T_d}\Big) -30 g_{1}^{2} \mbox{Tr}\Big({Y_{e}^{\dagger}  T_e}\Big) -20 g_{1}^{2} \mbox{Tr}\Big({Y_{u}^{\dagger}  T_u}\Big) -400 g_{3}^{2} \mbox{Tr}\Big({Y_{u}^{\dagger}  T_u}\Big) \nonumber \\ 
 &+150 \mbox{Tr}\Big(3{Y_d  Y_{d}^{\dagger}  T_d  Y_{d}^{\dagger}}+{Y_d  Y_{u}^{\dagger}  T_u  Y_{d}^{\dagger}}+{Y_e  Y_{e}^{\dagger}  T_e  Y_{e}^{\dagger}}+{Y_u  Y_{d}^{\dagger}  T_d  Y_{u}^{\dagger}}+3{Y_u  Y_{u}^{\dagger}  T_u  Y_{u}^{\dagger}}\Big) \Big)
\end{align}} 
\subsection{Bilinear Soft-Breaking Parameters}
{\allowdisplaybreaks  \begin{align} 
\Delta \beta_{B{\mu}}^{(1)} & =  
2 \lambda^* \Big(2 \mu T_{\lambda}  + 3 \lambda B{\mu} \Big)\\ 
\Delta  \beta_{B{\mu}}^{(2)} & =  
-\frac{1}{5} \lambda^* \Big(\lambda B{\mu} \Big(25 \mbox{Tr}\Big({Y_e  Y_{e}^{\dagger}}\Big)  -36 g_{1}^{2}-180 g_{2}^{2}   + 70 |\lambda|^2  + 75 \mbox{Tr}\Big({Y_d  Y_{d}^{\dagger}}+{Y_u  Y_{u}^{\dagger}}\Big) \Big)\nonumber \\ 
 &+2 \mu \Big(80 |\lambda|^2 T_{\lambda} +5 T_{\lambda} \Big(3 \mbox{Tr}\Big({Y_d  Y_{d}^{\dagger}}\Big)  + 3 \mbox{Tr}\Big({Y_u  Y_{u}^{\dagger}}\Big)  + \mbox{Tr}\Big({Y_e  Y_{e}^{\dagger}}\Big)\Big)\nonumber \\ 
 &+3 \lambda \Big(15 \mbox{Tr}\Big({Y_{d}^{\dagger}  T_d}\Big)  + 15 \mbox{Tr}\Big({Y_{u}^{\dagger}  T_u}\Big)  + 30 g_{2}^{2} M_2  + 5 \mbox{Tr}\Big({Y_{e}^{\dagger}  T_e}\Big)  + 6 g_{1}^{2} M_1 \Big)\Big)\Big)\\ 
\beta_{b_s}^{(1)} & =  
2 \lambda^* \Big(2 M_s T_{\lambda}  + \lambda b_s \Big)\\ 
\beta_{b_s}^{(2)} & =  
-\frac{2}{5} \lambda^* \Big(\lambda b_s \Big(10 |\lambda|^2  -15 g_{2}^{2}  + 15 \mbox{Tr}\Big({Y_d  Y_{d}^{\dagger}}\Big)  + 15 \mbox{Tr}\Big({Y_u  Y_{u}^{\dagger}}\Big)  -3 g_{1}^{2}  + 5 \mbox{Tr}\Big({Y_e  Y_{e}^{\dagger}}\Big) \Big)\nonumber \\ 
 &+2 M_s \Big(T_{\lambda} \Big(-15 g_{2}^{2}  + 15 \mbox{Tr}\Big({Y_d  Y_{d}^{\dagger}}\Big)  + 15 \mbox{Tr}\Big({Y_u  Y_{u}^{\dagger}}\Big)  + 20 |\lambda|^2  -3 g_{1}^{2}  + 5 \mbox{Tr}\Big({Y_e  Y_{e}^{\dagger}}\Big) \Big)\nonumber \\ 
 &+\lambda \Big(15 g_{2}^{2} M_2  + 15 \mbox{Tr}\Big({Y_{d}^{\dagger}  T_d}\Big)  + 15 \mbox{Tr}\Big({Y_{u}^{\dagger}  T_u}\Big)  + 3 g_{1}^{2} M_1  + 5 \mbox{Tr}\Big({Y_{e}^{\dagger}  T_e}\Big) \Big)\Big)\Big)
\end{align}} 
\subsection{Linear Soft-Breaking Parameters}
{\allowdisplaybreaks  \begin{align} 
\beta_{t_s}^{(1)} & =  
2 \lambda^* \Big(2 \xi_s T_{\lambda}  + \lambda t_s \Big) + 4 B{\mu}^* T_{\lambda}  + 4 \Big(m_{h_d}^2 + m_{h_u}^2\Big)\lambda \mu^* \\ 
\beta_{t_s}^{(2)} & =  
-\frac{2}{5} \Big(10 \lambda \lambda^{*,2} \Big(4 \xi_s T_{\lambda}  + \lambda t_s \Big)+\lambda^* \Big(20 \Big(2 m_{h_d}^2  + 2 m_{h_u}^2  + m_{s}^2\Big)\lambda^{2} \mu^* \nonumber \\ 
 &+\lambda t_s \Big(15 \mbox{Tr}\Big({Y_d  Y_{d}^{\dagger}}\Big)  -3 \Big(5 g_{2}^{2}  -5 \mbox{Tr}\Big({Y_u  Y_{u}^{\dagger}}\Big)  + g_{1}^{2}\Big) + 5 \mbox{Tr}\Big({Y_e  Y_{e}^{\dagger}}\Big) \Big)\nonumber \\ 
 &+2 \Big(T_{\lambda} \Big(20 \lambda B{\mu}^*  + \xi_s \Big(15 \mbox{Tr}\Big({Y_d  Y_{d}^{\dagger}}\Big)  -3 \Big(5 g_{2}^{2}  -5 \mbox{Tr}\Big({Y_u  Y_{u}^{\dagger}}\Big)  + g_{1}^{2}\Big) + 5 \mbox{Tr}\Big({Y_e  Y_{e}^{\dagger}}\Big) \Big)\Big)\nonumber \\ 
 &+\xi_s \lambda \Big(15 \mbox{Tr}\Big({Y_{d}^{\dagger}  T_d}\Big)  + 3 \Big(5 g_{2}^{2} M_2  + 5 \mbox{Tr}\Big({Y_{u}^{\dagger}  T_u}\Big)  + g_{1}^{2} M_1 \Big) + 5 \mbox{Tr}\Big({Y_{e}^{\dagger}  T_e}\Big) \Big)\Big)\Big)\nonumber \\ 
 &+2 \Big(B{\mu}^* \Big(T_{\lambda} \Big(15 \mbox{Tr}\Big({Y_d  Y_{d}^{\dagger}}\Big)  -3 \Big(5 g_{2}^{2}  -5 \mbox{Tr}\Big({Y_u  Y_{u}^{\dagger}}\Big)  + g_{1}^{2}\Big) + 5 \mbox{Tr}\Big({Y_e  Y_{e}^{\dagger}}\Big) \Big)\nonumber \\ 
 &+\lambda \Big(15 \mbox{Tr}\Big({Y_{d}^{\dagger}  T_d}\Big)  + 3 \Big(5 g_{2}^{2} M_2  + 5 \mbox{Tr}\Big({Y_{u}^{\dagger}  T_u}\Big)  + g_{1}^{2} M_1 \Big) + 5 \mbox{Tr}\Big({Y_{e}^{\dagger}  T_e}\Big) \Big)\Big)\nonumber \\ 
 &+\mu^* \Big(-3 g_{1}^{2} m_{h_d}^2 \lambda -15 g_{2}^{2} m_{h_d}^2 \lambda -3 g_{1}^{2} m_{h_u}^2 \lambda -15 g_{2}^{2} m_{h_u}^2 \lambda -6 g_{1}^{2} \lambda |M_1|^2 -30 g_{2}^{2} \lambda |M_2|^2 \nonumber \\ 
 &+20 \lambda |T_{\lambda}|^2 +3 g_{1}^{2} M_1 T_{\lambda} +15 g_{2}^{2} M_2 T_{\lambda} +30 m_{h_d}^2 \lambda \mbox{Tr}\Big({Y_d  Y_{d}^{\dagger}}\Big) +15 m_{h_u}^2 \lambda \mbox{Tr}\Big({Y_d  Y_{d}^{\dagger}}\Big) \nonumber \\ 
 &+10 m_{h_d}^2 \lambda \mbox{Tr}\Big({Y_e  Y_{e}^{\dagger}}\Big) +5 m_{h_u}^2 \lambda \mbox{Tr}\Big({Y_e  Y_{e}^{\dagger}}\Big) +15 m_{h_d}^2 \lambda \mbox{Tr}\Big({Y_u  Y_{u}^{\dagger}}\Big) +30 m_{h_u}^2 \lambda \mbox{Tr}\Big({Y_u  Y_{u}^{\dagger}}\Big) \nonumber \\ 
 &+15 T_{\lambda} \mbox{Tr}\Big({T_d^*  Y_{d}^{T}}\Big) +15 \lambda \mbox{Tr}\Big({T_d^*  T_{d}^{T}}\Big) +5 T_{\lambda} \mbox{Tr}\Big({T_e^*  Y_{e}^{T}}\Big) +5 \lambda \mbox{Tr}\Big({T_e^*  T_{e}^{T}}\Big) +15 T_{\lambda} \mbox{Tr}\Big({T_u^*  Y_{u}^{T}}\Big) \nonumber \\ 
 &+15 \lambda \mbox{Tr}\Big({T_u^*  T_{u}^{T}}\Big) +15 \lambda \mbox{Tr}\Big({Y_d  Y_{d}^{\dagger}  m_d^{2 *}}\Big) +15 \lambda \mbox{Tr}\Big({Y_d  m_q^{2 *}  Y_{d}^{\dagger}}\Big) +5 \lambda \mbox{Tr}\Big({Y_e  Y_{e}^{\dagger}  m_e^{2 *}}\Big) \nonumber \\ 
 &+5 \lambda \mbox{Tr}\Big({Y_e  m_l^{2 *}  Y_{e}^{\dagger}}\Big) +15 \lambda \mbox{Tr}\Big({Y_u  Y_{u}^{\dagger}  m_u^{2 *}}\Big) +15 \lambda \mbox{Tr}\Big({Y_u  m_q^{2 *}  Y_{u}^{\dagger}}\Big) \Big)\Big)\Big)\\ 
\beta_{t_{\bar{s}}}^{(1)} & =  
4 M_s B{\mu} \lambda^* \\ 
\beta_{t_{\bar{s}}}^{(2)} & =  
-\frac{4}{5} M_s \lambda^* \Big(B{\mu} \Big(10 |\lambda|^2  -15 g_{2}^{2}  + 15 \mbox{Tr}\Big({Y_d  Y_{d}^{\dagger}}\Big)  + 15 \mbox{Tr}\Big({Y_u  Y_{u}^{\dagger}}\Big)  -3 g_{1}^{2}  + 5 \mbox{Tr}\Big({Y_e  Y_{e}^{\dagger}}\Big) \Big)\nonumber \\ 
 &+\mu \Big(10 \lambda^* T_{\lambda}  + 15 g_{2}^{2} M_2  + 15 \mbox{Tr}\Big({Y_{d}^{\dagger}  T_d}\Big)  + 15 \mbox{Tr}\Big({Y_{u}^{\dagger}  T_u}\Big)  + 3 g_{1}^{2} M_1  + 5 \mbox{Tr}\Big({Y_{e}^{\dagger}  T_e}\Big) \Big)\Big)
\end{align}} 
\subsection{Soft-Breaking Scalar Masses} 
{\allowdisplaybreaks  \begin{align} 
\Delta \beta_{m_q^2}^{(2)} & =  
-2 T_{\lambda}^* \Big(\lambda \Big({Y_{d}^{\dagger}  T_d} + {Y_{u}^{\dagger}  T_u}\Big) + \Big({Y_{d}^{\dagger}  Y_d} + {Y_{u}^{\dagger}  Y_u}\Big)T_{\lambda} \Big)\nonumber \\ 
 &- \lambda^* \Big(2 \Big(2 m_{h_d}^2  + m_{h_u}^2 + m_{s}^2\Big)\lambda {Y_{d}^{\dagger}  Y_d} +2 \Big(2 m_{h_u}^2  + m_{h_d}^2 + m_{s}^2\Big)\lambda {Y_{u}^{\dagger}  Y_u} \nonumber \\ 
 &+\lambda {m_q^2  Y_{d}^{\dagger}  Y_d} +\lambda {m_q^2  Y_{u}^{\dagger}  Y_u} +2 \lambda {Y_{d}^{\dagger}  m_d^2  Y_d} +\lambda {Y_{d}^{\dagger}  Y_d  m_q^2} +2 \lambda {Y_{u}^{\dagger}  m_u^2  Y_u} \nonumber \\ 
 &+2 \lambda {T_{d}^{\dagger}  T_d} +2 \lambda {T_{u}^{\dagger}  T_u} +\lambda {Y_{u}^{\dagger}  Y_u  m_q^2} +2 {T_{d}^{\dagger}  Y_d} T_{\lambda} +2 {T_{u}^{\dagger}  Y_u} T_{\lambda} \Big)\\ 
\Delta \beta_{m_l^2}^{(2)} & =  
-2 T_{\lambda}^* \Big(\lambda {Y_{e}^{\dagger}  T_e}  + {Y_{e}^{\dagger}  Y_e} T_{\lambda} \Big)- \lambda^* \Big(2 \Big(2 m_{h_d}^2  + m_{h_u}^2 + m_{s}^2\Big)\lambda {Y_{e}^{\dagger}  Y_e} \nonumber \\ 
 &+\lambda \Big(2 {T_{e}^{\dagger}  T_e}  + 2 {Y_{e}^{\dagger}  m_e^2  Y_e}  + {m_l^2  Y_{e}^{\dagger}  Y_e} + {Y_{e}^{\dagger}  Y_e  m_l^2}\Big)+2 {T_{e}^{\dagger}  Y_e} T_{\lambda} \Big)\\ 
\Delta \beta_{m_{h_d}^2}^{(1)} & =  
2 \Big(\Big(m_{h_d}^2 + m_{h_u}^2 + m_{s}^2\Big)|\lambda|^2  + |T_{\lambda}|^2\Big)\\ 
\Delta \beta_{m_{h_d}^2}^{(2)} & =  
-6 \Big(2 \Big(m_{h_d}^2 + m_{h_u}^2 + m_{s}^2\Big)|\lambda|^4 +T_{\lambda}^* \Big(\lambda \mbox{Tr}\Big({Y_{u}^{\dagger}  T_u}\Big)  + T_{\lambda} \mbox{Tr}\Big({Y_u  Y_{u}^{\dagger}}\Big) \Big)\nonumber \\ 
 &+\lambda^* \Big(4 \lambda |T_{\lambda}|^2 +\Big(2 m_{h_u}^2  + m_{h_d}^2 + m_{s}^2\Big)\lambda \mbox{Tr}\Big({Y_u  Y_{u}^{\dagger}}\Big) +T_{\lambda} \mbox{Tr}\Big({T_u^*  Y_{u}^{T}}\Big) +\lambda \mbox{Tr}\Big({T_u^*  T_{u}^{T}}\Big) \nonumber \\ 
 & +\lambda \mbox{Tr}\Big({m_q^2  Y_{u}^{\dagger}  Y_u}\Big)+\lambda \mbox{Tr}\Big({m_u^2  Y_u  Y_{u}^{\dagger}}\Big) \Big)\Big)\\ 
\Delta \beta_{m_{h_u}^2}^{(1)} & =  
2 \Big(\Big(m_{h_d}^2 + m_{h_u}^2 + m_{s}^2\Big)|\lambda|^2  + |T_{\lambda}|^2\Big)\\ 
\Delta \beta_{m_{h_u}^2}^{(2)} & =  
-2 \Big(6 \Big(m_{h_d}^2 + m_{h_u}^2 + m_{s}^2\Big)|\lambda|^4 +T_{\lambda}^* \Big(\lambda \Big(3 \mbox{Tr}\Big({Y_{d}^{\dagger}  T_d}\Big)  + \mbox{Tr}\Big({Y_{e}^{\dagger}  T_e}\Big)\Big) \nonumber \\
& + T_{\lambda} \Big(3 \mbox{Tr}\Big({Y_d  Y_{d}^{\dagger}}\Big)  + \mbox{Tr}\Big({Y_e  Y_{e}^{\dagger}}\Big)\Big)\Big)+\lambda^* \Big(12 \lambda |T_{\lambda}|^2 \nonumber \\ 
 &+3 \Big(2 m_{h_d}^2  + m_{h_u}^2 + m_{s}^2\Big)\lambda \mbox{Tr}\Big({Y_d  Y_{d}^{\dagger}}\Big) +2 m_{h_d}^2 \lambda \mbox{Tr}\Big({Y_e  Y_{e}^{\dagger}}\Big) +m_{h_u}^2 \lambda \mbox{Tr}\Big({Y_e  Y_{e}^{\dagger}}\Big) \nonumber \\ 
 &+m_{s}^2 \lambda \mbox{Tr}\Big({Y_e  Y_{e}^{\dagger}}\Big) +3 T_{\lambda} \mbox{Tr}\Big({T_d^*  Y_{d}^{T}}\Big) +3 \lambda \mbox{Tr}\Big({T_d^*  T_{d}^{T}}\Big) +T_{\lambda} \mbox{Tr}\Big({T_e^*  Y_{e}^{T}}\Big) +\lambda \mbox{Tr}\Big({T_e^*  T_{e}^{T}}\Big) \nonumber \\ 
 &+3 \lambda \mbox{Tr}\Big({m_d^2  Y_d  Y_{d}^{\dagger}}\Big) +\lambda \mbox{Tr}\Big({m_e^2  Y_e  Y_{e}^{\dagger}}\Big) +\lambda \mbox{Tr}\Big({m_l^2  Y_{e}^{\dagger}  Y_e}\Big) +3 \lambda \mbox{Tr}\Big({m_q^2  Y_{d}^{\dagger}  Y_d}\Big) \Big)\Big)\\ 
\Delta \beta_{m_d^2}^{(2)} & =  
-2 \Big(2 T_{\lambda}^* \Big(\lambda {T_d  Y_{d}^{\dagger}}  + {Y_d  Y_{d}^{\dagger}} T_{\lambda} \Big)+\lambda^* \Big(2 \Big(2 m_{h_d}^2  + m_{h_u}^2 + m_{s}^2\Big)\lambda {Y_d  Y_{d}^{\dagger}} \nonumber \\ 
 &+\lambda \Big(2 {T_d  T_{d}^{\dagger}}  + 2 {Y_d  m_q^2  Y_{d}^{\dagger}}  + {m_d^2  Y_d  Y_{d}^{\dagger}} + {Y_d  Y_{d}^{\dagger}  m_d^2}\Big)+2 {Y_d  T_{d}^{\dagger}} T_{\lambda} \Big)\Big)\\ 
\Delta \beta_{m_u^2}^{(2)} & =  
-2 \Big(2 T_{\lambda}^* \Big(\lambda {T_u  Y_{u}^{\dagger}}  + {Y_u  Y_{u}^{\dagger}} T_{\lambda} \Big)+\lambda^* \Big(2 \Big(2 m_{h_u}^2  + m_{h_d}^2 + m_{s}^2\Big)\lambda {Y_u  Y_{u}^{\dagger}} \nonumber \\ 
 &+\lambda \Big(2 {T_u  T_{u}^{\dagger}}  + 2 {Y_u  m_q^2  Y_{u}^{\dagger}}  + {m_u^2  Y_u  Y_{u}^{\dagger}} + {Y_u  Y_{u}^{\dagger}  m_u^2}\Big)+2 {Y_u  T_{u}^{\dagger}} T_{\lambda} \Big)\Big)\\ 
\Delta \beta_{m_e^2}^{(2)} & =  
-2 \Big(2 T_{\lambda}^* \Big(\lambda {T_e  Y_{e}^{\dagger}}  + {Y_e  Y_{e}^{\dagger}} T_{\lambda} \Big)+\lambda^* \Big(2 \Big(2 m_{h_d}^2  + m_{h_u}^2 + m_{s}^2\Big)\lambda {Y_e  Y_{e}^{\dagger}} \nonumber \\ 
 &+\lambda \Big(2 {T_e  T_{e}^{\dagger}}  + 2 {Y_e  m_l^2  Y_{e}^{\dagger}}  + {m_e^2  Y_e  Y_{e}^{\dagger}} + {Y_e  Y_{e}^{\dagger}  m_e^2}\Big)+2 {Y_e  T_{e}^{\dagger}} T_{\lambda} \Big)\Big)\\ 
\beta_{m_{s}^2}^{(1)} & =  
4 \Big(\Big(m_{h_d}^2 + m_{h_u}^2 + m_{s}^2\Big)|\lambda|^2  + |T_{\lambda}|^2\Big)\\ 
\beta_{m_{s}^2}^{(2)} & =  
-\frac{4}{5} \Big(20 \Big(m_{h_d}^2 + m_{h_u}^2 + m_{s}^2\Big)|\lambda|^4 \nonumber \\ 
 &+T_{\lambda}^* \Big(T_{\lambda} \Big(15 \mbox{Tr}\Big({Y_d  Y_{d}^{\dagger}}\Big)  -3 \Big(5 g_{2}^{2}  -5 \mbox{Tr}\Big({Y_u  Y_{u}^{\dagger}}\Big)  + g_{1}^{2}\Big) + 5 \mbox{Tr}\Big({Y_e  Y_{e}^{\dagger}}\Big) \Big)\nonumber \\ 
 &+\lambda \Big(15 \mbox{Tr}\Big({Y_{d}^{\dagger}  T_d}\Big)  + 3 \Big(5 g_{2}^{2} M_2  + 5 \mbox{Tr}\Big({Y_{u}^{\dagger}  T_u}\Big)  + g_{1}^{2} M_1 \Big) + 5 \mbox{Tr}\Big({Y_{e}^{\dagger}  T_e}\Big) \Big)\Big)\nonumber \\ 
 &+\lambda^* \Big(-3 g_{1}^{2} m_{h_d}^2 \lambda -15 g_{2}^{2} m_{h_d}^2 \lambda -3 g_{1}^{2} m_{h_u}^2 \lambda -15 g_{2}^{2} m_{h_u}^2 \lambda -3 g_{1}^{2} m_{s}^2 \lambda -15 g_{2}^{2} m_{s}^2 \lambda \nonumber \\ 
 &+40 \lambda |T_{\lambda}|^2 +3 g_{1}^{2} M_1^* \Big(T_{\lambda}-2 M_1 \lambda\Big)+15 g_{2}^{2} M_2^* \Big(T_{\lambda}-2 M_2 \lambda \Big)+30 m_{h_d}^2 \lambda \mbox{Tr}\Big({Y_d  Y_{d}^{\dagger}}\Big) \nonumber \\ 
 &+15 m_{h_u}^2 \lambda \mbox{Tr}\Big({Y_d  Y_{d}^{\dagger}}\Big) +15 m_{s}^2 \lambda \mbox{Tr}\Big({Y_d  Y_{d}^{\dagger}}\Big) +10 m_{h_d}^2 \lambda \mbox{Tr}\Big({Y_e  Y_{e}^{\dagger}}\Big) +5 m_{h_u}^2 \lambda \mbox{Tr}\Big({Y_e  Y_{e}^{\dagger}}\Big) \nonumber \\ 
 &+5 m_{s}^2 \lambda \mbox{Tr}\Big({Y_e  Y_{e}^{\dagger}}\Big) +15 m_{h_d}^2 \lambda \mbox{Tr}\Big({Y_u  Y_{u}^{\dagger}}\Big) +30 m_{h_u}^2 \lambda \mbox{Tr}\Big({Y_u  Y_{u}^{\dagger}}\Big) +15 m_{s}^2 \lambda \mbox{Tr}\Big({Y_u  Y_{u}^{\dagger}}\Big) \nonumber \\ 
 &+5 T_{\lambda} \mbox{Tr}\Big(3{T_d^*  Y_{d}^{T}}+3{T_u^*  Y_{u}^{T}}+{T_e^*  Y_{e}^{T}}\Big) +5 \lambda \mbox{Tr}\Big(3{T_d^*  T_{d}^{T}}+{T_e^*  T_{e}^{T}}+3{T_u^*  T_{u}^{T}}\Big)\nonumber \\
 & +5 \lambda \mbox{Tr}\Big(3{m_d^2  Y_d  Y_{d}^{\dagger}}+{m_e^2  Y_e  Y_{e}^{\dagger}}+{m_l^2  Y_{e}^{\dagger}  Y_e}+3{m_q^2  Y_{d}^{\dagger}  Y_d}+3{m_q^2  Y_{u}^{\dagger}  Y_u}+3{m_u^2  Y_u  Y_{u}^{\dagger}}\Big) \Big)\Big)\\ 
\beta_{m_{\bar{s}}^2}^{(1)} & =  
0\\ 
\beta_{m_{\bar{s}}^2}^{(2)} & =  
0
\end{align}} 
\subsection{Vacuum expectation values}
{\allowdisplaybreaks  \begin{align} 
\Delta \beta_{v_d}^{(1)} & =  
- v_d |\lambda|^2 \\ 
\Delta  \beta_{v_d}^{(2)} & =  
\frac{3}{10} v_d |\lambda|^2 \Big(10 |\lambda|^2  + 10 \mbox{Tr}\Big({Y_u  Y_{u}^{\dagger}}\Big)  - \Big(5 g_{2}^{2}  + g_{1}^{2}\Big)\xi \Big)\\ 
\Delta  \beta_{v_u}^{(1)} & =  
- v_u |\lambda|^2 \\ 
\Delta  \beta_{v_u}^{(2)} & =  
\frac{1}{10} v_u |\lambda|^2 \Big(10 \mbox{Tr}\Big({Y_e  Y_{e}^{\dagger}}\Big)   + 30 |\lambda|^2   + 30 \mbox{Tr}\Big({Y_d  Y_{d}^{\dagger}}\Big)  -3( g_{1}^{2} + 5 g_{2}^{2}) \xi \Big)\\ 
\beta_{v_{s}}^{(1)} & =  
-2 v_{s} |\lambda|^2 \\ 
\beta_{v_{s}}^{(2)} & =  
\frac{2}{5} v_{s} |\lambda|^2 \Big(10 |\lambda|^2   -15 g_{2}^{2}  + 5 \left(3\mbox{Tr}\Big({Y_d  Y_{d}^{\dagger}}\Big)  + 3 \mbox{Tr}\Big({Y_u  Y_{u}^{\dagger}}\Big) + \mbox{Tr}\Big({Y_e  Y_{e}^{\dagger}}\Big)\right)  -3 g_{1}^{2} \Big)\\ 
\beta_{v_{\bar{s}}}^{(1)} & =  
0\\ 
\beta_{v_{\bar{s}}}^{(2)} & =  
0
\end{align}}

\section{Mass matrices}
\label{app:matrices}
\begin{itemize} 
\item {\bf Mass matrix for Charginos}, Basis: \( \left(\tilde{W}^-, \tilde{H}_d^-\right), \left(\tilde{W}^+, \tilde{H}_u^+\right) \) 
\begin{equation} 
m_{\tilde{\chi}^-} = \left( 
\begin{array}{cc}
M_2 &\frac{1}{\sqrt{2}} g_2 v_u \\ 
\frac{1}{\sqrt{2}} g_2 v_d  &\frac{1}{\sqrt{2}} v_{s} \lambda  + \mu\end{array} 
\right) 
 \end{equation} 
This matrix is diagonalized by \(U\) and \(V\) 
\begin{equation} 
U^* m_{\tilde{\chi}^-} V^{\dagger} = m^{dia}_{\tilde{\chi}^-} 
\end{equation} 

\item {\bf Mass matrix for CP odd Higgs}, Basis:$\left(\sigma_{d}, \sigma_{u}, \sigma_s, \sigma_{\bar{s}}\right)$
In Landau gauge the mass matrix is given by
\begin{equation} 
m^2_{A^0} = \left( 
\begin{array}{cccc}
m_{\sigma_{d}\sigma_{d}} &m_{\sigma_{u}\sigma_{d}} &\frac{1}{\sqrt{2}} v_u {\Re\Big(T_{\lambda}\Big)}  &- \frac{1}{\sqrt{2}} v_u {\Re\Big(\lambda M_s^* \Big)} \\ 
m_{\sigma_{d}\sigma_{u}} &m_{\sigma_{u}\sigma_{u}} &\frac{1}{\sqrt{2}} v_d {\Re\Big(T_{\lambda}\Big)}  &- \frac{1}{\sqrt{2}} v_d {\Re\Big(\lambda M_s^* \Big)} \\ 
\frac{1}{\sqrt{2}} v_u {\Re\Big(T_{\lambda}\Big)}  &\frac{1}{\sqrt{2}} v_d {\Re\Big(T_{\lambda}\Big)}  &m_{\sigma_s\sigma_s} &- {\Re\Big(b_s\Big)} \\ 
- \frac{1}{\sqrt{2}} v_u {\Re\Big(\lambda M_s^* \Big)}  &- \frac{1}{\sqrt{2}} v_d {\Re\Big(\lambda M_s^* \Big)}  &- {\Re\Big(b_s\Big)}  &m_{\bar{s}}^2 + |M_s|^2\end{array} 
\right) 
 \end{equation} 
with 
\begin{align} 
m_{\sigma_{d}\sigma_{d}} &= |\mu|^2  + \sqrt{2} v_{s} \Re(\lambda \mu^*)+ \Big(v_{s}^{2} + v_{u}^{2}\Big)\frac{|\lambda|^2}{2}+ \frac{1}{8} \Big(g_{1}^{2} + g_{2}^{2}\Big)\Big(v_{d}^{2}- v_{u}^{2} \Big) + m_{h_d}^2\\ 
m_{\sigma_{d}\sigma_{u}} &= \frac{1}{4} \Big(4 {\Re\Big(B{\mu}\Big)}  + 4 {\Re\Big(\lambda \xi_s^* \Big)}  + \sqrt{2} \Big(2 v_{\bar{s}} {\Re\Big(\lambda M_s^* \Big)}  + 2 v_{s} {\Re\Big(T_{\lambda}\Big)} \Big)\Big)\\ 
m_{\sigma_{u}\sigma_{u}} &=  |\mu|^2  + \sqrt{2} v_{s} \Re(\lambda \mu^*) + \Big(v_{d}^{2} + v_{s}^{2}\Big)\frac{|\lambda|^2}{2} -\frac{1}{8} \Big(g_{1}^{2} + g_{2}^{2}\Big)\Big(v_{d}^{2}- v_{u}^{2} \Big) + m_{h_u}^2\\ 
m_{\sigma_s\sigma_s} &= \frac{1}{2} \Big(v_{d}^{2} + v_{u}^{2}\Big)|\lambda|^2  + m_{s}^2 + |M_s|^2
\end{align} 
The gauge fixing part is the same as in the MSSM:
\begin{equation} 
m^2 (\xi_{Z}) = \frac{1}{4} \left( 
\begin{array}{cccc}
v_d^2 &-v_d v_u &0 &0\\ 
-v_d v_u & v_u^2 &0 &0\\ 
0 &0 &0 &0\\ 
0 &0 &0 &0\end{array} 
\right) \Big(g_1 \sin\Theta_W   + g_2 \cos\Theta_W  \Big)^{2}
 \end{equation} 
This matrix is diagonalized by \(Z^A\): 
\begin{equation} 
Z^A m^2_{A^0} Z^{A,\dagger} = m^{dia}_{2,A^0} 
\end{equation} 
\item {\bf Mass matrix for Charged Higgs}, Basis: \( \left(H_d^-, H_u^{+,*}\right), \left(H_d^{-,*}, H_u^+\right) \) 
\begin{equation} 
m^2_{H^-} = \left( 
\begin{array}{cc}
m_{H_d^-H_d^{-,*}} &m^*_{H_u^{+,*}H_d^{-,*}}\\ 
m_{H_d^-H_u^+} &m_{H_u^{+,*}H_u^+}\end{array} 
\right)
 \end{equation} 
with 
\begin{align} 
m_{H_d^-H_d^{-,*}} &=  |\mu|^2  + \sqrt{2} v_{s} \Re(\lambda \mu^*) + v^2_{s} \frac{|\lambda|^2}{2}+ \frac{1}{8} \Big(g_{1}^{2} \Big(v_{d}^{2}- v_{u}^{2} \Big) + g_{2}^{2} \Big(v_{d}^{2} + v_{u}^{2}\Big)\Big) + m_{h_d}^2\\ 
m_{H_d^-H_u^+} &= \frac{1}{4} g_{2}^{2} v_d v_u  + \frac{1}{\sqrt{2}} v_{s} T_{\lambda}  + \lambda \Big(-\frac{1}{2} v_d v_u \lambda^*  + \frac{1}{\sqrt{2}} v_{\bar{s}} M_s^*  + \xi_s^*\Big) + B{\mu}\\ 
m_{H_u^{+,*}H_u^+} &=  |\mu|^2  + \sqrt{2} v_{s} \Re(\lambda \mu^*) + v^2_{s} \frac{|\lambda|^2}{2}+ \frac{1}{8} \Big(g_{1}^{2} \Big(v_{u}^{2}- v_{d}^{2}  \Big) + g_{2}^{2} \Big(v_{d}^{2} + v_{u}^{2}\Big)\Big) + m_{h_u}^2
\end{align} 
and
\begin{equation} 
m^2 (\xi_{W^-}) = \frac{1}{4} g_{2}^{2}\left( 
\begin{array}{cc}
 v_{d}^{2}  &-v_d v_u \\ 
- v_d v_u  & v_{u}^{2} \end{array} 
\right) 
 \end{equation} 
This matrix is diagonalized by \(Z^+\): 
\begin{equation} 
Z^+ m^2_{H^-} Z^{+,\dagger} = m^{dia}_{2,H^-} 
\end{equation} 
\end{itemize}

\section{Vertices}
\label{app:vertices}
We give here the difference of Higgs and neutralino vertices in comparison to the MSSM. The shown expressions are understood as
\begin{equation}
\Delta \Gamma_{fields} = \Gamma_{fields}^{\text{DiracNMSSM}} - \Gamma_{fields}^{\text{MSSM}}
\end{equation}
Chiral vertices are parametrized by
\begin{equation}
\Gamma^L_{fields} P_L + \Gamma^R_{fields} P_R 
\end{equation}
with the projection operators $P_L$ and $P_R$. 

\subsection{Interactions with Fermions}
\begin{align} 
\Delta& \Gamma^L_{\tilde{\chi}^+_{{i}}\tilde{\chi}^-_{{j}}A^0_{{k}}}  =   \,
\frac{1}{\sqrt{2}} \lambda U^*_{j 2} V^*_{i 2} Z_{{k 3}}^{A} \\ 
\Delta& \Gamma^R_{\tilde{\chi}^+_{{i}}\tilde{\chi}^-_{{j}}A^0_{{k}}}  =   \,- \frac{1}{\sqrt{2}} \lambda^* U_{{i 2}} V_{{j 2}} Z_{{k 3}}^{A}  
\\ 
\Delta& \Gamma^L_{\tilde{\chi}^0_{{i}}\tilde{\chi}^0_{{j}}A^0_{{k}}}  =   \,
- \frac{1}{\sqrt{2}} \lambda \Big(N^*_{i 3} \Big(N^*_{j 4} Z_{{k 3}}^{A}  + N^*_{j 5} Z_{{k 2}}^{A} \Big) + N^*_{i 4} \Big(N^*_{j 3} Z_{{k 3}}^{A}  + N^*_{j 5} Z_{{k 1}}^{A} \Big) + N^*_{i 5} \Big(N^*_{j 3} Z_{{k 2}}^{A}  + N^*_{j 4} Z_{{k 1}}^{A} \Big)\Big)\\ 
\Delta& \Gamma^R_{\tilde{\chi}^0_{{i}}\tilde{\chi}^0_{{j}}A^0_{{k}}}  =   \,\frac{1}{\sqrt{2}} \lambda^* \Big(Z_{{k 1}}^{A} \Big(N_{{i 4}} N_{{j 5}}  + N_{{i 5}} N_{{j 4}} \Big) + Z_{{k 2}}^{A} \Big(N_{{i 3}} N_{{j 5}}  + N_{{i 5}} N_{{j 3}} \Big) + Z_{{k 3}}^{A} \Big(N_{{i 3}} N_{{j 4}}  + N_{{i 4}} N_{{j 3}} \Big)\Big) 
\\ 
\Delta& \Gamma^L_{\tilde{\chi}^0_{{i}}\tilde{\chi}^-_{{j}}H^+_{{k}}}  =   \,
-i \lambda U^*_{j 2} N^*_{i 5} Z_{{k 2}}^{+} \\ 
\Delta& \Gamma^R_{\tilde{\chi}^0_{{i}}\tilde{\chi}^-_{{j}}H^+_{{k}}}  =   \,-i \lambda^* V_{{j 2}} N_{{i 5}} Z_{{k 1}}^{+}  
\\ 
\Delta& \Gamma^L_{\tilde{\chi}^+_{{i}}\tilde{\chi}^-_{{j}}h_{{k}}}  =   \,
-i \frac{1}{\sqrt{2}} \lambda U^*_{j 2} V^*_{i 2} Z_{{k 3}}^{H} \\ 
\Delta& \Gamma^R_{\tilde{\chi}^+_{{i}}\tilde{\chi}^-_{{j}}h_{{k}}}  =   \,-i \frac{1}{\sqrt{2}} \lambda^* U_{{i 2}} V_{{j 2}} Z_{{k 3}}^{H}   \\
\Delta& \Gamma^L_{\tilde{\chi}^0_{{i}}\tilde{\chi}^0_{{j}}h_{{k}}}  =   \,
i \frac{1}{\sqrt{2}} \lambda \Big(N^*_{i 3} \Big(N^*_{j 4} Z_{{k 3}}^{H}  + N^*_{j 5} Z_{{k 2}}^{H} \Big) + N^*_{i 4} \Big(N^*_{j 3} Z_{{k 3}}^{H}  + N^*_{j 5} Z_{{k 1}}^{H} \Big) + N^*_{i 5} \Big(N^*_{j 3} Z_{{k 2}}^{H}  + N^*_{j 4} Z_{{k 1}}^{H} \Big)\Big)\\ 
\Delta& \Gamma^R_{\tilde{\chi}^0_{{i}}\tilde{\chi}^0_{{j}}h_{{k}}}  =   \,i \frac{1}{\sqrt{2}} \lambda^* \Big(Z_{{k 1}}^{H} \Big(N_{{i 4}} N_{{j 5}}  + N_{{i 5}} N_{{j 4}} \Big) + Z_{{k 2}}^{H} \Big(N_{{i 3}} N_{{j 5}}  + N_{{i 5}} N_{{j 3}} \Big) + Z_{{k 3}}^{H} \Big(N_{{i 3}} N_{{j 4}}  + N_{{i 4}} N_{{j 3}} \Big)\Big) 
\\ 
\Delta& \Gamma^L_{\tilde{\chi}^+_{{i}}\tilde{\chi}^0_{{j}}H^-_{{k}}}  =  \,-i \lambda V^*_{i 2} N^*_{j 5} Z^{+,*}_{k 1} \\ 
\Delta& \Gamma^R_{\tilde{\chi}^+_{{i}}\tilde{\chi}^0_{{j}}H^-_{{k}}}  =  \,-i \lambda^* Z^{+,*}_{k 2} U_{{i 2}} N_{{j 5}}  
\end{align} 
\subsection{Three scalar interactions}
\begin{align} 
\Delta &\Gamma_{A^0_{{i}}A^0_{{j}}h_{{k}}}  =  \, 
-\frac{i}{4} \Big(\lambda^* \Big(4 \lambda Z_{{i 3}}^{A} Z_{{j 3}}^{A} \Big(v_d Z_{{k 1}}^{H}  + v_u Z_{{k 2}}^{H} \Big)+2 Z_{{i 1}}^{A} Z_{{j 1}}^{A} \Big(\Big(2 v_{s} \lambda  + \sqrt{2} \mu \Big)Z_{{k 3}}^{H}  + 2 v_u \lambda Z_{{k 2}}^{H} \Big)\nonumber \\ 
 &+Z_{{i 2}}^{A} \Big(2 Z_{{j 2}}^{A} \Big(2 v_d \lambda Z_{{k 1}}^{H}  + \Big(2 v_{s} \lambda  + \sqrt{2} \mu \Big)Z_{{k 3}}^{H} \Big) + \sqrt{2} M_s \Big(Z_{{j 1}}^{A} Z_{{k 4}}^{H}  - Z_{{j 4}}^{A} Z_{{k 1}}^{H} \Big)\Big)\nonumber \\ 
 &- \sqrt{2} M_s \Big(Z_{{i 1}}^{A} \Big(- Z_{{j 2}}^{A} Z_{{k 4}}^{H}  + Z_{{j 4}}^{A} Z_{{k 2}}^{H} \Big) + Z_{{i 4}}^{A} \Big(Z_{{j 1}}^{A} Z_{{k 2}}^{H}  + Z_{{j 2}}^{A} Z_{{k 1}}^{H} \Big)\Big)\Big)\nonumber \\ 
 &+\sqrt{2} \Big(T_{\lambda}^* \Big(\Big(Z_{{i 1}}^{A} Z_{{j 2}}^{A}  + Z_{{i 2}}^{A} Z_{{j 1}}^{A} \Big)Z_{{k 3}}^{H}  + Z_{{i 3}}^{A} \Big(Z_{{j 1}}^{A} Z_{{k 2}}^{H}  + Z_{{j 2}}^{A} Z_{{k 1}}^{H} \Big) + Z_{{j 3}}^{A} \Big(Z_{{i 1}}^{A} Z_{{k 2}}^{H}  + Z_{{i 2}}^{A} Z_{{k 1}}^{H} \Big)\Big)\nonumber \\ 
 &+T_{\lambda} \Big(\Big(Z_{{i 1}}^{A} Z_{{j 2}}^{A}  + Z_{{i 2}}^{A} Z_{{j 1}}^{A} \Big)Z_{{k 3}}^{H}  + Z_{{i 3}}^{A} \Big(Z_{{j 1}}^{A} Z_{{k 2}}^{H}  + Z_{{j 2}}^{A} Z_{{k 1}}^{H} \Big) + Z_{{j 3}}^{A} \Big(Z_{{i 1}}^{A} Z_{{k 2}}^{H}  + Z_{{i 2}}^{A} Z_{{k 1}}^{H} \Big)\Big)\nonumber \\ 
 &+\lambda \Big(2 \mu^* \Big(Z_{{i 1}}^{A} Z_{{j 1}}^{A}  + Z_{{i 2}}^{A} Z_{{j 2}}^{A} \Big)Z_{{k 3}}^{H} +M_s^* \Big(\Big(Z_{{i 1}}^{A} Z_{{j 2}}^{A}  + Z_{{i 2}}^{A} Z_{{j 1}}^{A} \Big)Z_{{k 4}}^{H}  - \Big(Z_{{i 1}}^{A} Z_{{j 4}}^{A}  + Z_{{i 4}}^{A} Z_{{j 1}}^{A} \Big)Z_{{k 2}}^{H} \nonumber \\ 
 & - \Big(Z_{{i 2}}^{A} Z_{{j 4}}^{A}  + Z_{{i 4}}^{A} Z_{{j 2}}^{A} \Big)Z_{{k 1}}^{H} \Big)\Big)\Big)\Big) 
\\ 
\Delta &\Gamma_{A^0_{{i}}\tilde{d}_{{j \beta}}\tilde{d}^*_{{k \gamma}}}  =  \, 
\frac{1}{2} \delta_{\beta \gamma} \Big(- \lambda \sum_{b=1}^{3}\sum_{a=1}^{3}Y^*_{d,{a b}} Z^{D,*}_{j 3 + a}  Z_{{k b}}^{D}   + \lambda^* \sum_{b=1}^{3}Z^{D,*}_{j b} \sum_{a=1}^{3}Y_{d,{a b}} Z_{{k 3 + a}}^{D}   \Big)\Big(v_{s} Z_{{i 2}}^{A}  + v_u Z_{{i 3}}^{A} \Big) 
\\ 
\Delta &\Gamma_{A^0_{{i}}\tilde{e}_{{j}}\tilde{e}^*_{{k}}}  =  \, 
\frac{1}{2} \Big(- \lambda \sum_{b=1}^{3}\sum_{a=1}^{3}Y^*_{e,{a b}} Z^{E,*}_{j 3 + a}  Z_{{k b}}^{E}   + \lambda^* \sum_{b=1}^{3}Z^{E,*}_{j b} \sum_{a=1}^{3}Y_{e,{a b}} Z_{{k 3 + a}}^{E}   \Big)\Big(v_{s} Z_{{i 2}}^{A}  + v_u Z_{{i 3}}^{A} \Big) 
\\ 
\Delta &\Gamma_{A^0_{{i}}\tilde{u}_{{j \beta}}\tilde{u}^*_{{k \gamma}}}  =  \, 
\frac{1}{2} \delta_{\beta \gamma} \Big(- \lambda \sum_{b=1}^{3}\sum_{a=1}^{3}Y^*_{u,{a b}} Z^{U,*}_{j 3 + a}  Z_{{k b}}^{U}   + \lambda^* \sum_{b=1}^{3}Z^{U,*}_{j b} \sum_{a=1}^{3}Y_{u,{a b}} Z_{{k 3 + a}}^{U}   \Big)\Big(v_d Z_{{i 3}}^{A}  + v_{s} Z_{{i 1}}^{A} \Big) 
\\ 
\Delta &\Gamma_{h_{{i}}h_{{j}}h_{{k}}}  =  \, 
\frac{i}{4} \Big(- \lambda^* \Big(2 Z_{{i 1}}^{H} \Big(2 \lambda Z_{{j 2}}^{H} \Big(v_d Z_{{k 2}}^{H}  + v_u Z_{{k 1}}^{H} \Big)+Z_{{j 3}}^{H} \Big(2 v_d \lambda Z_{{k 3}}^{H}  + \Big(2 v_{s} \lambda  + \sqrt{2} \mu \Big)Z_{{k 1}}^{H} \Big)\nonumber \\ 
 &+Z_{{j 1}}^{H} \Big(\Big(2 v_{s} \lambda  + \sqrt{2} \mu \Big)Z_{{k 3}}^{H}  + 2 v_u \lambda Z_{{k 2}}^{H} \Big)\Big)+2 Z_{{i 3}}^{H} \Big(\sqrt{2} \mu \Big(Z_{{j 1}}^{H} Z_{{k 1}}^{H}  + Z_{{j 2}}^{H} Z_{{k 2}}^{H} \Big)\nonumber \\ 
 &+2 \lambda \Big(Z_{{j 1}}^{H} \Big(v_d Z_{{k 3}}^{H}  + v_{s} Z_{{k 1}}^{H} \Big) + Z_{{j 2}}^{H} \Big(v_{s} Z_{{k 2}}^{H}  + v_u Z_{{k 3}}^{H} \Big) + Z_{{j 3}}^{H} \Big(v_d Z_{{k 1}}^{H}  + v_u Z_{{k 2}}^{H} \Big)\Big)\Big)\nonumber \\ 
 &+Z_{{i 2}}^{H} \Big(- \sqrt{2} M_s Z_{{j 4}}^{H} Z_{{k 1}}^{H} +2 Z_{{j 3}}^{H} \Big(\Big(2 v_{s} \lambda  + \sqrt{2} \mu \Big)Z_{{k 2}}^{H}  + 2 v_u \lambda Z_{{k 3}}^{H} \Big)\nonumber \\ 
 &+2 Z_{{j 2}}^{H} \Big(2 v_d \lambda Z_{{k 1}}^{H}  + \Big(2 v_{s} \lambda  + \sqrt{2} \mu \Big)Z_{{k 3}}^{H} \Big)+Z_{{j 1}}^{H} \Big(4 v_d \lambda Z_{{k 2}}^{H}  + 4 v_u \lambda Z_{{k 1}}^{H}  - \sqrt{2} M_s Z_{{k 4}}^{H} \Big)\Big)\nonumber \\ 
 &- \sqrt{2} M_s \Big(Z_{{i 1}}^{H} \Big(Z_{{j 2}}^{H} Z_{{k 4}}^{H}  + Z_{{j 4}}^{H} Z_{{k 2}}^{H} \Big) + Z_{{i 4}}^{H} \Big(Z_{{j 1}}^{H} Z_{{k 2}}^{H}  + Z_{{j 2}}^{H} Z_{{k 1}}^{H} \Big)\Big)\Big)\nonumber \\ 
 &+\sqrt{2} \Big(\Big(T_{\lambda}^* + T_{\lambda}\Big)\Big(Z_{{i 1}}^{H} \Big(Z_{{j 2}}^{H} Z_{{k 3}}^{H}  + Z_{{j 3}}^{H} Z_{{k 2}}^{H} \Big) + Z_{{i 2}}^{H} \Big(Z_{{j 1}}^{H} Z_{{k 3}}^{H}  + Z_{{j 3}}^{H} Z_{{k 1}}^{H} \Big) + Z_{{i 3}}^{H} \Big(Z_{{j 1}}^{H} Z_{{k 2}}^{H}  + Z_{{j 2}}^{H} Z_{{k 1}}^{H} \Big)\Big)\nonumber \\ 
 &+\lambda \Big(-2 \mu^* \Big(\Big(Z_{{i 1}}^{H} Z_{{j 1}}^{H}  + Z_{{i 2}}^{H} Z_{{j 2}}^{H} \Big)Z_{{k 3}}^{H}  + Z_{{i 3}}^{H} \Big(Z_{{j 1}}^{H} Z_{{k 1}}^{H}  + Z_{{j 2}}^{H} Z_{{k 2}}^{H} \Big) + Z_{{j 3}}^{H} \Big(Z_{{i 1}}^{H} Z_{{k 1}}^{H}  + Z_{{i 2}}^{H} Z_{{k 2}}^{H} \Big)\Big)\nonumber \\ 
 &+M_s^* \Big(\Big(Z_{{i 1}}^{H} Z_{{j 2}}^{H}  + Z_{{i 2}}^{H} Z_{{j 1}}^{H} \Big)Z_{{k 4}}^{H}  + Z_{{i 4}}^{H} \Big(Z_{{j 1}}^{H} Z_{{k 2}}^{H}  + Z_{{j 2}}^{H} Z_{{k 1}}^{H} \Big) + Z_{{j 4}}^{H} \Big(Z_{{i 1}}^{H} Z_{{k 2}}^{H}  + Z_{{i 2}}^{H} Z_{{k 1}}^{H} \Big)\Big)\Big)\Big)\Big) 
\\ 
\Delta &\Gamma_{h_{{i}}H^-_{{j}}H^+_{{k}}}  =  \, 
-\frac{i}{2} \Big(\sqrt{2} \Big(\lambda \mu^* Z_{{i 3}}^{H} \Big(Z^{+,*}_{j 1} Z_{{k 1}}^{+}  + Z^{+,*}_{j 2} Z_{{k 2}}^{+} \Big) + T_{\lambda}^* Z^{+,*}_{j 2} Z_{{i 3}}^{H} Z_{{k 1}}^{+}  \nonumber \\
 & + Z^{+,*}_{j 1} \Big(\lambda M_s^* Z_{{i 4}}^{H} + T_{\lambda} Z_{{i 3}}^{H} \Big)Z_{{k 2}}^{+} \Big) +\lambda^* \Big(Z^{+,*}_{j 1} \Big(\Big(2 v_{s} \lambda  + \sqrt{2} \mu \Big)Z_{{i 3}}^{H} Z_{{k 1}}^{+}  - \lambda \Big(v_d Z_{{i 2}}^{H}  + v_u Z_{{i 1}}^{H} \Big)Z_{{k 2}}^{+} \Big)\nonumber \\ 
 &+Z^{+,*}_{j 2} \Big(2 v_{s} \lambda Z_{{i 3}}^{H} Z_{{k 2}}^{+}  + \sqrt{2} M_s Z_{{i 4}}^{H} Z_{{k 1}}^{+}  + \sqrt{2} \mu Z_{{i 3}}^{H} Z_{{k 2}}^{+}  - v_d \lambda Z_{{i 2}}^{H} Z_{{k 1}}^{+}  - v_u \lambda Z_{{i 1}}^{H} Z_{{k 1}}^{+} \Big)\Big)\Big) 
\\ 
\Delta &\Gamma_{h_{{i}}\tilde{d}_{{j \beta}}\tilde{d}^*_{{k \gamma}}}  =  \, 
\frac{i}{2} \delta_{\beta \gamma} \Big(\lambda \sum_{b=1}^{3}\sum_{a=1}^{3}Y^*_{d,{a b}} Z^{D,*}_{j 3 + a}  Z_{{k b}}^{D}   + \lambda^* \sum_{b=1}^{3}Z^{D,*}_{j b} \sum_{a=1}^{3}Y_{d,{a b}} Z_{{k 3 + a}}^{D}   \Big)\Big(v_{s} Z_{{i 2}}^{H}  + v_u Z_{{i 3}}^{H} \Big) 
\\ 
\Delta &\Gamma_{h_{{i}}\tilde{e}_{{j}}\tilde{e}^*_{{k}}}  =  \, 
\frac{i}{2} \Big(\lambda \sum_{b=1}^{3}\sum_{a=1}^{3}Y^*_{e,{a b}} Z^{E,*}_{j 3 + a}  Z_{{k b}}^{E}   + \lambda^* \sum_{b=1}^{3}Z^{E,*}_{j b} \sum_{a=1}^{3}Y_{e,{a b}} Z_{{k 3 + a}}^{E}   \Big)\Big(v_{s} Z_{{i 2}}^{H}  + v_u Z_{{i 3}}^{H} \Big) 
\\ 
\Delta &\Gamma_{h_{{i}}\tilde{u}_{{j \beta}}\tilde{u}^*_{{k \gamma}}}  =  \, 
\frac{i}{2} \delta_{\beta \gamma} \Big(\lambda \sum_{b=1}^{3}\sum_{a=1}^{3}Y^*_{u,{a b}} Z^{U,*}_{j 3 + a}  Z_{{k b}}^{U}   + \lambda^* \sum_{b=1}^{3}Z^{U,*}_{j b} \sum_{a=1}^{3}Y_{u,{a b}} Z_{{k 3 + a}}^{U}   \Big)\Big(v_d Z_{{i 3}}^{H}  + v_{s} Z_{{i 1}}^{H} \Big) 
\\ 
\Delta &\Gamma_{H^-_{{i}}\tilde{u}_{{j \beta}}\tilde{d}^*_{{k \gamma}}}  =  \, 
i \frac{1}{\sqrt{2}} v_{s} \delta_{\beta \gamma} \Big(\lambda Z^{+,*}_{i 1} \sum_{b=1}^{3}\sum_{a=1}^{3}Y^*_{u,{a b}} Z^{U,*}_{j 3 + a}  Z_{{k b}}^{D}   + \lambda^* Z^{+,*}_{i 2} \sum_{b=1}^{3}Z^{U,*}_{j b} \sum_{a=1}^{3}Y_{d,{a b}} Z_{{k 3 + a}}^{D}   \Big) 
\\ 
\Delta &\Gamma_{H^-_{{i}}\tilde{\nu}_{{j}}\tilde{e}^*_{{k}}}  =  \, 
i \frac{1}{\sqrt{2}} v_{s} \lambda^* Z^{+,*}_{i 2} \sum_{b=1}^{3}Z^{V,*}_{j b} \sum_{a=1}^{3}Y_{e,{a b}} Z_{{k 3 + a}}^{E}    
\\ 
\Delta &\Gamma_{\tilde{d}_{{i \alpha}}H^+_{{j}}\tilde{u}^*_{{k \gamma}}}  =  \, 
i \frac{1}{\sqrt{2}} v_{s} \delta_{\alpha \gamma} \Big(\lambda \sum_{b=1}^{3}\sum_{a=1}^{3}Y^*_{d,{a b}} Z^{D,*}_{i 3 + a}  Z_{{k b}}^{U}  Z_{{j 2}}^{+}  + \lambda^* \sum_{b=1}^{3}Z^{D,*}_{i b} \sum_{a=1}^{3}Y_{u,{a b}} Z_{{k 3 + a}}^{U}   Z_{{j 1}}^{+} \Big) 
\\ 
\Delta &\Gamma_{\tilde{e}_{{i}}H^+_{{j}}\tilde{\nu}^*_{{k}}}  = \, 
i \frac{1}{\sqrt{2}} v_{s} \lambda \sum_{b=1}^{3}\sum_{a=1}^{3}Y^*_{e,{a b}} Z^{E,*}_{i 3 + a}  Z_{{k b}}^{V}  Z_{{j 2}}^{+}  
\end{align} 
\subsection{Four scalar interactions}
\begin{align} 
\Delta &\Gamma_{A^0_{{i}}A^0_{{j}}A^0_{{k}}A^0_{{l}}}  =  \, 
-i |\lambda|^2 \times \nonumber \\
 & \Big(Z_{{i 3}}^{A} \Big(Z_{{j 1}}^{A} \Big(Z_{{k 1}}^{A} Z_{{l 3}}^{A}  + Z_{{k 3}}^{A} Z_{{l 1}}^{A} \Big) + Z_{{j 2}}^{A} \Big(Z_{{k 2}}^{A} Z_{{l 3}}^{A}  + Z_{{k 3}}^{A} Z_{{l 2}}^{A} \Big) + Z_{{j 3}}^{A} \Big(Z_{{k 1}}^{A} Z_{{l 1}}^{A}  + Z_{{k 2}}^{A} Z_{{l 2}}^{A} \Big)\Big)\nonumber \\ 
 &+Z_{{i 2}}^{A} \Big(Z_{{j 1}}^{A} \Big(Z_{{k 1}}^{A} Z_{{l 2}}^{A}  + Z_{{k 2}}^{A} Z_{{l 1}}^{A} \Big) + Z_{{j 2}}^{A} \Big(Z_{{k 1}}^{A} Z_{{l 1}}^{A}  + Z_{{k 3}}^{A} Z_{{l 3}}^{A} \Big) + Z_{{j 3}}^{A} \Big(Z_{{k 2}}^{A} Z_{{l 3}}^{A}  + Z_{{k 3}}^{A} Z_{{l 2}}^{A} \Big)\Big)\nonumber \\ 
 &+Z_{{i 1}}^{A} \Big(Z_{{j 1}}^{A} \Big(Z_{{k 2}}^{A} Z_{{l 2}}^{A}  + Z_{{k 3}}^{A} Z_{{l 3}}^{A} \Big) + Z_{{j 2}}^{A} \Big(Z_{{k 1}}^{A} Z_{{l 2}}^{A}  + Z_{{k 2}}^{A} Z_{{l 1}}^{A} \Big) + Z_{{j 3}}^{A} \Big(Z_{{k 1}}^{A} Z_{{l 3}}^{A}  + Z_{{k 3}}^{A} Z_{{l 1}}^{A} \Big)\Big)\Big) 
\\ 
\Delta &\Gamma_{A^0_{{i}}A^0_{{j}}h_{{k}}h_{{l}}}  =  \, 
-i |\lambda|^2 \Big(Z_{{i 3}}^{A} Z_{{j 3}}^{A} \Big(Z_{{k 1}}^{H} Z_{{l 1}}^{H}  + Z_{{k 2}}^{H} Z_{{l 2}}^{H} \Big)+Z_{{i 2}}^{A} Z_{{j 2}}^{A} \Big(Z_{{k 1}}^{H} Z_{{l 1}}^{H}  + Z_{{k 3}}^{H} Z_{{l 3}}^{H} \Big)\nonumber \\ 
 &\hspace{3cm} +Z_{{i 1}}^{A} Z_{{j 1}}^{A} \Big(Z_{{k 2}}^{H} Z_{{l 2}}^{H}  + Z_{{k 3}}^{H} Z_{{l 3}}^{H} \Big)\Big) 
\\ 
\Delta &\Gamma_{A^0_{{i}}A^0_{{j}}H^-_{{k}}H^+_{{l}}}  =  \, 
-\frac{i}{2} |\lambda|^2 \Big(Z^{+,*}_{k 1} \Big(2 Z_{{i 3}}^{A} Z_{{j 3}}^{A} Z_{{l 1}}^{+}  + \Big(Z_{{i 1}}^{A} Z_{{j 2}}^{A}  + Z_{{i 2}}^{A} Z_{{j 1}}^{A} \Big)Z_{{l 2}}^{+} \Big)\nonumber \\ 
 &\hspace{3cm} +Z^{+,*}_{k 2} \Big(2 Z_{{i 3}}^{A} Z_{{j 3}}^{A} Z_{{l 2}}^{+}  + Z_{{i 1}}^{A} Z_{{j 2}}^{A} Z_{{l 1}}^{+}  + Z_{{i 2}}^{A} Z_{{j 1}}^{A} Z_{{l 1}}^{+} \Big)\Big) 
\\ 
\Delta &\Gamma_{A^0_{{i}}A^0_{{j}}\tilde{d}_{{k \gamma}}\tilde{d}^*_{{l \delta}}}  =  \, 
-\frac{i}{2} \delta_{\gamma \delta} \Big(\lambda \sum_{b=1}^{3}\sum_{a=1}^{3}Y^*_{d,{a b}} Z^{D,*}_{k 3 + a}  Z_{{l b}}^{D}   + \lambda^* \sum_{b=1}^{3}Z^{D,*}_{k b} \sum_{a=1}^{3}Y_{d,{a b}} Z_{{l 3 + a}}^{D}   \Big)\Big(Z_{{i 2}}^{A} Z_{{j 3}}^{A}  + Z_{{i 3}}^{A} Z_{{j 2}}^{A} \Big) 
\\ 
\Delta &\Gamma_{A^0_{{i}}A^0_{{j}}\tilde{e}_{{k}}\tilde{e}^*_{{l}}}  =  \, 
-\frac{i}{2} \Big(\lambda \sum_{b=1}^{3}\sum_{a=1}^{3}Y^*_{e,{a b}} Z^{E,*}_{k 3 + a}  Z_{{l b}}^{E}   + \lambda^* \sum_{b=1}^{3}Z^{E,*}_{k b} \sum_{a=1}^{3}Y_{e,{a b}} Z_{{l 3 + a}}^{E}   \Big)\Big(Z_{{i 2}}^{A} Z_{{j 3}}^{A}  + Z_{{i 3}}^{A} Z_{{j 2}}^{A} \Big) 
\\ 
\Delta &\Gamma_{A^0_{{i}}A^0_{{j}}\tilde{u}_{{k \gamma}}\tilde{u}^*_{{l \delta}}}  =  \, 
-\frac{i}{2} \delta_{\gamma \delta} \Big(\lambda \sum_{b=1}^{3}\sum_{a=1}^{3}Y^*_{u,{a b}} Z^{U,*}_{k 3 + a}  Z_{{l b}}^{U}   + \lambda^* \sum_{b=1}^{3}Z^{U,*}_{k b} \sum_{a=1}^{3}Y_{u,{a b}} Z_{{l 3 + a}}^{U}   \Big)\Big(Z_{{i 1}}^{A} Z_{{j 3}}^{A}  + Z_{{i 3}}^{A} Z_{{j 1}}^{A} \Big) 
\\ 
\Delta &\Gamma_{A^0_{{i}}h_{{j}}H^-_{{k}}H^+_{{l}}}  =  \, 
-\frac{1}{2} |\lambda|^2 \Big(Z_{{i 1}}^{A} Z_{{j 2}}^{H}  + Z_{{i 2}}^{A} Z_{{j 1}}^{H} \Big)\Big(- Z^{+,*}_{k 1} Z_{{l 2}}^{+}  + Z^{+,*}_{k 2} Z_{{l 1}}^{+} \Big) 
\\ 
\Delta &\Gamma_{A^0_{{i}}H^-_{{j}}\tilde{u}_{{k \gamma}}\tilde{d}^*_{{l \delta}}}  =  \, 
\frac{1}{\sqrt{2}} \delta_{\gamma \delta} \Big(- \lambda Z^{+,*}_{j 1} \sum_{b=1}^{3}\sum_{a=1}^{3}Y^*_{u,{a b}} Z^{U,*}_{k 3 + a}  Z_{{l b}}^{D}   + \lambda^* Z^{+,*}_{j 2} \sum_{b=1}^{3}Z^{U,*}_{k b} \sum_{a=1}^{3}Y_{d,{a b}} Z_{{l 3 + a}}^{D}   \Big)Z_{{i 3}}^{A}  
\\ 
\Delta &\Gamma_{A^0_{{i}}H^-_{{j}}\tilde{\nu}_{{k}}\tilde{e}^*_{{l}}}  =  \, 
\frac{1}{\sqrt{2}} \lambda^* Z^{+,*}_{j 2} \sum_{b=1}^{3}Z^{V,*}_{k b} \sum_{a=1}^{3}Y_{e,{a b}} Z_{{l 3 + a}}^{E}   Z_{{i 3}}^{A}  
\\ 
\Delta &\Gamma_{A^0_{{i}}\tilde{d}_{{j \beta}}H^+_{{k}}\tilde{u}^*_{{l \delta}}}  =  \, 
\frac{1}{\sqrt{2}} \delta_{\beta \delta} Z_{{i 3}}^{A} \Big(- \lambda \sum_{b=1}^{3}\sum_{a=1}^{3}Y^*_{d,{a b}} Z^{D,*}_{j 3 + a}  Z_{{l b}}^{U}  Z_{{k 2}}^{+}  + \lambda^* \sum_{b=1}^{3}Z^{D,*}_{j b} \sum_{a=1}^{3}Y_{u,{a b}} Z_{{l 3 + a}}^{U}   Z_{{k 1}}^{+} \Big) 
\\ 
\Delta &\Gamma_{A^0_{{i}}\tilde{e}_{{j}}H^+_{{k}}\tilde{\nu}^*_{{l}}}  =  \, 
- \frac{1}{\sqrt{2}} \lambda \sum_{b=1}^{3}\sum_{a=1}^{3}Y^*_{e,{a b}} Z^{E,*}_{j 3 + a}  Z_{{l b}}^{V}  Z_{{i 3}}^{A} Z_{{k 2}}^{+}  
\\ 
\Delta &\Gamma_{h_{{i}}h_{{j}}h_{{k}}h_{{l}}}  =  \, 
-i |\lambda|^2 \times \nonumber \\
 & \Big(Z_{{i 3}}^{H} \Big(Z_{{j 1}}^{H} \Big(Z_{{k 1}}^{H} Z_{{l 3}}^{H}  + Z_{{k 3}}^{H} Z_{{l 1}}^{H} \Big) + Z_{{j 2}}^{H} \Big(Z_{{k 2}}^{H} Z_{{l 3}}^{H}  + Z_{{k 3}}^{H} Z_{{l 2}}^{H} \Big) + Z_{{j 3}}^{H} \Big(Z_{{k 1}}^{H} Z_{{l 1}}^{H}  + Z_{{k 2}}^{H} Z_{{l 2}}^{H} \Big)\Big)\nonumber \\ 
 &+Z_{{i 2}}^{H} \Big(Z_{{j 1}}^{H} \Big(Z_{{k 1}}^{H} Z_{{l 2}}^{H}  + Z_{{k 2}}^{H} Z_{{l 1}}^{H} \Big) + Z_{{j 2}}^{H} \Big(Z_{{k 1}}^{H} Z_{{l 1}}^{H}  + Z_{{k 3}}^{H} Z_{{l 3}}^{H} \Big) + Z_{{j 3}}^{H} \Big(Z_{{k 2}}^{H} Z_{{l 3}}^{H}  + Z_{{k 3}}^{H} Z_{{l 2}}^{H} \Big)\Big)\nonumber \\ 
 &+Z_{{i 1}}^{H} \Big(Z_{{j 1}}^{H} \Big(Z_{{k 2}}^{H} Z_{{l 2}}^{H}  + Z_{{k 3}}^{H} Z_{{l 3}}^{H} \Big) + Z_{{j 2}}^{H} \Big(Z_{{k 1}}^{H} Z_{{l 2}}^{H}  + Z_{{k 2}}^{H} Z_{{l 1}}^{H} \Big) + Z_{{j 3}}^{H} \Big(Z_{{k 1}}^{H} Z_{{l 3}}^{H}  + Z_{{k 3}}^{H} Z_{{l 1}}^{H} \Big)\Big)\Big) 
\\ 
\Delta &\Gamma_{h_{{i}}h_{{j}}H^-_{{k}}H^+_{{l}}}  =  \, 
\frac{i}{2} |\lambda|^2 \Big(Z^{+,*}_{k 1} \Big(-2 Z_{{i 3}}^{H} Z_{{j 3}}^{H} Z_{{l 1}}^{+}  + \Big(Z_{{i 1}}^{H} Z_{{j 2}}^{H}  + Z_{{i 2}}^{H} Z_{{j 1}}^{H} \Big)Z_{{l 2}}^{+} \Big)\nonumber \\ 
 & \hspace{3cm} +Z^{+,*}_{k 2} \Big(-2 Z_{{i 3}}^{H} Z_{{j 3}}^{H} Z_{{l 2}}^{+}  + Z_{{i 1}}^{H} Z_{{j 2}}^{H} Z_{{l 1}}^{+}  + Z_{{i 2}}^{H} Z_{{j 1}}^{H} Z_{{l 1}}^{+} \Big)\Big) 
\\ 
\Delta &\Gamma_{h_{{i}}h_{{j}}\tilde{d}_{{k \gamma}}\tilde{d}^*_{{l \delta}}}  =  \, 
\frac{i}{2} \delta_{\gamma \delta} \Big(\lambda \sum_{b=1}^{3}\sum_{a=1}^{3}Y^*_{d,{a b}} Z^{D,*}_{k 3 + a}  Z_{{l b}}^{D}   + \lambda^* \sum_{b=1}^{3}Z^{D,*}_{k b} \sum_{a=1}^{3}Y_{d,{a b}} Z_{{l 3 + a}}^{D}   \Big)\Big(Z_{{i 2}}^{H} Z_{{j 3}}^{H}  + Z_{{i 3}}^{H} Z_{{j 2}}^{H} \Big) 
\\ 
\Delta &\Gamma_{h_{{i}}h_{{j}}\tilde{e}_{{k}}\tilde{e}^*_{{l}}}  =  \, 
\frac{i}{2} \Big(\lambda \sum_{b=1}^{3}\sum_{a=1}^{3}Y^*_{e,{a b}} Z^{E,*}_{k 3 + a}  Z_{{l b}}^{E}   + \lambda^* \sum_{b=1}^{3}Z^{E,*}_{k b} \sum_{a=1}^{3}Y_{e,{a b}} Z_{{l 3 + a}}^{E}   \Big)\Big(Z_{{i 2}}^{H} Z_{{j 3}}^{H}  + Z_{{i 3}}^{H} Z_{{j 2}}^{H} \Big) 
\\ 
\Delta &\Gamma_{h_{{i}}h_{{j}}\tilde{u}_{{k \gamma}}\tilde{u}^*_{{l \delta}}}  =  \, 
\frac{i}{2} \delta_{\gamma \delta} \Big(\lambda \sum_{b=1}^{3}\sum_{a=1}^{3}Y^*_{u,{a b}} Z^{U,*}_{k 3 + a}  Z_{{l b}}^{U}   + \lambda^* \sum_{b=1}^{3}Z^{U,*}_{k b} \sum_{a=1}^{3}Y_{u,{a b}} Z_{{l 3 + a}}^{U}   \Big)\Big(Z_{{i 1}}^{H} Z_{{j 3}}^{H}  + Z_{{i 3}}^{H} Z_{{j 1}}^{H} \Big) 
\\ 
\Delta &\Gamma_{h_{{i}}H^-_{{j}}\tilde{u}_{{k \gamma}}\tilde{d}^*_{{l \delta}}}  =  \, 
i \frac{1}{\sqrt{2}} \delta_{\gamma \delta} \Big(\lambda Z^{+,*}_{j 1} \sum_{b=1}^{3}\sum_{a=1}^{3}Y^*_{u,{a b}} Z^{U,*}_{k 3 + a}  Z_{{l b}}^{D}   + \lambda^* Z^{+,*}_{j 2} \sum_{b=1}^{3}Z^{U,*}_{k b} \sum_{a=1}^{3}Y_{d,{a b}} Z_{{l 3 + a}}^{D}   \Big)Z_{{i 3}}^{H}  
\\ 
\Delta &\Gamma_{h_{{i}}H^-_{{j}}\tilde{\nu}_{{k}}\tilde{e}^*_{{l}}}  =  \, 
i \frac{1}{\sqrt{2}} \lambda^* Z^{+,*}_{j 2} \sum_{b=1}^{3}Z^{V,*}_{k b} \sum_{a=1}^{3}Y_{e,{a b}} Z_{{l 3 + a}}^{E}   Z_{{i 3}}^{H}  
\\ 
\Delta &\Gamma_{h_{{i}}\tilde{d}_{{j \beta}}H^+_{{k}}\tilde{u}^*_{{l \delta}}}  =  \, 
i \frac{1}{\sqrt{2}} \delta_{\beta \delta} Z_{{i 3}}^{H} \Big(\lambda \sum_{b=1}^{3}\sum_{a=1}^{3}Y^*_{d,{a b}} Z^{D,*}_{j 3 + a}  Z_{{l b}}^{U}  Z_{{k 2}}^{+}  + \lambda^* \sum_{b=1}^{3}Z^{D,*}_{j b} \sum_{a=1}^{3}Y_{u,{a b}} Z_{{l 3 + a}}^{U}   Z_{{k 1}}^{+} \Big) 
\\ 
\Delta &\Gamma_{h_{{i}}\tilde{e}_{{j}}H^+_{{k}}\tilde{\nu}^*_{{l}}}  =  \, 
i \frac{1}{\sqrt{2}} \lambda \sum_{b=1}^{3}\sum_{a=1}^{3}Y^*_{e,{a b}} Z^{E,*}_{j 3 + a}  Z_{{l b}}^{V}  Z_{{i 3}}^{H} Z_{{k 2}}^{+}  
\\ 
\Delta &\Gamma_{H^-_{{i}}H^-_{{j}}H^+_{{k}}H^+_{{l}}}  =  \, 
-i |\lambda|^2 \Big(Z^{+,*}_{i 1} Z^{+,*}_{j 2}  + Z^{+,*}_{i 2} Z^{+,*}_{j 1} \Big)\Big(Z_{{k 1}}^{+} Z_{{l 2}}^{+}  + Z_{{k 2}}^{+} Z_{{l 1}}^{+} \Big) 
\end{align}

\section{The Higgs sector of the DiracNMSSM at the loop level}
\label{app:HiggsLoop}
 We give here some details of the calculation which are carried out by a combination of the public tools \SARAH and \SPheno. 

To calculate the Higgs mass at the one-loop level, first the one-loop corrections to the tadpoles equations are needed. These are a sum of loops 
involving massive vector bosons and ghosts, standard model fermions, SUSY fermions, sfermions and Higgs fields:
\begin{equation}
\delta \Theta^{(1)}_{i} = \delta \Theta^{(1),V}_{i}  + \delta \Theta^{(1),f}_{i} + \delta \Theta^{(1),\tilde{f}}_{i} +\delta \Theta^{(1),\chi}_{i} + \delta \Theta^{(1),\phi}_{i}
\end{equation}
The explicit expression for all contributions read
\begin{align} 
\delta \Theta^{(1),V}_{i} = & \, +2 {A_0\Big(m^2_{Z}\Big)}{\Gamma_{\check{h}_{{i}},Z,Z}} +{A_0\Big(m^2_{\eta^-}\Big)} {\Gamma_{\check{h}_{{i}},\bar{\eta^-},\eta^-}} +{A_0\Big(m^2_{\eta^+}\Big)} {\Gamma_{\check{h}_{{i}},\bar{\eta^+},\eta^+}} \nonumber \\ 
 &+{A_0\Big(m^2_{\eta^Z}\Big)} {\Gamma_{\check{h}_{{i}},\bar{\eta^Z},\eta^Z}} +4  {A_0\Big(m^2_{W^-}\Big)}{\Gamma_{\check{h}_{{i}},W^+,W^-}} \\
\delta \Theta^{(1),\phi}_{i} = &- \sum_{a=1}^{2}{A_0\Big(m^2_{H^-_{{a}}}\Big)} {\Gamma_{\check{h}_{{i}},H^+_{{a}},H^-_{{a}}}}  -\frac{1}{2} \sum_{a=1}^{4}{A_0\Big(m^2_{A^0_{{a}}}\Big)} {\Gamma_{\check{h}_{{i}},A^0_{{a}},A^0_{{a}}}} -\frac{1}{2} \sum_{a=1}^{4}{A_0\Big(m^2_{h_{{a}}}\Big)} {\Gamma_{\check{h}_{{i}},h_{{a}},h_{{a}}}}   \\
\delta \Theta^{(1),\chi}_{i} = &+2 \sum_{a=1}^{2}{A_0\Big(m^2_{\tilde{\chi}^-_{{a}}}\Big)} m_{\tilde{\chi}^-_{{a}}} \Big({\Gamma^L_{\check{h}_{{i}},\tilde{\chi}^+_{{a}},\tilde{\chi}^-_{{a}}}} + {\Gamma^R_{\check{h}_{{i}},\tilde{\chi}^+_{{a}},\tilde{\chi}^-_{{a}}}}\Big)  +\sum_{a=1}^{6}{A_0\Big(m^2_{\tilde{\chi}^0_{{a}}}\Big)} m_{\tilde{\chi}^0_{{a}}} \Big({\Gamma^L_{\check{h}_{{i}},\tilde{\chi}^0_{{a}},\tilde{\chi}^0_{{a}}}} + {\Gamma^R_{\check{h}_{{i}},\tilde{\chi}^0_{{a}},\tilde{\chi}^0_{{a}}}}\Big)  \\
\delta \Theta^{(1),\tilde{f}}_{i} = & - \sum_{a=1}^{3}{A_0\Big(m^2_{\tilde{\nu}_{{a}}}\Big)} {\Gamma_{\check{h}_{{i}},\tilde{\nu}^*_{{a}},\tilde{\nu}_{{a}}}}   -3 \sum_{a=1}^{6}{A_0\Big(m^2_{\tilde{d}_{{a}}}\Big)} {\Gamma_{\check{h}_{{i}},\tilde{d}^*_{{a}},\tilde{d}_{{a}}}}  - \sum_{a=1}^{6}{A_0\Big(m^2_{\tilde{e}_{{a}}}\Big)} {\Gamma_{\check{h}_{{i}},\tilde{e}^*_{{a}},\tilde{e}_{{a}}}}  \nonumber \\ 
 &-3 \sum_{a=1}^{6}{A_0\Big(m^2_{\tilde{u}_{{a}}}\Big)} {\Gamma_{\check{h}_{{i}},\tilde{u}^*_{{a}},\tilde{u}_{{a}}}}  \\
\delta \Theta^{(1),f}_{i} =  &+6 \sum_{a=1}^{3}{A_0\Big(m^2_{d_{{a}}}\Big)} m_{d_{{a}}} \Big({\Gamma^L_{\check{h}_{{i}},\bar{d}_{{a}},d_{{a}}}} + {\Gamma^R_{\check{h}_{{i}},\bar{d}_{{a}},d_{{a}}}}\Big) +2 \sum_{a=1}^{3}{A_0\Big(m^2_{e_{{a}}}\Big)} m_{e_{{a}}} \Big({\Gamma^L_{\check{h}_{{i}},\bar{e}_{{a}},e_{{a}}}} + {\Gamma^R_{\check{h}_{{i}},\bar{e}_{{a}},e_{{a}}}}\Big) \nonumber \\ 
 &+6 \sum_{a=1}^{3}{A_0\Big(m^2_{u_{{a}}}\Big)} m_{u_{{a}}} \Big({\Gamma^L_{\check{h}_{{i}},\bar{u}_{{a}},u_{{a}}}} + {\Gamma^R_{\check{h}_{{i}},\bar{u}_{{a}},u_{{a}}}}\Big) 
\end{align} 
with $\delta \Theta^{(1)}_1 = \delta \Theta^{(1)}_d$, $\delta \Theta^{(1)}_2 = \delta \Theta^{(1)}_u$, $\delta \Theta^{(1)}_3 = \delta \Theta^{(1)}_s$, and 
$\delta \Theta^{(1)}_4 = \delta \Theta^{(1)}_{\bar{s}}$, and $\check{h} = (\phi_d, \phi_u, \phi_s, \phi_{\bar{s}})^T$. Here, chiral vertices are parametrized by $\Gamma^L P_L + \Gamma^R P_R$ with the projection operators $P_{L,R}$ and non-chiral vertices by $\Gamma$. The subscripts denote the involved particles.
All necessary vertices and mass matrices for the DiracNMSSM are given in appendix~\ref{app:matrices}--\ref{app:vertices}. Note, the rotation matrix corresponding to the external scalar has to be replaced by the identity matrix in these calculations. The finite part of the Passarino-Veltman integral $A_0$ is given by
\begin{equation}
A_0(m^2)\ = m^2\left(1 -\ln{\frac{m^2}{Q^2}}\right) 
\end{equation}
with the renormalisation scale $Q$ which is chosen to be the average of the stop masses \\

The one-loop corrections to the tadpoles are applied to find the new values of $\mu, b\mu, t_s, t_{\bar{s}}$ which the vacuum conditions
\begin{equation}
\label{eq:onelooptad}
 \Theta_i + \delta \Theta^{(1)}_i = 0 \, , \hspace{1cm} i=d,u,s,\bar{s}
\end{equation}
The second step is to calculate the self-energy matrix of the CP even Higgs
\begin{equation}
\Pi^{h}_{i,j}(p^2) = \Pi^{h,V}_{i,j}(p^2) + \Pi^{h,\phi}_{i,j}(p^2) + \Pi^{h,V\phi}_{i,j}(p^2)  + \Pi^{h,\tilde{f}}_{i,j}(p^2) + \Pi^{h,f}_{i,j}(p^2) + \Pi^{h,\chi}_{i,j}(p^2) 
\end{equation}
The corrections stemming from vector bosons and ghosts are given
\begin{align} 
\Pi^{h,V}_{i,j}(p^2) &= +2  {B_0\Big(p^{2},m^2_{Z},m^2_{Z}\Big)}{\Gamma^*_{\check{h}_{{j}},Z,Z}} {\Gamma_{\check{h}_{{i}},Z,Z}} +4 + {B_0\Big(p^{2},m^2_{W^-},m^2_{W^-}\Big)}{\Gamma^*_{\check{h}_{{j}},W^+,W^-}} {\Gamma_{\check{h}_{{i}},W^+,W^-}} \nonumber \\ 
 &- {B_0\Big(p^{2},m^2_{\eta^-},m^2_{\eta^-}\Big)} {\Gamma_{\check{h}_{{i}},\bar{\eta^-},\eta^-}} {\Gamma_{\check{h}_{{j}},\bar{\eta^-},\eta^-}} - {B_0\Big(p^{2},m^2_{\eta^+},m^2_{\eta^+}\Big)} {\Gamma_{\check{h}_{{i}},\bar{\eta^+},\eta^+}} {\Gamma_{\check{h}_{{j}},\bar{\eta^+},\eta^+}} \nonumber \\ 
 &- {B_0\Big(p^{2},m^2_{\eta^Z},m^2_{\eta^Z}\Big)} {\Gamma_{\check{h}_{{i}},\bar{\eta^Z},\eta^Z}} {\Gamma_{\check{h}_{{j}},\bar{\eta^Z},\eta^Z}} +2 {A_0\Big(m^2_{Z}\Big)}{\Gamma_{\check{h}_{{i}},\check{h}_{{j}},Z,Z}} +4  {A_0\Big(m^2_{W^-}\Big)}{\Gamma_{\check{h}_{{i}},\check{h}_{{j}},W^+,W^-}} 
\end{align} 
The corrections from Higgs scalars in the loops read 
\begin{align} 
\Pi^{h,\phi}_{i,j}(p^2) &=- \sum_{a=1}^{2}{A_0\Big(m^2_{H^-_{{a}}}\Big)} {\Gamma_{\check{h}_{{i}},\check{h}_{{j}},H^+_{{a}},H^-_{{a}}}}  +\sum_{a=1}^{2}\sum_{b=1}^{2}{B_0\Big(p^{2},m^2_{H^-_{{a}}},m^2_{H^-_{{b}}}\Big)} {\Gamma^*_{\check{h}_{{j}},H^+_{{a}},H^-_{{b}}}} {\Gamma_{\check{h}_{{i}},H^+_{{a}},H^-_{{b}}}} \nonumber \\
&-\frac{1}{2} \sum_{a=1}^{4}{A_0\Big(m^2_{A^0_{{a}}}\Big)} {\Gamma_{\check{h}_{{i}},\check{h}_{{j}},A^0_{{a}},A^0_{{a}}}}  -\frac{1}{2} \sum_{a=1}^{4}{A_0\Big(m^2_{h_{{a}}}\Big)} {\Gamma_{\check{h}_{{i}},\check{h}_{{j}},h_{{a}},h_{{a}}}}  \nonumber \\ 
 &+\frac{1}{2} \sum_{a=1}^{4}\sum_{b=1}^{4}{B_0\Big(p^{2},m^2_{A^0_{{a}}},m^2_{A^0_{{b}}}\Big)} {\Gamma^*_{\check{h}_{{j}},A^0_{{a}},A^0_{{b}}}} {\Gamma_{\check{h}_{{i}},A^0_{{a}},A^0_{{b}}}}  \nonumber \\ 
 &+\sum_{a=1}^{4}\sum_{b=1}^{4}{B_0\Big(p^{2},m^2_{h_{{a}}},m^2_{A^0_{{b}}}\Big)} {\Gamma^*_{\check{h}_{{j}},h_{{a}},A^0_{{b}}}} {\Gamma_{\check{h}_{{i}},h_{{a}},A^0_{{b}}}} +\frac{1}{2} \sum_{a=1}^{4}\sum_{b=1}^{4}{B_0\Big(p^{2},m^2_{h_{{a}}},m^2_{h_{{b}}}\Big)} {\Gamma^*_{\check{h}_{{j}},h_{{a}},h_{{b}}}} {\Gamma_{\check{h}_{{i}},h_{{a}},h_{{b}}}}  \nonumber \\ 
\end{align} 
The mixed contributions involving scalars and vector bosons are given by
\begin{align}
\Pi^{h,V\phi}_{i,j}(p^2) &=+2 \sum_{b=1}^{2}{\Gamma^*_{\check{h}_{{j}},W^+,H^-_{{b}}}} {\Gamma_{\check{h}_{{i}},W^+,H^-_{{b}}}} {F_0\Big(p^{2},m^2_{H^-_{{b}}},m^2_{W^-}\Big)}  +\sum_{b=1}^{4}{\Gamma^*_{\check{h}_{{j}},Z,A^0_{{b}}}} {\Gamma_{\check{h}_{{i}},Z,A^0_{{b}}}} {F_0\Big(p^{2},m^2_{A^0_{{b}}},m^2_{Z}\Big)}  
\end{align}
The corrections due to charginos and neutralinos read
\begin{align} 
\Pi^{h,\chi}_{i,j}(p^2) &=-2 \sum_{a=1}^{2}m_{\tilde{\chi}^-_{{a}}} \sum_{b=1}^{2}{B_0\Big(p^{2},m^2_{\tilde{\chi}^-_{{a}}},m^2_{\tilde{\chi}^-_{{b}}}\Big)} m_{\tilde{\chi}^-_{{b}}} \Big({\Gamma^{L*}_{\check{h}_{{j}},\tilde{\chi}^+_{{a}},\tilde{\chi}^-_{{b}}}} {\Gamma^R_{\check{h}_{{i}},\tilde{\chi}^+_{{a}},\tilde{\chi}^-_{{b}}}}  + {\Gamma^{R*}_{\check{h}_{{j}},\tilde{\chi}^+_{{a}},\tilde{\chi}^-_{{b}}}} {\Gamma^L_{\check{h}_{{i}},\tilde{\chi}^+_{{a}},\tilde{\chi}^-_{{b}}}} \Big)  \nonumber \\ 
 &+\sum_{a=1}^{2}\sum_{b=1}^{2}{G_0\Big(p^{2},m^2_{\tilde{\chi}^-_{{a}}},m^2_{\tilde{\chi}^-_{{b}}}\Big)} \Big({\Gamma^{L*}_{\check{h}_{{j}},\tilde{\chi}^+_{{a}},\tilde{\chi}^-_{{b}}}} {\Gamma^L_{\check{h}_{{i}},\tilde{\chi}^+_{{a}},\tilde{\chi}^-_{{b}}}}  + {\Gamma^{R*}_{\check{h}_{{j}},\tilde{\chi}^+_{{a}},\tilde{\chi}^-_{{b}}}} {\Gamma^R_{\check{h}_{{i}},\tilde{\chi}^+_{{a}},\tilde{\chi}^-_{{b}}}} \Big) \nonumber \\
 &- \sum_{a=1}^{6}m_{\tilde{\chi}^0_{{a}}} \sum_{b=1}^{6}{B_0\Big(p^{2},m^2_{\tilde{\chi}^0_{{a}}},m^2_{\tilde{\chi}^0_{{b}}}\Big)} m_{\tilde{\chi}^0_{{b}}} \Big({\Gamma^{L*}_{\check{h}_{{j}},\tilde{\chi}^0_{{a}},\tilde{\chi}^0_{{b}}}} {\Gamma^R_{\check{h}_{{i}},\tilde{\chi}^0_{{a}},\tilde{\chi}^0_{{b}}}}  + {\Gamma^{R*}_{\check{h}_{{j}},\tilde{\chi}^0_{{a}},\tilde{\chi}^0_{{b}}}} {\Gamma^L_{\check{h}_{{i}},\tilde{\chi}^0_{{a}},\tilde{\chi}^0_{{b}}}} \Big)  \nonumber \\ 
 &+\frac{1}{2} \sum_{a=1}^{6}\sum_{b=1}^{6}{G_0\Big(p^{2},m^2_{\tilde{\chi}^0_{{a}}},m^2_{\tilde{\chi}^0_{{b}}}\Big)} \Big({\Gamma^{L*}_{\check{h}_{{j}},\tilde{\chi}^0_{{a}},\tilde{\chi}^0_{{b}}}} {\Gamma^L_{\check{h}_{{i}},\tilde{\chi}^0_{{a}},\tilde{\chi}^0_{{b}}}}  + {\Gamma^{R*}_{\check{h}_{{j}},\tilde{\chi}^0_{{a}},\tilde{\chi}^0_{{b}}}} {\Gamma^R_{\check{h}_{{i}},\tilde{\chi}^0_{{a}},\tilde{\chi}^0_{{b}}}} \Big) \nonumber \\ 
 \end{align} 
The corrections due to Sfermions are
\begin{align} 
\Pi^{h,\tilde{f}}_{i,j}(p^2) &=- \sum_{a=1}^{3}{A_0\Big(m^2_{\tilde{\nu}_{{a}}}\Big)} {\Gamma_{\check{h}_{{i}},\check{h}_{{j}},\tilde{\nu}^*_{{a}},\tilde{\nu}_{{a}}}}  +\sum_{a=1}^{3}\sum_{b=1}^{3}{B_0\Big(p^{2},m^2_{\tilde{\nu}_{{a}}},m^2_{\tilde{\nu}_{{b}}}\Big)} {\Gamma^*_{\check{h}_{{j}},\tilde{\nu}^*_{{a}},\tilde{\nu}_{{b}}}} {\Gamma_{\check{h}_{{i}},\tilde{\nu}^*_{{a}},\tilde{\nu}_{{b}}}} \nonumber \\ 
 &-3 \sum_{a=1}^{6}{A_0\Big(m^2_{\tilde{d}_{{a}}}\Big)} {\Gamma_{\check{h}_{{i}},\check{h}_{{j}},\tilde{d}^*_{{a}},\tilde{d}_{{a}}}}  - \sum_{a=1}^{6}{A_0\Big(m^2_{\tilde{e}_{{a}}}\Big)} {\Gamma_{\check{h}_{{i}},\check{h}_{{j}},\tilde{e}^*_{{a}},\tilde{e}_{{a}}}}  \nonumber \\ 
 &-3 \sum_{a=1}^{6}{A_0\Big(m^2_{\tilde{u}_{{a}}}\Big)} {\Gamma_{\check{h}_{{i}},\check{h}_{{j}},\tilde{u}^*_{{a}},\tilde{u}_{{a}}}}  +3 \sum_{a=1}^{6}\sum_{b=1}^{6}{B_0\Big(p^{2},m^2_{\tilde{d}_{{a}}},m^2_{\tilde{d}_{{b}}}\Big)} {\Gamma^*_{\check{h}_{{j}},\tilde{d}^*_{{a}},\tilde{d}_{{b}}}} {\Gamma_{\check{h}_{{i}},\tilde{d}^*_{{a}},\tilde{d}_{{b}}}}  \nonumber \\ 
 &+\sum_{a=1}^{6}\sum_{b=1}^{6}{B_0\Big(p^{2},m^2_{\tilde{e}_{{a}}},m^2_{\tilde{e}_{{b}}}\Big)} {\Gamma^*_{\check{h}_{{j}},\tilde{e}^*_{{a}},\tilde{e}_{{b}}}} {\Gamma_{\check{h}_{{i}},\tilde{e}^*_{{a}},\tilde{e}_{{b}}}} +3 \sum_{a=1}^{6}\sum_{b=1}^{6}{B_0\Big(p^{2},m^2_{\tilde{u}_{{a}}},m^2_{\tilde{u}_{{b}}}\Big)} {\Gamma^*_{\check{h}_{{j}},\tilde{u}^*_{{a}},\tilde{u}_{{b}}}} {\Gamma_{\check{h}_{{i}},\tilde{u}^*_{{a}},\tilde{u}_{{b}}}}  \nonumber \\ 
\end{align}
And those stemming from SM fermions are
\begin{align}  
\Pi^{h,f}_{i,j}(p^2) &= -6 \sum_{a=1}^{3}m_{d_{{a}}} \sum_{b=1}^{3}{B_0\Big(p^{2},m^2_{d_{{a}}},m^2_{d_{{b}}}\Big)} m_{d_{{b}}} \Big({\Gamma^{L*}_{\check{h}_{{j}},\bar{d}_{{a}},d_{{b}}}} {\Gamma^R_{\check{h}_{{i}},\bar{d}_{{a}},d_{{b}}}}  + {\Gamma^{R*}_{\check{h}_{{j}},\bar{d}_{{a}},d_{{b}}}} {\Gamma^L_{\check{h}_{{i}},\bar{d}_{{a}},d_{{b}}}} \Big) \nonumber \\
& +3 \sum_{a=1}^{3}\sum_{b=1}^{3}{G_0\Big(p^{2},m^2_{d_{{a}}},m^2_{d_{{b}}}\Big)} \Big({\Gamma^{L*}_{\check{h}_{{j}},\bar{d}_{{a}},d_{{b}}}} {\Gamma^L_{\check{h}_{{i}},\bar{d}_{{a}},d_{{b}}}}  + {\Gamma^{R*}_{\check{h}_{{j}},\bar{d}_{{a}},d_{{b}}}} {\Gamma^R_{\check{h}_{{i}},\bar{d}_{{a}},d_{{b}}}} \Big) \nonumber \\ 
 &-2 \sum_{a=1}^{3}m_{e_{{a}}} \sum_{b=1}^{3}{B_0\Big(p^{2},m^2_{e_{{a}}},m^2_{e_{{b}}}\Big)} m_{e_{{b}}} \Big({\Gamma^{L*}_{\check{h}_{{j}},\bar{e}_{{a}},e_{{b}}}} {\Gamma^R_{\check{h}_{{i}},\bar{e}_{{a}},e_{{b}}}}  + {\Gamma^{R*}_{\check{h}_{{j}},\bar{e}_{{a}},e_{{b}}}} {\Gamma^L_{\check{h}_{{i}},\bar{e}_{{a}},e_{{b}}}} \Big)  \nonumber \\ 
 &+\sum_{a=1}^{3}\sum_{b=1}^{3}{G_0\Big(p^{2},m^2_{e_{{a}}},m^2_{e_{{b}}}\Big)} \Big({\Gamma^{L*}_{\check{h}_{{j}},\bar{e}_{{a}},e_{{b}}}} {\Gamma^L_{\check{h}_{{i}},\bar{e}_{{a}},e_{{b}}}}  + {\Gamma^{R*}_{\check{h}_{{j}},\bar{e}_{{a}},e_{{b}}}} {\Gamma^R_{\check{h}_{{i}},\bar{e}_{{a}},e_{{b}}}} \Big)\nonumber \\ 
 &-6 \sum_{a=1}^{3}m_{u_{{a}}} \sum_{b=1}^{3}{B_0\Big(p^{2},m^2_{u_{{a}}},m^2_{u_{{b}}}\Big)} m_{u_{{b}}} \Big({\Gamma^{L*}_{\check{h}_{{j}},\bar{u}_{{a}},u_{{b}}}} {\Gamma^R_{\check{h}_{{i}},\bar{u}_{{a}},u_{{b}}}}  + {\Gamma^{R*}_{\check{h}_{{j}},\bar{u}_{{a}},u_{{b}}}} {\Gamma^L_{\check{h}_{{i}},\bar{u}_{{a}},u_{{b}}}} \Big)  \nonumber \\ 
 &+3 \sum_{a=1}^{3}\sum_{b=1}^{3}{G_0\Big(p^{2},m^2_{u_{{a}}},m^2_{u_{{b}}}\Big)} \Big({\Gamma^{L*}_{\check{h}_{{j}},\bar{u}_{{a}},u_{{b}}}} {\Gamma^L_{\check{h}_{{i}},\bar{u}_{{a}},u_{{b}}}}  + {\Gamma^{R*}_{\check{h}_{{j}},\bar{u}_{{a}},u_{{b}}}} {\Gamma^R_{\check{h}_{{i}},\bar{u}_{{a}},u_{{b}}}} \Big) \nonumber \\ 
\end{align} 
The Passarino-Veltman integral $B_0$ can be expressed by
\begin{equation}
 B_0(p, m_1, m_2)  = - \ln\left(\frac{p^2}{Q^2}\right) - f_B(x_+) - f_B(x_-) \, , 
\end{equation}
with $f_B(x)  = \ln(1-x) - x\ln(1-x^{-1})-1$, 
$ x_{\pm}\ =\ \frac{s \pm \sqrt{s^2 - 4p^2(m_1^2-i\varepsilon)}}{2p^2}$, 
and $s=p^2-m_2^2+m_1^2$.
All appearing integrals can be expressed in terms of $A_0$ and $B_0$
\begin{align}
B_1(p, m_1,m_2)  =& {\frac{1}{2p^2}}\Big( A_0(m_2) - A_0(m_1) + (p^2 +m_1^2 -m_2^2) B_0(p, m_1, m_2)\Big) \\
B_{22}(p, m_1,m_2) =& \frac{1}{6} \Big(\frac{1}{2}\big(A_0(m_1)+A_0(m_2)\big)
+\left(m_1^2+m_2^2-\frac{1}{2}p^2\right)B_0(p,m_1,m_2)\nonumber \\ 
&+ \frac{m_2^2-m_1^2}{2p^2} \big[A_0(m_2)-A_0(m_1)-(m_2^2-m_1^2) B_0(p,m_1,m_2)\big] \nonumber\\ 
&  +  m_1^2 + m_2^2 -\frac{1}{3}p^2\,\Big) \\
F_0(p,m_1,m_2) =& A_0(m_1)-2A_0(m_2)- (2p^2+2m^2_1-m^2_2) B_0(p,m_1,m_2)\ , \\ 
G_0(p,m_1,m_2) =& (p^2-m_1^2-m_2^2)B_0(p,m_1,m_2)-A_0(m_1)-A_0(m_2)\ .
\end{align}

In our numerical analysis the one-loop scalar Higgs masses are then calculated by taking the real part of the poles
of the corresponding propagator matrices
\begin{equation}
\mathrm{Det}\left[ p^2_i \mathbf{1} - m^{2,h}_{1L}(p^2) \right] = 0,
\label{eq:propagator}
\end{equation}
where
\begin{equation}
 m^{2,h}_{1L}(p^2) = \tilde{m}^{2,h}_T -  \Pi_{hh}(p^2) .
\end{equation}
Here, \(\tilde{m}^{2,h}_T\) is the tree-level mass matrix from eq.\
(\ref{eq:HiggsTreeMass}) where the parameters obtained from the one-loop
tadpole equations are inserted. Equation (\ref{eq:propagator}) has to be solved
for each eigenvalue $p^2=m^2_i$. The same procedure is also applied for
the pseudo scalar Higgs bosons.

\subsubsection*{Dominant Two-loop corrections}
In addition to the full correction at one-loop we have also added the dominant two-loop corrections due to stops known from the MSSM
presented in \cite{Brignole:2001jy,Brignole:2002bz,Dedes:2002dy,Dedes:2003km}. Because of the presence of these corrections, the renormalisation conditions eq.~(\ref{eq:onelooptad}) are modified to
\begin{align}
 \Theta_i + \delta \Theta^{(1)}_i + \delta \Theta^{(2)}_i = 0 \, , \hspace{1cm} i=d,u  \\
 \Theta_i + \delta \Theta^{(1)}_i = 0 \, , \hspace{1cm} i=s,\bar{s}
\end{align}
and the 2-loop self-energy of the CP even Higgs takes the form
\begin{equation}
\Pi^{(2L)} = \left(\begin{array}{cc} \Pi^{(2L),MSSM} & {\bf 0} \\ {\bf 0} & {\bf 0} \end{array}\right) 
\end{equation}
Here, ${\bf 0}$ is a $2\times 2$ matrix carrying only 0's, and $\Pi^{(2L),MSSM}$ are the two-loop MSSM self-energy contributions to the Higgs.
\end{appendix}

\bibliography{NMSSM}
\bibliographystyle{ArXiv}

\end{document}